%% file: main.tex
\documentclass[a4paper,11pt]{article}
\usepackage{jheppub}
\usepackage{graphicx}
\usepackage{tabularx,booktabs}
\usepackage{subcaption}
\usepackage{dcolumn}
\usepackage{bm}
\usepackage{hyperref} 
\usepackage{orcidlink}
\usepackage{multirow}
\usepackage{float}
\usepackage{tikz}
\usetikzlibrary{decorations.pathmorphing,arrows.meta}

\begin{document}

\title{Sideband Structure of Axion Electrodynamics}
\author[a,b,1]{Run-Min Yao\note{Corresponding author.}\orcidlink{0000-0002-7807-6713}}
\author[b,c]{Xiao-Jun Bi\orcidlink{0000-0002-5334-9754}}
\author[c]{Peng-Fei Yin\orcidlink{0000-0001-6514-5196}}
\author[a,b,d,1]{Qing-Guo Huang\orcidlink{0000-0003-1584-345X}}

\emailAdd{yaorunmin@ucas.ac.cn}
\emailAdd{bixj@ihep.ac.cn}
\emailAdd{yinpf@ihep.ac.cn}
\emailAdd{huangqg@itp.ac.cn}

\affiliation[a]{School of Fundamental Physics and Mathematical Sciences, Hangzhou Institute for Advanced
Study, UCAS, Hangzhou 310024, China}
\affiliation[b]{University of Chinese Academy of Sciences, Beijing 100049, China}
\affiliation[c]{Key Laboratory of Particle Astrophysics, Institute of High Energy Physics, Chinese Academy of Sciences, Beijing, China}
\affiliation[d]{Institute of Theoretical Physics, Chinese Academy
of Sciences}

\date{\today}

\abstract{We develop a Floquet--Bloch sideband formulation of the linearized Maxwell--axion system in a coherent periodic axion background. Linearizing around prescribed magnetic and axion fields, we show that the pump generates a sideband ladder of photon and axion branches. Near an isolated folded degeneracy, this ladder reduces to a two-mode crossing whose algebra is fixed by the symplectic signatures of the colliding modes. In temporal fixed-momentum evolution, same-Krein-sign collisions give stable avoided crossings, whereas opposite-sign collisions give parametric instabilities, unifying the axion-photon difference channel with the Mathieu and Masaki-Aoki-Soda resonances. In stationary fixed-frequency transfer, the corresponding flux signatures distinguish bounded forward conversion from forward-backward stop bands and distributed reflection. Ray projection of a temporal pump gives a related but local WKB description of driven forward mixing, with an effective wavenumber distinct from the true axion momentum. External-field diagrams reproduce the sideband selection rules, and full temporal monodromy calculations verify the instability topology and finite-coupling shifts.}

\maketitle

\input{content}

\begin{acknowledgments}
This work is supported by the National Natural Science Foundation of China under Grants Nos. 12547110, 12475065, 12250010, 12447105, and 12575113.
\end{acknowledgments}
\input{appendix}

\bibliographystyle{JHEP}
\bibliography{ref}

\end{document}

%% file: content.tex
\input{sections/introduction}
\clearpage
\input{notation_table}
\clearpage
\input{sections/linearized_model}
\input{sections/evolution_symplectic}
\input{sections/floquet_bloch}
\input{sections/temporal_fixed_k}
\input{sections/spatial_fixed_frequency}
\input{sections/phenomenological_classification}
\input{sections/numerical_illustrations}
\input{sections/conclusions}

%% file: sections/introduction.tex
\section{Introduction}

Axions and axion-like particles are well-motivated extensions of the Standard Model, driven in particular by the strong CP problem~\cite{Peccei:1977hh,Weinberg:1977ma,Wilczek:1977pj,Kim:1979if,Shifman:1979if,Zhitnitsky:1980tq,Dine:1981rt,DiLuzio:2020wdo} and by the possibility that light pseudoscalars constitute some or all of the dark matter~\cite{Preskill:1982cy,Abbott:1982af,Dine:1982ah,Turner:1985si,Marsh:2015xka,Adams:2022pbo}. Their electromagnetic coupling,
\begin{equation}
    \mathcal L_{\rm int}=-\frac{1}{4}g_{a\gamma}aF_{\mu\nu}\widetilde F^{\mu\nu}
    =g_{a\gamma}a\,\mathbf E\cdot\mathbf B,
\end{equation}
allows electromagnetic waves and magnetized media to probe axionic degrees of freedom~\cite{Sikivie:1983ip,Raffelt:1987im,Wilczek:1987mv,Graham:2015ouw,Irastorza:2018dyq,CAST:2017uph,ADMX:2019uok}. At the level of wave propagation, this interaction leads to axion electrodynamics: electromagnetic perturbations propagate in a spacetime-dependent axion background, while axion perturbations can mix with photons in an external magnetic field. When the axion field is coherent, it is intrinsically periodic, with a temporal oscillation set by the axion mass and, for a moving component, a true spatial Fourier scale set by its momentum~\cite{Marsh:2015xka}.

The physical question being posed determines the natural formulation of this common system. In a slowly varying background, geometric optics gives birefringence, polarization phase modulation, and polarization-dependent ray deflection~\cite{Carroll:1989vb,Carroll:1998zi,Harari:1992ea,McDonald:2019wou,Plascencia:2017kca,Blas:2019qqp,Mcdonald:2020hjm}; cosmological applications use the same limit to search for oscillatory or anisotropic CMB birefringence~\cite{Liu:2016dcg,Fedderke:2019ajk,Minami:2020odp,Jain:2022jrp}. At fixed frequency, magnetized propagation is formulated as axion-photon conversion through level crossing or phase matching~\cite{Raffelt:1987im,Pshirkov:2007st,Hook:2018iia,Raffelt:1996wa,Kuster:2008zz,Primakoff:1951iae,Sikivie:1983ip}. At fixed momentum, a homogeneous oscillating background is a temporal pump that can produce stimulated decay or parametric amplification~\cite{Tkachev:2014dpa,Hertzberg:2018zte,Caputo:2018ljp,Caputo:2018vmy,Arza:2018dcy,Masaki:2019ggg}. A stationary periodic medium instead calls for Bloch waves and a spatial transfer matrix. These formulations differ in their conserved label, evolution variable, and boundary or initial data, although they begin from the same linearized equations.

Periodic axion backgrounds reveal a common sideband structure behind these descriptions. The pump supplies discrete spacetime four-momentum, replicating the free photon and axion branches into a ladder whose folded crossings identify possible resonances. A homogeneous temporal pump gives a fixed-\(k\) Floquet initial-value problem in which stability is diagnosed by temporal growth or bounded evolution. A static spatial modulation gives a fixed-\(\omega\) transfer problem and determines conversion, transmission, and reflection. Evaluating a temporally oscillating pump along a propagation ray instead gives a local WKB envelope problem. Along that ray, the pump phase acts as an effective spatial drive only within the ray approximation. Thus similar local crossing equations can describe different observables; the conserved label, evolution variable, and boundary or initial data select their physical meaning.

\smallskip
\noindent\textbf{Main results and organization.}

This paper formulates the linearized transverse Maxwell-axion system as a Floquet-Bloch sideband problem and uses the real symplectic structure of the corresponding one-dimensional reductions to classify isolated sideband crossings. The colliding branches carry a conserved symplectic signature: the temporal Krein sign in a fixed-\(k\) Floquet problem or the spatial flux sign in a fixed-\(\omega\) transfer problem. For an isolated crossing with a nonzero projected coupling, same-sign branches give a compact normal form and bounded conversion or avoided crossings, whereas opposite-sign branches give a non-compact normal form and temporal instability tongues or spatial stop bands, depending on the evolution problem. After removal of a common phase, these two cases have \(SU(2)\)-type and \(SU(1,1)\)-type algebra, respectively.

The temporal specialization places the photon Mathieu instability and the Masaki--Aoki--Soda (MAS) sum-frequency resonance~\cite{Masaki:2019ggg} in the class of opposite-Krein-sign instabilities, while same-sign axion-photon crossings give bounded beating. The spatial specialization gives the corresponding fixed-\(\omega\) transfer classification: forward-forward flux crossings produce bounded spatial conversion, while forward-backward crossings produce Bragg stop bands and evanescent Bloch behavior. The analysis also separates genuine stationary spatial phase matching, controlled by the wavenumber \(Q\) of a static modulation, from ray-projected forward conversion, controlled by the effective wavenumber \(Q_{\rm eff}=m_a/v_g\), where \(v_g\) is the group velocity. When an equal-time static proxy is introduced, the true axion momentum \(Q_{\rm real}=m_a v_a\), where \(v_a\) is the axion velocity, sets its spatial scale; the underlying freely moving axion component does not define a stationary fixed-\(\omega\) grating.

Our analysis assumes prescribed axion and magnetic backgrounds, a local real plasma response, lossless propagation, and isolated near-resonant crossings. The ray-projected construction additionally excludes plasma cutoffs and turning points. Within this regime, the signature classification encompasses ordinary axion-photon conversion, photon Mathieu instability, the MAS instability, driven forward conversion, and stationary spatial Bragg reflection. Full real-block monodromy calculations illustrate the temporal instability channels without rotating-wave or sideband truncation. The stationary spatial equations retain the full transfer problem; only the isolated-crossing classification and ray-projected reduction are local.

Section~\ref{sec:model} defines the linearized Maxwell-axion model and assumptions. Section~\ref{sec:evolution} introduces the one-dimensional reduction and its symplectic structure. Section~\ref{sec:floquetBloch} constructs the Floquet-Bloch ladder, local signature normal form, and external-field interpretation. Sections~\ref{sec:floquet} and~\ref{sec:transfer} specialize the framework to temporal fixed-\(k\) Floquet evolution and fixed-\(\omega\) spatial transfer. Section~\ref{sec:pheno} summarizes the phenomenological classification, including the \(Q_{\rm real}\) versus \(Q_{\rm eff}\) distinction and the WKB mapping to inhomogeneous plasma conversion. Section~\ref{sec:numerics} presents temporal monodromy illustrations, and Section~\ref{sec:conclusions} concludes. The appendices provide explicit real-block symplectic constructions, pump-frame comments, the sideband-chain Schur complement and its positive-positive and MAS specializations, and the Jacobi--Anger reorganization of the temporal ladder.


%% file: notation_table.tex
\begin{table}[p]
\centering
\caption{Principal notation used throughout the paper.}
\label{tab:notation}
\small
\begin{tabularx}{\textwidth}{llX}
\toprule
Symbol & Dimensions & Meaning \\
\midrule
$g\equiv g_{a\gamma}$ & $\mathrm{mass}^{-1}$ & Axion-photon coupling \\
$\omega_{\rm pl}$ & $\mathrm{mass}$ & Plasma frequency (effective photon mass) \\
$m_a$ & $\mathrm{mass}$ & Axion mass \\
$\bar{\mathbf B}$ & $\mathrm{mass}^2$ & External magnetic field \\
$\bar a(t,\mathbf x)$ & $\mathrm{mass}$ & Background axion field \\
$\Omega_{\rm p}$ & $\mathrm{mass}$ & Pump frequency \\
$Q_{\rm p}$ & $\mathrm{mass}$ & Pump spatial momentum \\
$K^\mu=(\Omega_{\rm p},Q_{\rm p})$ & $\mathrm{mass}$ & Pump four-wavevector for a spacetime-periodic background \\
$Q_{\rm real}$ & $\mathrm{mass}$ & True spatial wavenumber of a nonrelativistic axion component, $Q_{\rm real}=m_a v_a$ \\
$Q_{\rm eff}$ & $\mathrm{mass}$ & Ray-projected pump wavenumber for forward WKB propagation, $Q_{\rm eff}\simeq m_a/v_g$ \\
\midrule
$s$ & $\mathrm{mass}^{-1}$ & Abstract evolution variable ($t$ for fixed-$k$ Floquet stability, $z$ for fixed-$\omega$ transfer) \\
$\lambda_\eta$ & --- & Dimensionless Fourier label conjugate to the dimensionless coordinate $\eta$ \\
$\varepsilon,\kappa$ & $\mathrm{mass}$ & Spacetime Bloch labels in the covariant Floquet-Bloch expansion \\
$\varepsilon_F$ & $\mathrm{mass}$ & Temporal quasi-frequency in the homogeneous fixed-$k$ problem \\
$n$ & --- & Integer sideband rung label \\
$N$ & --- & Signed harmonic difference between two near-resonant rungs; $|N|$ is the sideband order \\
$\sigma_i=\pm1$ & --- & Frequency sign in temporal problems, or propagation/flux label in spatial transfer problems \\
\midrule
$\Delta_t$ & $\mathrm{mass}$ & Generic temporal detuning in quasi-frequency units \\
$\Delta_{ij}^{(\sigma_i,\sigma_j;N)}$ & $\mathrm{mass}$ & Temporal channel detuning for species $i,j\in\{a,\gamma\}$ and signs $\sigma_i,\sigma_j$ \\
$\Delta_z$ & $\mathrm{mass}$ & Spatial phase-mismatch or Bragg detuning in fixed-frequency transfer \\
$\Xi_{ij}^{(\sigma_i,\sigma_j;N)}$ & $\mathrm{mass}^2$ & Channel-specific shell-space endpoint matrix element before action normalization \\
$\Sigma_i^{(\sigma_i,\sigma_j;N)}$ & $\mathrm{mass}^2$ & Diagonal shell-space self-energy correction from eliminated off-resonant rungs \\
$\mathcal G_B(n)$ & $\mathrm{mass}^2$ & Temporal same-rung magnetic bridge between \(a_n\) and the Cartesian photon coefficient \(A_{x,n}\) \\
$\tau,\tau_R$ & $\mathrm{mass}^2$ & Directed Cartesian photon hopping and its real symmetric representative after rung rephasing \\
$\mathcal B_z$ & $\mathrm{mass}^2$ & Spatial same-rung magnetic bridge between \(a_n\) and \(A_{x,n}\) \\
$\mu_{\rm eff}$ & $\mathrm{mass}$ & Generic action-normalized temporal slow-envelope coupling \\
$\mu_{ij}^{(\sigma_i,\sigma_j;N)}$ & $\mathrm{mass}$ & Channel-specific temporal action-normalized endpoint coupling \\
$g_z$ & $\mathrm{mass}$ & Spatially normalized endpoint coupling in the $z$-evolution problem \\
\bottomrule
\end{tabularx}
\end{table}

%% file: sections/linearized_model.tex
\section{Linearized Maxwell-axion model}
\label{sec:model}

We work in natural units $\hbar=c=1$ and restrict the analysis to the transverse photon subspace in radiation gauge. The longitudinal plasma response and the scalar potential lie outside the near-resonant subspace of interest and are omitted from the dynamical equations that follow.

The linearized Maxwell-axion system is studied under the following assumptions, which together guarantee a real symplectic structure for lossless propagation:
\begin{itemize}
    \item The external magnetic field $\bar{\mathbf B}$ is treated as a prescribed classical background; we neglect backreaction of the perturbations on the background fields.
    \item The background axion field $\bar a(t,\mathbf x)$ is similarly a prescribed classical field.
    \item The plasma response enters only through a local, real effective mass squared $\omega_{\rm pl}^2$, appropriate for a cold, collisionless, unmagnetized plasma after projecting onto the transverse channel.
    \item The medium is taken to be lossless and free of nonlocal dielectric response. In this regime the equations of motion admit a real symplectic (or, upon appropriate reduction, Hamiltonian) formulation.
    \item Whenever a ray-projected WKB reduction is used, propagation is assumed to occur away from plasma cutoffs and turning points. The exact temporal monodromy and stationary spatial transfer constructions do not require this WKB assumption.
\end{itemize}

Absorptive media require a non-Hermitian formulation; nonlocal response requires an extended dispersive description; and near-cutoff propagation requires a turning-point treatment beyond WKB. These extensions lie beyond the scope of this work.

With the above assumptions, linearization of the axion-coupled Maxwell equations~\cite{Wilczek:1987mv,Raffelt:1987im} yields
\begin{align}
\partial_t^2\delta\mathbf A-\nabla^2\delta\mathbf A+
\omega_{\rm pl}^2\delta\mathbf A
&=
g_{a\gamma}\bar{\mathbf B}\,\partial_t\delta a
+g_{a\gamma}\partial_t\bar a\,\nabla\times\delta\mathbf A
-g_{a\gamma}\nabla\bar a\times\partial_t\delta\mathbf A,
\label{eq:maxwell}\\
\partial_t^2\delta a-\nabla^2\delta a+m_a^2\delta a
&=-\,g_{a\gamma}\bar{\mathbf B}\cdot\partial_t\delta\mathbf A.
\label{eq:axion}
\end{align}
The left-hand sides describe free propagation of the transverse photon perturbation $\delta\mathbf A$ and the axion perturbation $\delta a$. The right-hand sides contain the axion-photon interaction terms: the coupling proportional to $\bar{\mathbf B}$ is the standard Primakoff vertex~\cite{Primakoff:1951iae,Sikivie:1983ip,Raffelt:1987im}, while the terms involving $\partial_t\bar a$ and $\nabla\bar a$ arise from the spacetime gradients of the periodic axion background.

The scalar potential and longitudinal electric field omitted here are constrained variables rather than independent transverse wave degrees of freedom. In a cold, collisionless, unmagnetized plasma, their singularities reside in the longitudinal response, whereas the resonances studied below involve transverse photon branches and the propagating axion perturbation. Away from longitudinal plasma resonances, solving Gauss's law induces only non-resonant local corrections in the parameter range considered here.

This reduction can fail near a longitudinal plasmon--axion resonance~\cite{Caputo:2020quz}, in magnetized or anisotropic/nonlocal dielectric media~\cite{Hook:2018iia,Millar:2021gzs,Witte:2021arp}, or when a longitudinal mode becomes phase matched with the sideband ladder.

We take the propagation direction to be along the \(z\)-axis and restrict the prescribed magnetic field to a transverse direction, \(\bar{\mathbf B}=B\hat{\mathbf x}\). The photon component \(A_x\), aligned with \(\bar{\mathbf B}\), couples to \(\delta a\) through the Primakoff vertex. The orthogonal transverse component \(A_y\) couples to \(A_x\) through the axion background. All subsequent analysis is carried out in this closed transverse channel. A longitudinal magnetic component would require retaining the constrained longitudinal sector and lies outside the present model. For notational simplicity we write \(g\equiv g_{a\gamma}\) from this point onward.

After polarization reduction, Eqs.~\eqref{eq:maxwell}--\eqref{eq:axion} are partial differential equations in $(t,z)$. Different physical questions posed to this system correspond to different mathematical reductions of the same underlying dynamics. If the axion background $\bar a$ varies slowly compared to the wave periods, a WKB or geometric-optics expansion extracts helicity-dependent refractive indices and polarization rotation, the traditional language of axion birefringence~\cite{Carroll:1989vb,Carroll:1998zi,Harari:1992ea,McDonald:2019wou}. For spatially varying backgrounds, the same geometric-optics framework also describes polarization-dependent ray deflection~\cite{Plascencia:2017kca,Blas:2019qqp,Mcdonald:2020hjm}. For coherent ultralight backgrounds, the polarization-rotation viewpoint underlies CMB birefringence and washout searches~\cite{Liu:2016dcg,Fedderke:2019ajk,Minami:2020odp,Jain:2022jrp}. In a monochromatic boundary-value problem, the prescribed magnetic field produces the usual \(N=0\) Primakoff mixing between propagating axion and photon perturbations~\cite{Primakoff:1951iae,Raffelt:1987im,Sikivie:1983ip}; a separate static spatial profile of \(\bar a\) adds the periodic sideband couplings studied below.

This paper focuses on the regime in which $\bar a$ acts as a coherent periodic pump. We denote the uncoupled fixed-\(k\) axion and photon frequencies by \(\omega_a\) and \(\omega_\gamma\), and their on-shell fixed-\(\omega\) wavenumbers by \(k_a\) and \(k_\gamma\).

In the decoupled-photon limit, where the magnetic-field-induced mixing between the photon and axion perturbations is neglected, \eqref{eq:maxwell} reduces to a Mathieu equation. The photon sector then exhibits parametric instability under the condition $2\omega_\gamma\simeq N m_a$~\cite{Tkachev:2014dpa,Hertzberg:2018zte,Caputo:2018ljp,Caputo:2018vmy,Arza:2018dcy}. Masaki, Aoki, and Soda~\cite{Masaki:2019ggg} extended this analysis to the fully coupled axion--photon system in a magnetic field and identified an instability tongue governed by the sum-frequency condition $\omega_a+\omega_\gamma\simeq N m_a$. In the ray-projected fixed-frequency problem studied in our previous work~\cite{Yao:2026yez}, we examined the complementary difference-frequency resonance $k_a-k_\gamma\simeq N m_a$, which produces stable driven mixing rather than exponential growth. The present paper places these phenomena in a single Floquet-Bloch framework. The pump-induced sideband ladder identifies the resonant crossings, while their symplectic signatures and conserved labels distinguish parametric instability, sum-frequency coupling, and driven resonance.

The periodic axion background is parameterized, for a single coherent component propagating along $z$, as
\begin{equation}
    \bar a(t,z)=a_0\cos(\Omega_{\rm p} t-Q_{\rm p}z+\varphi),\qquad
    \Omega_{\rm p}^2-Q_{\rm p}^2=m_a^2,
    \label{eq:pump}
\end{equation}
where $\Omega_{\rm p}$ and $Q_{\rm p}$ denote the frequency and spatial momentum of the pump. The velocity parameter $v\equiv Q_{\rm p}/\Omega_{\rm p}$ controls the relative weight of temporal and spatial modulation in the pump. The on-shell condition in Eq.~\eqref{eq:pump} applies to a freely propagating massive axion component. Later spatial transfer calculations also use static spatially periodic media and ray-projected effective periods as auxiliary reductions, giving separate stationary or WKB realizations of the same local crossing algebra.

In each regime, the residual symmetry supplies a conserved quantity that reduces the $(t,z)$ partial differential equations to a one-dimensional problem in an evolution variable $s$. The homogeneous temporal problem conserves \(k\), whereas an exact stationary spatial medium conserves \(\omega\). Ray projection of a temporal pump instead gives a local WKB spatial problem.

%% file: sections/evolution_symplectic.tex
\section{Evolution variable and symplectic structure}
\label{sec:evolution}

Equations~\eqref{eq:maxwell}--\eqref{eq:axion} are partial differential equations in $(t,z)$. Extracting physical observables requires reducing the dynamics to a one-dimensional problem. We select an evolution variable $s$ and identify the conserved label supplied by the residual symmetry of the system.

\subsection{The \texorpdfstring{$\xi,\eta$}{xi, eta} coordinates and the conserved label}

The single-phase pump~\eqref{eq:pump} breaks the full spacetime translation symmetry, preserving invariance along the direction orthogonal to the phase
\begin{equation}
    \xi\equiv\Omega_{\rm p}t-Q_{\rm p}z .
\end{equation}
We introduce the complementary dimensionless coordinate
\begin{equation}
    \eta\equiv Q_{\rm p}t-\Omega_{\rm p}z,
\end{equation}
which spans the continuous symmetry direction. With this convention both $\xi$ and $\eta$ are phases. The chain rule
\begin{equation}
    \partial_t=\Omega_{\rm p}\partial_\xi+Q_{\rm p}\partial_\eta,\qquad
    \partial_z=-Q_{\rm p}\partial_\xi-\Omega_{\rm p}\partial_\eta
\end{equation}
converts the spacetime derivatives into derivatives along $\xi$ and $\eta$. Introducing the dimensionless Fourier labels \(\nu\) and \(\lambda_\eta\) conjugate to \(\xi\) and \(\eta\), respectively, a plane wave $e^{-i\omega t+ikz}$ can be written as $e^{-i\nu\xi-i\lambda_\eta\eta}$, which gives
\begin{equation}
    \omega=\nu\Omega_{\rm p}+\lambda_\eta Q_{\rm p},\qquad
    k=\nu Q_{\rm p}+\lambda_\eta\Omega_{\rm p}.
    \label{eq:nulambda}
\end{equation}
Because the background is independent of $\eta$, $\lambda_\eta$ is conserved in any reduction that respects the symmetry. Multiplying $\lambda_\eta$ by the appropriate scale gives the dimensional conserved quantity for a chosen slicing. For a homogeneous temporal pump,
\begin{equation}
    Q_{\rm p}=0\;\Rightarrow\; \xi=\Omega_{\rm p}t,
    \qquad \eta=-\Omega_{\rm p}z,
    \qquad k=\Omega_{\rm p}\lambda_\eta,
\end{equation}
so conservation of $\lambda_\eta$ is equivalent to conservation of the physical wave number $k$. For a strictly static spatial modulation, treated as an auxiliary periodic medium,
\begin{equation}
    \Omega_{\rm p}=0\;\Rightarrow\; \xi=-Q_{\rm p}z,
    \qquad \eta=Q_{\rm p}t,
    \qquad \omega=Q_{\rm p}\lambda_\eta,
\end{equation}
so conservation of $\lambda_\eta$ is equivalent to conservation of the physical frequency $\omega$. Equation~\eqref{eq:nulambda} therefore expresses the conserved labels $k$ and $\omega$ of the two principal reductions through one symmetry label while retaining their dimensional conversion to laboratory observables. Selecting $s$ then determines the physical problem: temporal evolution at fixed $k$, spatial evolution at fixed frequency, or a general traveling-wave Floquet problem at fixed $\lambda_\eta$, which is not pursued here.

\subsection{Parent \texorpdfstring{$s$}{s}-evolution system and symplectic form}

For the homogeneous temporal and stationary spatial reductions used in the main text, selecting $s$ and factoring out the conserved Fourier label gives a second-order system of the form
\begin{equation}
    \frac{\mathrm d^2 q}{\mathrm d s^2}+G\frac{\mathrm d q}{\mathrm d s}+V(s)q=0,
    \label{eq:parent}
\end{equation}
where $q\in\mathbb R^n$ is a vector of real field coordinates, \(G\) is constant, and $V(s)$ inherits the periodicity of the background. When the medium is lossless and the external fields are real, the coefficient matrices satisfy
\begin{equation}
    G^{\mathsf T}=-G,\qquad V^{\mathsf T}=V.
    \label{eq:GVconditions}
\end{equation}
Under these conditions, the system~\eqref{eq:parent} admits a real quadratic Lagrangian
\begin{equation}
    L_s=\frac12\dot q^{\mathsf T}\dot q-\frac12q^{\mathsf T}G\dot q-\frac12q^{\mathsf T}V(s)q,
    \qquad \dot{q}\equiv\frac{\mathrm d q}{\mathrm d s},
\end{equation}
with canonical momentum
\begin{equation}
    p=\dot q+\frac12Gq.
\end{equation}
Defining the phase-space variable $X=(q,p)^{\mathsf T}$ and the standard symplectic matrix
\begin{equation}
    J=\begin{pmatrix}0&I\\-I&0\end{pmatrix},
\end{equation}
the first-order form is
\begin{equation}
    \frac{\mathrm d X}{\mathrm d s}=M(s)X,\qquad
    M(s)=JK(s),\qquad
    K(s)^{\mathsf T}=K(s).
    \label{eq:HamiltonianForm}
\end{equation}
Let $\Phi_s(s)$ denote the fundamental matrix of~\eqref{eq:HamiltonianForm},
\begin{equation}
    \frac{\mathrm d\Phi_s}{\mathrm ds}=M(s)\Phi_s,\qquad
    \Phi_s(0)=I_{2n}.
\end{equation}
If the coefficients are periodic with period $L_s$ in the chosen evolution variable, the one-period map is
\begin{equation}
    F_s\equiv \Phi_s(L_s)
    =\mathcal P_s\exp\!\left[\int_0^{L_s}M(s)\,\mathrm ds\right],
    \label{eq:monodromy}
\end{equation}
where $\mathcal P_s$ denotes ordering along $s$. Since $M(s)$ lies in the symplectic Lie algebra $\mathfrak{sp}(2n,\mathbb R)$, the map satisfies $F_s\in\operatorname{Sp}(2n,\mathbb R)$. For temporal evolution $s=t$, $L_s=T$, this is the Floquet monodromy matrix. For spatial evolution $s=z$, $L_s=L$, it is the one-cell transfer matrix. For two complexified phase-space solutions \(X=(q,p)^{\mathsf T}\) and \(Y=(\tilde q,\tilde p)^{\mathsf T}\), the conserved Hermitian symplectic form is
\begin{equation}
    h(X,Y)\equiv \mathrm i X^\dagger J Y ,
    \label{eq:HermitianSymplecticForm}
\end{equation}
and \(h(u,u)\) assigns a real symplectic norm to an eigenmode \(u\) of \(F_s\). For temporal monodromy eigenmodes, its sign is the Krein signature~\cite{Chernyavsky:2017krein}; for propagating spatial-transfer branches, the same form is proportional to the conserved flux, up to an overall convention-dependent sign. Only relative signs enter the crossing classification. Section~\ref{sec:floquetBloch} uses these temporal and spatial signatures to derive the compact or non-compact normal forms near an isolated crossing.

The variable-\(G_z(z)\) extension relevant to more general spatial reductions is given in Appendix~\ref{app:block}.

%% file: sections/floquet_bloch.tex
\section{Floquet-Bloch decomposition and crossing classification}
\label{sec:floquetBloch}

Section~\ref{sec:evolution} established the real symplectic structure of the reduced $s$-evolution problem. We now construct its modes in a periodic axion background using a Floquet-Bloch expansion~\cite{Shirley:1965,Sambe:1973cnm,Eckardt:2016lof}, derive the selection rules for folded degeneracies, and show how the Krein (temporal) or flux (spatial) signatures determine the local crossing algebra.

\subsection{Covariant Floquet-Bloch expansion}

The single-phase pump of Section~\ref{sec:model} has the covariant form
\begin{equation}
    \bar a(x)=a_0\cos(K_\mu x^\mu),\qquad
    K^\mu=(\Omega_{\rm p},Q_{\rm p}),
    \label{eq:pumpCov}
\end{equation}
where a freely propagating massive axion component satisfies $K_\mu K^\mu=m_a^2$. A static spatially periodic medium has $K^\mu=(0,Q)$ and corresponds to a distinct stationary reduction. The background is invariant under the discrete spacetime translations
\begin{equation}
    K_\mu\Delta x^\mu=2\pi\ell,\qquad \ell\in\mathbb Z.
\end{equation}
A field in such a background admits the Floquet-Bloch expansion
\begin{equation}
    \Psi(t,z)=e^{-i\varepsilon t+i\kappa z}
    \sum_{n\in\mathbb Z}\Psi_n\,e^{-in\Omega_{\rm p} t+in Q_{\rm p}z},
    \label{eq:FloquetBloch}
\end{equation}
where $\varepsilon$ and $\kappa$ are the spacetime Bloch labels. The $n$th sideband rung carries the four-momentum
\begin{equation}
    (\omega_n,k_n)=(\varepsilon+n\Omega_{\rm p},\;\kappa+n Q_{\rm p}).
    \label{eq:rung}
\end{equation}
At this level $\varepsilon$ and $\kappa$ are labels; they acquire the interpretation of a quasi-frequency or quasi-wavenumber only after a specific slicing (fixed $k$ or fixed external frequency) is chosen.

\subsection{Sideband ladder and selection rules}

The gradient couplings $\partial_t\bar a$ and $\nabla\bar a$ produce terms proportional to $e^{\pm i(\Omega_{\rm p} t-Q_{\rm p}z)}$. In the notation of Eq.~\eqref{eq:rung}, the \(n\)th rung therefore couples to its nearest neighbors \(n\pm1\); iterating these couplings generates the spacetime sideband ladder
\begin{equation}
    (\omega_n,k_n)=(\varepsilon+n\Omega_{\rm p},\;\kappa+n Q_{\rm p}),\qquad n\in\mathbb Z.
    \label{eq:ladder}
\end{equation}
Any rung can serve as the reference Fourier mode: shifting \((\varepsilon,\kappa)\) by an integer multiple of \(K^\mu\) leaves the physical ladder unchanged. For an on-shell massive axion component, the timelike pump generates this ladder; the special case $Q_{\rm p}=0$ gives homogeneous temporal sidebands $\omega_n=\varepsilon+n\Omega_{\rm p}$ at fixed $k=\kappa$. More generally, a single timelike pump has a pump-rest-frame representative in which the sideband index is purely temporal. Appendix~\ref{app:boost} gives the boost and explains why it relates ladder representations rather than boundary-value problems. A purely spatial ladder $k_n=\kappa+nQ$ at fixed external frequency describes a stationary medium with $K^\mu=(0,Q)$ and is posed as a boundary-value transfer problem. A spatial reduction obtained by WKB ray projection is instead local and inherits the ray approximation.

Substituting the Floquet-Bloch expansion~\eqref{eq:FloquetBloch} into the equations of motion, each Fourier amplitude $\Psi_n$ satisfies
\begin{equation}
    D(\omega_n,k_n)\,\Psi_n + \sum_{m\neq n}\mathcal C_{nm}\Psi_m = 0,
    \label{eq:ladderEq}
\end{equation}
where $D(\omega,k)$ is the free dispersion function ($D_\gamma=-\omega^2+k^2+\omega_{\rm pl}^2$ for photons, $D_a=-\omega^2+k^2+m_a^2$ for the axion) and $\mathcal C_{nm}$ are coupling coefficients proportional to the pump amplitude and to the appropriate external frequency or wavenumber. In weak coupling, eliminating a well-separated rung produces corrections of order $O(|\mathcal C|^2/\Delta D)$. Non-perturbative mode mixing requires two diagonal shell denominators to become small at the same Bloch label:
\begin{equation}
    D(\omega_n,k_n)\simeq 0,\qquad D(\omega_m,k_m)\simeq 0,\qquad n\neq m.
    \label{eq:doubleOnShell}
\end{equation}
This is the folded-zone form of a near-degeneracy: two different sideband copies of free dispersion branches intersect at the same \((\varepsilon,\kappa)\). Since the two rungs differ by \((n-m)K^\mu\), Eq.~\eqref{eq:doubleOnShell} immediately gives an on-shell momentum-matching rule. Same-frequency-sign intersections give difference-frequency matching, while positive/negative-frequency intersections are conventionally rewritten as sum-frequency matching. We write \(N\equiv n-m\) for the signed harmonic difference and use \(|N|\) for the sideband order. Reversing the pump insertion changes the sign of \(N\); in the sum-frequency channels below we choose the orientation for which \(N>0\). The possible matchings fall into three classes:
\begin{align}
    p_f^\mu-p_i^\mu &\simeq N K^\mu,
    &&\text{difference-frequency / driven conversion},
    \label{eq:diffSel}\\
    p_{\gamma,1}^\mu+p_{\gamma,2}^\mu &\simeq N K^\mu,
    &&\text{single-species sum-frequency channel},
    \label{eq:sameSel}\\
    p_a^\mu+p_\gamma^\mu &\simeq N K^\mu,
    &&\text{cross-species sum-frequency channel}.
    \label{eq:crossSel}
\end{align}
Here $p_{\gamma,1}$ and $p_{\gamma,2}$ may be distinct folded branches of the same species; they coincide in the degenerate parametric channel. For a real field, however, the negative-frequency coefficient at spatial Fourier label \(+k\) is the complex conjugate of a positive-frequency excitation at \(-k\). Thus, for a homogeneous temporal pump with zero spatial momentum, the single-species sum-frequency rule describes a physical photon pair with momenta \((+k,-k)\), even though the fixed-\(k\) Floquet calculation represents it as a positive/negative-frequency collision at one conserved Fourier label.

Equations~\eqref{eq:diffSel}--\eqref{eq:crossSel} constitute the \emph{selection rules}: the condition for two rungs to enter a non-perturbative resonance. Their concrete form depends on the slicing used to define the one-dimensional evolution problem. The temporal fixed-$k$ realization is developed below, where these covariant selection rules become the familiar frequency-matching conditions for avoided crossings and parametric instabilities.

We separate the dimensionless pump amplitude from the frequency scale supplied by the pump:
\begin{equation}
    \alpha_a\equiv g a_0,\qquad
    \epsilon_0\equiv g a_0\,\Omega_{\rm p}=\alpha_a\Omega_{\rm p},
    \label{eq:epsilon0}
\end{equation}
where $\alpha_a$ is dimensionless whereas $\epsilon_0$ has dimensions of frequency. The actual entries of the second-order shell matrix carry dimensions of frequency squared and are of order $\epsilon_0$ times a characteristic frequency or wavenumber of the mode being scattered. The perturbative Floquet-Bloch expansion is therefore controlled by the smallness of the relevant off-diagonal matrix elements compared with the shell separations, schematically $|\mathcal C_{nm}|\ll |D_n-D_m|$.
\subsection{Local two-mode reduction}
\label{subsec:localTwoMode}

Near an isolated folded crossing, the infinite sideband problem reduces systematically to the two nearly resonant envelopes. Let \(c_R=(u,v)^{\mathsf T}\) denote the coordinates of the full ladder eigenvector projected onto the two resonant endpoint modes, and let \(c_O\) collect all off-resonant amplitudes. After the Floquet-Bloch expansion, the algebraic ladder can be partitioned as
\begin{equation}
    \begin{pmatrix}
        \mathcal L_{RR} & \mathcal L_{RO}\\
        \mathcal L_{OR} & \mathcal L_{OO}
    \end{pmatrix}
    \begin{pmatrix}
        c_R\\ c_O
    \end{pmatrix}
    =0 .
\end{equation}
The reduced amplitudes are distinct from the original Floquet-rung coefficients. Each \(\Psi_n\) in Eq.~\eqref{eq:FloquetBloch} contains all field components on rung \(n\), whereas \(u\) and \(v\) are the two action-normalized endpoint coordinates selected from the full ladder. For example, in the temporal fixed-\(k\) positive-positive axion-photon crossing below, the endpoints are the axion component \(a_{n_a}\) and a photon component \(A_{\gamma,n_\gamma}\), satisfying
\begin{equation}
    \varepsilon_F+n_a m_a\simeq+\omega_a(k),
    \qquad
    \varepsilon_F+n_\gamma m_a\simeq+\omega_\gamma(k).
\end{equation}
In the weak-coupling limit, \(u\) and \(v\) reduce to action-normalized versions of these components, while all other photon, axion, polarization, and sideband components belong to \(c_O\). At finite coupling, the eliminated amplitudes dress the endpoints perturbatively through the Schur complement.
The crossing assumption means that \(\mathcal L_{RR}\) is small, while the off-resonant block \(\mathcal L_{OO}\) is invertible at the same Bloch label. Eliminating \(c_O\) gives the Schur-complement equation
\begin{equation}
    \mathcal L_{\rm eff}c_R=0,\qquad
    \mathcal L_{\rm eff}
    =
    \mathcal L_{RR}
    -
    \mathcal L_{RO}\mathcal L_{OO}^{-1}\mathcal L_{OR}.
    \label{eq:localSchur}
\end{equation}
Writing the two retained diagonal shell denominators as \(D_1,D_2\), we define the shell-space hierarchy by
\begin{equation}
    \mathcal L_{\rm eff}
    \equiv
    \begin{pmatrix}
    D_1-\Sigma_1 & \Xi_{12}\\
    \widetilde\Xi_{21} & D_2-\Sigma_2
    \end{pmatrix}.
    \label{eq:endpointHierarchy}
\end{equation}
Thus \(\Sigma_i\) is the diagonal self-energy correction generated by eliminated off-resonant rungs, while \(\Xi_{12}\) and \(\widetilde\Xi_{21}\) are the off-diagonal endpoint matrix elements in the second-order shell equation. They have dimensions of frequency squared in the temporal problem. Linearizing the endpoint denominators and dividing by their action slopes converts \(\Xi\) into the first-order coupling \(\mu_{\rm eff}\), which has dimensions of frequency. In weak coupling, \(\Xi\) and \(\Sigma_i\) expand into sideband hops and off-resonant denominators. The co-rotating organization developed in Appendix~\ref{app:JacobiAnger} instead resums the same-branch photon dressing into Bessel harmonics, but retains separate branch-changing pump terms and requires the same resonant-subspace projection.

To recover an evolution equation from this algebraic shell problem, one expands each endpoint denominator to first order in the spectral label conjugate to \(s\). Dividing by these slopes is the action or flux normalization of the two endpoint coordinates. After also expanding in the detuning \(\Delta\) from the crossing, the local evolution takes the form
\begin{equation}
    \mathrm i\partial_s c_R
    =
    S h_{\rm eff} c_R,\qquad
    S=\operatorname{diag}(\sigma_1,\sigma_2),
    \label{eq:signatureNormalForm}
\end{equation}
where \(s\) is the evolution coordinate, \(\sigma_i=\pm1\) are the temporal Krein or spatial flux signs of the two colliding modes, and \(h_{\rm eff}=h_{\rm eff}^\dagger\) contains \(\Delta\) and the action-normalized coupling \(\mu_{\rm eff}\). If \(\sigma_1=\sigma_2\), the signature matrix is definite and the reduced algebra is compact. If \(\sigma_1=-\sigma_2\), the signature matrix is indefinite and the same near-degenerate coupling gives a non-compact normal form. The temporal and spatial subsections below determine the corresponding spectra and physical interpretations after the evolution problem has been selected.

Let \(\mathcal U_R(s,s_0)\) denote the propagator of the reduced endpoint system, with \(s_0\) the initial evolution coordinate:
\begin{equation}
    c_R(s)=\mathcal U_R(s,s_0)c_R(s_0).
\end{equation}
The group distinction follows directly from the metric preserved by this propagator. For a same-sign pair, the common sign can be absorbed and Eq.~\eqref{eq:signatureNormalForm} becomes
\begin{equation}
    \mathrm i\partial_s c_R=h_{\rm eff}c_R,\qquad
    \mathcal U_R^\dagger(s,s_0)\mathcal U_R(s,s_0)=I .
    \label{eq:SU2Metric}
\end{equation}
After removing the common phase, \(\mathcal U_R(s,s_0)\in SU(2)\). The compact evolution preserves
\(c_R^\dagger c_R=|u|^2+|v|^2\) and acts as a norm-preserving rotation between the two endpoint amplitudes.

For an opposite-sign pair, choose \(S=\operatorname{diag}(1,-1)\). The propagator instead satisfies
\begin{equation}
    \mathcal U_R^\dagger(s,s_0)S\mathcal U_R(s,s_0)=S ,
    \label{eq:SU11Metric}
\end{equation}
so, after removing the common phase, \(\mathcal U_R(s,s_0)\in SU(1,1)\). This non-compact evolution preserves
\begin{equation}
    c_R^\dagger S c_R=|u|^2-|v|^2,
\end{equation}
but not the positive norm. Because \(SU(1,1)\) is non-compact, its evolution can be hyperbolic rather than a bounded rotation. The observable meaning of that non-compact behavior is fixed only after the evolution coordinate and boundary data have been specified.

We use the following convention when writing these specializations. Same-sign crossings are displayed directly in the Schr\"odinger-like form \( \mathrm i\partial_s c_R=h_{\rm eff}c_R\), with the detuning on the diagonal of \(h_{\rm eff}\). For opposite-sign crossings we use the negative-norm endpoint as a conjugate envelope in the temporal problem, or as the backward-flux envelope in the spatial problem. With this choice Eq.~\eqref{eq:signatureNormalForm} is equivalently written as a Bogoliubov transfer equation,
\begin{equation}
    \partial_s
    \binom{u}{v}
    =
    \begin{pmatrix}
    -\mathrm i\Delta/2 & \mu_{\rm eff}\\
    \mu_{\rm eff}^* & +\mathrm i\Delta/2
    \end{pmatrix}
    \binom{u}{v},
    \label{eq:bogoliubovConvention}
\end{equation}
after fixing the phase convention for the off-diagonal coupling. Thus the signature remains the matrix \(S=\operatorname{diag}(+1,-1)\) in Eq.~\eqref{eq:signatureNormalForm}; the explicit factors of \(\mathrm i\) in Eq.~\eqref{eq:bogoliubovConvention} are a display convention for the slowly varying envelopes.

\subsection{Temporal Krein sign}

For the homogeneous temporal fixed-$k$ problem, \(Q_{\rm p}=0\) and \(\Omega_{\rm p}=m_a\). The spacetime Bloch label \(\kappa\) is then the conserved physical momentum \(k\), while the remaining label \(\varepsilon\) becomes the temporal quasi-frequency. In the temporal sections we denote this quasi-frequency by \(\varepsilon_F\), defined modulo \(m_a\). The one-period map \(F_t(k)\) is the temporal monodromy matrix introduced in Eq.~\eqref{eq:monodromy}. Its eigenvalues are the Floquet multipliers:
\begin{equation}
    F_t(k)U=\lambda U,\qquad
    \lambda=\exp(-\mathrm i\varepsilon_F T),
    \label{eq:floquetMultiplierDef}
\end{equation}
where \(U\) is the full phase-space Floquet eigenvector and \(T\) is the pump period. Stability of a branch is diagnosed by whether its multipliers lie on the unit circle. The Hermitian symplectic form~\eqref{eq:HermitianSymplecticForm} evaluated on a Floquet eigenmode yields its Krein signature. A simple Floquet multiplier $\lambda$ on the unit circle carries a nonzero Krein norm,
\begin{equation}
    \operatorname{sgn}_K(\lambda;U)=\operatorname{sgn}\bigl(\mathrm i U^\dagger J U\bigr).
\end{equation}

The positive/negative labels used in the temporal channel notation refer first to the frequency sign of the uncoupled oscillator component. A positive-frequency component has phase \(e^{-\mathrm i\omega t}\), while a negative-frequency component has phase \(e^{+\mathrm i\omega t}\), up to the sideband factor that folds it into the common quasi-frequency zone. For a standard lossless canonical oscillator these two components carry opposite Krein signatures: the action-normalized positive-frequency amplitude has positive symplectic norm, and the conjugate negative-frequency amplitude has negative symplectic norm. Thus, in the homogeneous temporal fixed-\(k\) problem, the superscripts \((+,+)\) and \((+,-)\) may be read equivalently as same-sign and opposite-sign Krein pairings after canonical projection. The frequency sign identifies the branch in the Floquet ladder; the Krein sign is the invariant symplectic signature that controls whether the reduced two-mode algebra is compact or non-compact.

There are therefore three related, but distinct, eigenvectors in the temporal discussion below. The vector \((\Psi_n)\) is the infinite Floquet-ladder eigenvector in Fourier space. The vector \(c_R=(u,v)^{\mathsf T}\) is its projection onto the resonant endpoint subspace after the off-resonant components have been eliminated and the endpoint oscillators have been action-normalized. The vector \(U\) is the corresponding phase-space Floquet eigenvector of the real monodromy matrix. The reduced vector \(c_R\) gives the leading resonant coordinates of \(U\), while the eliminated ladder components give perturbative dressing and self-energy shifts. Thus the local two-mode equations determine the nearby quasi-frequency shifts and endpoint mixing coefficients, but the actual stability diagnostic remains the multiplier and Krein norm of the full phase-space eigenvector.

Krein theory states that an isolated multiplier on the unit circle cannot leave it by itself; a collision with another multiplier is required. If the colliding modes carry the same Krein sign, the restricted symplectic form is definite and the local two-mode problem is compact: the multipliers remain on the unit circle after the collision, producing an avoided crossing with bounded amplitude exchange. If the colliding modes carry opposite Krein sign, the restricted form is indefinite and generic coupling splits the pair into reciprocal hyperbolic multipliers, signaling temporal exponential growth~\cite{Chernyavsky:2017krein}. This is the symplectic origin of the $(+,+)$ versus $(+,-)$ classification for temporal problems.

Near an uncoupled crossing at quasi-frequency \(\varepsilon_*\), define the bare collision multiplier and the nearby coupled multipliers by
\begin{equation}
    \lambda_*\equiv\exp(-\mathrm i\varepsilon_*T),
    \qquad
    \lambda_r=\lambda_*\exp(-\mathrm i\delta\varepsilon_r T).
    \label{eq:bareCollisionMultiplier}
\end{equation}
Here the eigenvalues of the reduced normal form give the local quasi-frequency shifts \(\delta\varepsilon_r\).
For a same-sign pair these shifts remain real and the multipliers stay on the unit circle. For an opposite-sign pair they become imaginary inside the instability interval, producing a reciprocal growing/decaying pair. Section~\ref{sec:floquet} gives the channel-specific detunings and couplings.

\subsection{Spatial Wronskian and flux sign}

For the spatial fixed-$\omega$ transfer problem, the corresponding sign follows from the conserved Wronskian current rather than from a temporal multiplier norm. For a single decoupled spatial channel, let \(u(z)\) denote the scalar wave amplitude after fixing the external frequency, let a prime denote \(\partial_z\), and form the spatial phase-space vector \(y=(u,u')^{\mathsf T}\). With
\begin{equation}
    J_z=\begin{pmatrix}0&I\\-I&0\end{pmatrix},
\end{equation}
the Wronskian is the same antisymmetric symplectic bilinear written in scalar notation,
\begin{equation}
    W[u^*,u]=y^\dagger J_z y=u^*u'-(u^*)'u.
\end{equation}
The real spatial flux \(\mathcal F\) is the standard normalization of this current,
\begin{equation}
    \mathcal F=\frac{1}{2\mathrm i}W[u^*,u]=\operatorname{Im}(u^*u').
\end{equation}
For the convention \(h_z(y,y)=\mathrm i y^\dagger J_z y\) used in the canonical phase space, \(h_z=-2\mathcal F\). Thus the Hermitian symplectic form and the physical flux differ by a fixed overall minus sign; only their relative signs enter the crossing classification. We label forward and backward branches by the sign of \(\mathcal F\). For real, lossless spatial coefficients this current is conserved along \(z\). A spatial $(+,-)$ collision is therefore a forward/backward flux collision. It can produce real spatial exponents in the transfer matrix, corresponding to evanescent decay or distributed reflection along $z$ within a stationary scattering calculation.

\subsection{Physical interpretation of the crossing types}

The compact/non-compact algebra is shared locally by the temporal and spatial problems, but the physical interpretation is governed by the choice of evolution variable:
\begin{align}
    s=t:\quad (+,-)&\;\Rightarrow\; \text{positive/negative-frequency coupling}\notag\\
    &\qquad\;\rightarrow\; \text{temporal parametric instability},
    \label{eq:tempInterpret}\\[4pt]
    s=z:\quad (+,-)&\;\Rightarrow\; \text{forward/backward-flux coupling}\notag\\
    &\qquad\;\rightarrow\; \text{Bragg reflection or stop-band formation}.
    \label{eq:spatInterpret}
\end{align}
Conversely, a $(+,+)$ collision in either problem produces bounded conversion or an avoided crossing. Before specializing these normal forms to temporal and spatial evolution, we give a complementary external-field interpretation of the three selection rules.

\subsection{Diagrammatic interpretation}

The selection rules also have an external-field interpretation. Splitting the axion field into a prescribed coherent component and a quantum perturbation,
\begin{equation}
    a(x)=\bar a(x)+\delta a(x),
\end{equation}
the interaction \( -g a F_{\mu\nu}\widetilde F^{\mu\nu}/4\) contains two types of vertex. The term with \(\delta a\) is the ordinary quantum axion-photon interaction, whereas \(\bar a(x)\) appears as an external classical insertion rather than a quantum propagator. Equivalently, the coherent background is the large-occupation-number limit of an axion coherent state, in which a classical Fourier component replaces the absorption or emission of a pump quantum. For a periodic background,
\begin{equation}
    \bar a(x)=\frac{a_0}{2}\left(e^{\mathrm iK\cdot x}+e^{-\mathrm iK\cdot x}\right),
\end{equation}
where \(K^\mu=(\Omega_{\rm p},Q_{\rm p})\). The Fourier components at \(+K^\mu\) and \(-K^\mu\) change the four-momentum label of a quantum line by \(\pm K^\mu\). The periodic field therefore dresses the line into the Floquet harmonic tower
\begin{equation}
    p^\mu\longrightarrow p^\mu+nK^\mu,
\end{equation}
which is the diagrammatic counterpart of the Floquet-Bloch ladder. The integer \(n\) labels harmonics of one classical background, not independent scattering processes. Appendix~\ref{app:JacobiAnger} gives a complementary co-rotating organization: same-branch photon hops are resummed into Bessel-dressed magnetic vertices, while the positive-negative-frequency pump block remains explicit. In the weak-modulation, large-detuning overlap, expanding the Bessel weights reproduces the product of sideband hops and off-resonant denominators. Near a folded crossing, the resonant endpoint subspace must still be isolated and resummed.

The diagrams determine local four-momentum bookkeeping and endpoint connectivity, not the observable. Growth, conversion, and reflection follow only after the evolution variable and boundary data have been specified. A static spatial Fourier component \(K^\mu=(0,Q)\) conserves frequency and shifts momentum by \(NQ\), whereas a homogeneous temporal pump \(K^\mu=(m_a,0)\) conserves \(k\) and shifts frequency by \(Nm_a\).

\begin{figure}[t]
\centering
\resizebox{0.94\textwidth}{!}{%
\begin{tikzpicture}[
    every node/.style={font=\small},
    qline/.style={thick},
    photon/.style={thick,decorate,decoration={snake,amplitude=1.1pt,segment length=5pt}},
    axion/.style={thick,dashed},
    pump/.style={thick,dashed},
    insertion/.style={circle,draw,inner sep=1.8pt,fill=white},
    >=Stealth,
    scale=0.92
]
\path[use as bounding box] (-0.75,-4.95) rectangle (10.95,1.75);
\newcommand{\xcross}[2]{\draw (#1-0.09,#2-0.09) -- (#1+0.09,#2+0.09); \draw (#1-0.09,#2+0.09) -- (#1+0.09,#2-0.09);}
\draw[axion] (0,0.15) -- (1.25,0.15);
\draw[photon] (1.25,0.15) -- (4.2,0.15);
\draw[photon] (1.25,0.15) -- (1.25,1.02);
\xcross{1.25}{1.02}
\node[above] at (1.25,1.13) {\(B\)};
\node[left] at (1.18,0.62) {\(q_B=0\)};
\draw[pump] (3.0,0.15) -- (3.0,-0.68);
\node[insertion] at (3.0,-0.82) {};
\xcross{3.0}{-0.82}
\node[right] at (3.13,-0.82) {\(\bar a\)};
\node[right] at (3.05,-0.34) {\(K\)};
\node[left] at (0,0.15) {\(\delta a\)};
\node[right] at (4.2,0.15) {\(\gamma\)};
\node[below] at (2.05,-1.22) {(a) \(p_f-p_i\simeq NK\)};

\draw[photon] (6.15,0.15) -- (10.35,0.15);
\draw[pump] (8.25,0.15) -- (8.25,-0.68);
\node[insertion] at (8.25,-0.82) {};
\xcross{8.25}{-0.82}
\node[right] at (8.38,-0.82) {\(\bar a\)};
\node[right] at (8.3,-0.34) {\(K\)};
\node[left] at (6.15,0.15) {\(\gamma_+\)};
\node[right] at (10.35,0.15) {\(\gamma_-\)};
\node[below] at (8.25,-1.22) {(b) \(p_{\gamma,1}+p_{\gamma,2}\simeq NK\)};

\draw[axion] (2.7,-3.15) -- (3.95,-3.15);
\draw[photon] (3.95,-3.15) -- (7.45,-3.15);
\draw[photon] (3.95,-3.15) -- (3.95,-2.28);
\xcross{3.95}{-2.18}
\node[above] at (3.95,-2.17) {\(B\)};
\node[left] at (3.88,-2.68) {\(q_B=0\)};
\draw[pump] (5.95,-3.15) -- (5.95,-3.98);
\node[insertion] at (5.95,-4.12) {};
\xcross{5.95}{-4.12}
\node[right] at (6.08,-4.12) {\(\bar a\)};
\node[right] at (6.0,-3.62) {\(K\)};
\node[left] at (2.7,-3.15) {\(\delta a_+\)};
\node[right] at (7.45,-3.15) {\(\gamma_-\)};
\node[below] at (5.1,-4.52) {(c) \(p_a+p_\gamma\simeq NK\)};
\end{tikzpicture}
}
\vspace{0.8em}
    \caption{External-field diagrams for the three sideband selection rules. Dashed horizontal lines denote quantum axion perturbations, wavy horizontal lines denote photons, vertical \(B\) legs denote the prescribed static and uniform magnetic-field component with \(q_B^\mu=0\), and vertical \(\bar a\) legs denote Fourier components of the periodic \(c\)-number background. The real background contains both signs of \(K^\mu\); \(|N|\) is the order of the dressed harmonic connecting the resonant endpoints. Panel (a) is external-field-assisted positive-frequency conversion, panel (b) is the photon positive/negative collision underlying the Mathieu channel, and panel (c) is the mixed axion-photon positive/negative collision underlying the MAS channel.}
\label{fig:ExternalFieldDiagrams}
\end{figure}
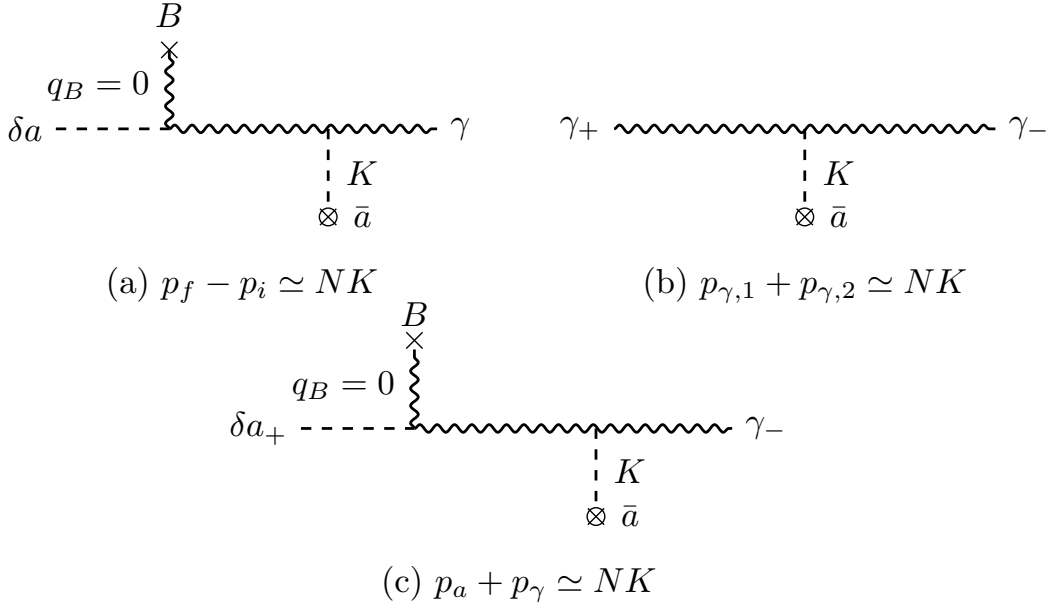

Figure~\ref{fig:ExternalFieldDiagrams} summarizes some effects in an external-field diagrammatic language. The \(B\) and \(\bar a\) legs are not propagating quantum particles; they denote Fourier components of the prescribed magnetic field and the periodic \(c\)-number axion background. In the homogeneous temporal setup, \(q_B^\mu\) denotes the magnetic-field four-wavevector and vanishes, whereas the axion background has support at \(\pm K^\mu\). The signed harmonic difference is \(N\), and \(|N|\) labels the order of the background-dressed harmonic. A structured magnetic field would contribute a nonzero \(q_B^\mu\) to the matching rule.

The three panels instantiate Eqs.~\eqref{eq:diffSel}--\eqref{eq:crossSel}. Panel (a) connects same-sign temporal branches or co-propagating spatial branches through one magnetic insertion and a background-dressed photon harmonic. Panels (b) and (c) connect opposite-sign endpoints, giving respectively the single-species Mathieu/Bragg and cross-species MAS/Bragg topologies.

%% file: sections/temporal_fixed_k.tex
\section{Temporal \texorpdfstring{fixed-$k$}{fixed-k} Floquet problem}
\label{sec:floquet}

\subsection{Homogeneous temporal pump and real block}

We now specialize to the temporally homogeneous limit $Q_{\rm p}=0$ of the pump~\eqref{eq:pump}, in which spatial translation symmetry is preserved and the wave number $k$ is globally conserved. The pump reduces to
\begin{equation}
    \bar a(t)=a_0\cos(m_a t),\qquad
    \dot{\bar a}(t)=-a_0 m_a\sin(m_a t).
\end{equation}
For a single Fourier mode we write
\begin{equation}
    \delta\bm A(t,z)=\bm A_k(t)e^{\mathrm i kz},\qquad
    \delta a(t,z)=a_k(t)e^{\mathrm i kz}.
\end{equation}
This complex Fourier representation fixes the conserved momentum label and gives the most compact equations of motion.

Define the effective frequencies
\begin{equation}
    \omega_\gamma^2\equiv k^2+\omega_{\rm pl}^2,\qquad
    \omega_a^2\equiv k^2+m_a^2,
\end{equation}
and the time-dependent pump-induced polarization coupling
\begin{equation}
    \sigma(t)\equiv g k\,\dot{\bar a}(t).
\end{equation}
In Cartesian components the fixed-$k$ equations of motion are
\begin{align}
    \ddot A_x+\omega_\gamma^2 A_x &= gB\,\dot a-\mathrm i\sigma(t)A_y,
    \label{eq:complexFixedKx}\\
    \ddot A_y+\omega_\gamma^2 A_y &= \mathrm i\sigma(t)A_x,
    \label{eq:complexFixedKy}\\
    \ddot a+\omega_a^2 a &= -gB\,\dot A_x.
    \label{eq:complexFixedKa}
\end{align}
For the real monodromy problem, the \(\pm k\) Fourier pair is combined into standing waves. Appendix~\ref{app:block} gives the field-level construction; its two parity sectors reduce to the equivalent real blocks
\begin{equation}
    q_1=(X_c,\,Y_s,\,\alpha_c)^{\mathsf T},\qquad
    q_2=(X_s,\,-Y_c,\,\alpha_s)^{\mathsf T}.
\end{equation}
Each block satisfies a second-order system of the parent form~\eqref{eq:parent}:
\begin{equation}
    \ddot q_1+G_R\dot q_1+V_R(t)q_1=0,
\end{equation}
with
\begin{equation}
    G_R=
    \begin{pmatrix}
    0&0&-gB\\
    0&0&0\\
    gB&0&0
    \end{pmatrix},\qquad
    V_R(t)=
    \begin{pmatrix}
    \omega_\gamma^2 & \sigma(t) & 0\\
    \sigma(t) & \omega_\gamma^2 & 0\\
    0 & 0 & \omega_a^2
    \end{pmatrix}.
    \label{eq:GRVR}
\end{equation}
The matrices satisfy $G_R^{\mathsf T}=-G_R$ and $V_R^{\mathsf T}=V_R$, as required by Eq.~\eqref{eq:GVconditions}. The temporal monodromy matrix \(F_t(k)=\Phi_t(T;k)\), defined as in Eq.~\eqref{eq:monodromy}, therefore belongs to $\operatorname{Sp}(6,\mathbb R)$, and its Floquet multipliers determine the stability of the fixed-$k$ mode. The sideband ladder below is the Floquet-Fourier representation of this eigenvalue problem: it exposes the resonant branches and perturbative endpoint couplings, while the monodromy multipliers provide the full stability diagnostic.

\subsection{Temporal ladder recursion}

In our previous analysis of forward driven conversion~\cite{Yao:2026yez}, a co-rotating frame reorganized the periodic polarization rotation into Bessel harmonics. Here we keep the individual Floquet rungs visible because the fixed-\(k\) problem also contains positive-negative-frequency mixing, which produces the Mathieu and MAS channels. The pump gives nearest-neighbor photon hops, the magnetic field gives same-rung axion-photon mixing, and the endpoint frequency signs determine the local monodromy type. Appendix~\ref{app:JacobiAnger} rewrites the temporal system in a doubled first-order basis and compares the rung-by-rung and co-rotating organizations.

We construct the ladder directly in the Cartesian polarization basis. Define the modulation amplitude
\begin{equation}
    \epsilon\equiv g a_0 k m_a,
\end{equation}
and resolve the time-periodic coefficients into Floquet rungs. For \(Q_{\rm p}=0\), the generic Bloch label \(\varepsilon\) of Section~\ref{sec:floquetBloch} becomes the temporal quasi-frequency \(\varepsilon_F\). We choose the relative rung phases
\begin{align}
    A_x(t) &=-\mathrm i e^{-\mathrm i\varepsilon_F t}
             \sum_{n\in\mathbb Z}A_{x,n}e^{-\mathrm i n m_a t},\\
    A_y(t) &=e^{-\mathrm i\varepsilon_F t}
             \sum_{n\in\mathbb Z}A_{y,n}e^{-\mathrm i n m_a t},\\
    a(t)   &=e^{-\mathrm i\varepsilon_F t}
             \sum_{n\in\mathbb Z}a_ne^{-\mathrm i n m_a t}.
    \label{eq:CartesianTemporalExpansion}
\end{align}
The relative factor of \(-\mathrm i\) is a phase convention that makes the Cartesian rung connectivity explicit; it has no effect on the spectrum. The quasi-frequency \(\varepsilon_F\) is defined modulo \(m_a\), and the \(n\)th rung carries
\begin{equation}
    \omega_n=\varepsilon_F+n m_a.
\end{equation}
Defining the shell denominators
\begin{equation}
    D_\gamma(n)\equiv\omega_\gamma^2-(\varepsilon_F+n m_a)^2,\qquad
    D_a(n)\equiv\omega_a^2-(\varepsilon_F+n m_a)^2,
\end{equation}
we write the photon hopping as
\begin{equation}
    \tau\equiv\frac{\epsilon}{2\mathrm i},
    \qquad
    \tau_R\equiv|\tau|=\frac{|\epsilon|}{2}
    =\frac{|g a_0 k m_a|}{2},
\end{equation}
and define the same-rung Cartesian magnetic bridge by
\begin{equation}
    \mathcal G_B(n)\equiv gB(\varepsilon_F+n m_a),
    \qquad
    \mathcal G_B\equiv\mathcal G_B(0)=gB\,\varepsilon_F .
    \label{eq:GBtauDefs}
\end{equation}
The Cartesian equations~\eqref{eq:complexFixedKx}--\eqref{eq:complexFixedKa} then give
\begin{subequations}
\label{eq:CartesianTemporalRec}
\begin{align}
    D_\gamma(n)A_{x,n}
    -\tau\bigl(A_{y,n-1}-A_{y,n+1}\bigr)
    &=\mathcal G_B(n)a_n,\\
    D_\gamma(n)A_{y,n}
    -\tau\bigl(A_{x,n-1}-A_{x,n+1}\bigr)
    &=0,\\
    D_a(n)a_n&=\mathcal G_B(n)A_{x,n}.
\end{align}
\end{subequations}
Thus the full ladder separates into two rung-parity sectors,
\begin{equation}
    \mathcal C_0=\{a_{2\ell},A_{x,2\ell},A_{y,2\ell+1}\},
    \qquad
    \mathcal C_1=\{a_{2\ell+1},A_{x,2\ell+1},A_{y,2\ell}\},
    \qquad \ell\in\mathbb Z .
    \label{eq:temporalParityChains}
\end{equation}
The cross-species resonance anchored at \(a_0\) lies in \(\mathcal C_0\), whose photon path alternates as
\begin{equation}
    a_0\;\longleftrightarrow\;A_{x,0}
    \;\longleftrightarrow\;A_{y,-1}
    \;\longleftrightarrow\;A_{x,-2}
    \;\longleftrightarrow\;\cdots .
    \label{eq:selectedParityChain}
\end{equation}
The complementary sector is dynamically invariant and does not enter this selected endpoint reduction. In the unrephased recursion, the two directed entries on a photon link are complex conjugates, with magnitude \(\tau_R\). Rung rephasings convert each selected scalar link to the real symmetric hopping \(\tau_R\), which is the convention used in the Schur complements below. Endpoint phases remain convention dependent, whereas the self-energy shifts and multiplier splittings do not.

\begin{figure}[!t]
    \centering
    \begin{subfigure}{0.60\textwidth}
        \centering
        \includegraphics[width=\linewidth]{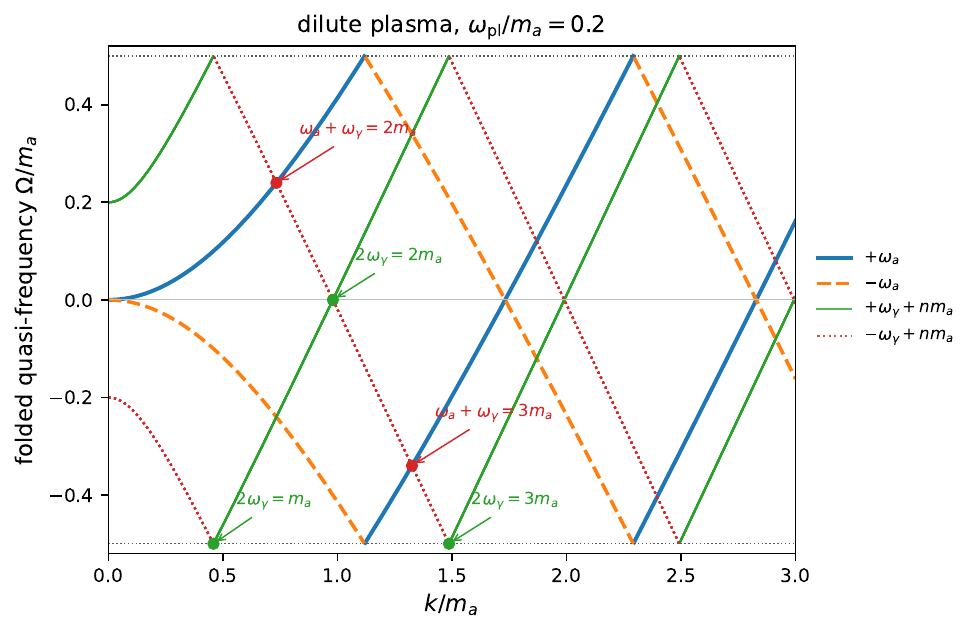}
        \caption{Dilute-plasma hierarchy.}
    \end{subfigure}
    \par\medskip
    \begin{subfigure}{0.60\textwidth}
        \centering
        \includegraphics[width=\linewidth]{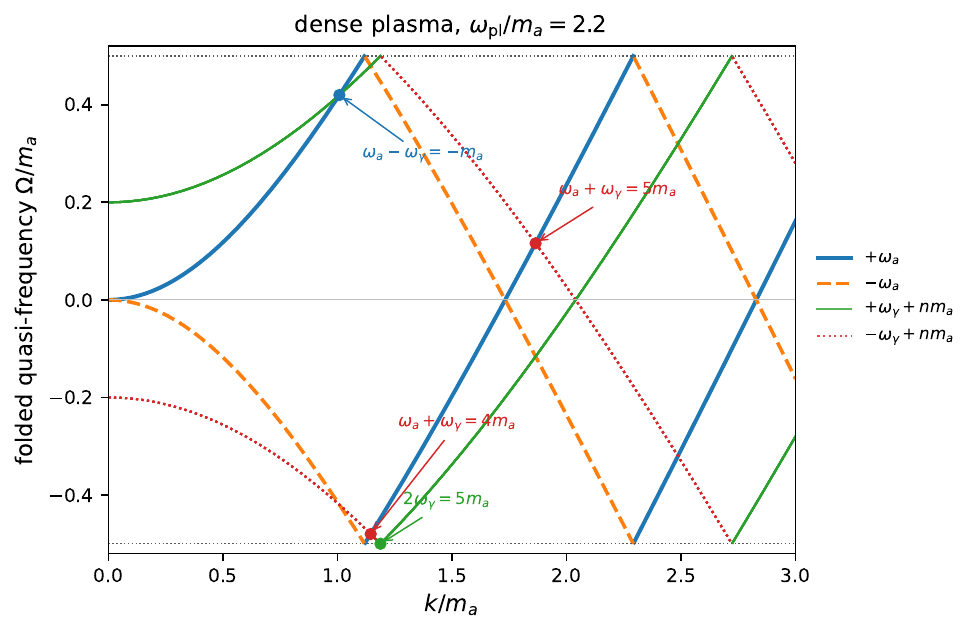}
        \caption{Dense-plasma hierarchy.}
    \end{subfigure}
    \caption{Folded quasi-frequency shells for the temporal fixed-$k$ problem. The plotted branches are the folded zeros of \(D_\gamma(n)\) and \(D_a(n)\); axion copies \(\pm\omega_a+n m_a\) collapse onto the displayed \(\pm\omega_a\) branches modulo \(m_a\). Panel (a) has $\omega_{\rm pl}/m_a=0.2$, where the accessible finite-order collisions are the opposite-sign Mathieu conditions $2\omega_\gamma=N m_a$ (green) and the opposite-sign MAS conditions $\omega_a+\omega_\gamma=N m_a$ (red). Panel (b) has $\omega_{\rm pl}/m_a=2.2$, where the same-sign condition $\omega_a-\omega_\gamma=-m_a$ becomes accessible (blue), together with visible sum-frequency collisions. Finite coupling determines the local splitting and produces higher-order self-energy shifts relative to the bare folded crossing.}
    \label{fig:FoldedShells}
\end{figure}

Figure~\ref{fig:FoldedShells} folds the zeros
\(\varepsilon_F+n m_a=\pm\omega_{\gamma,a}(k)\) into one temporal Floquet zone, separating algebraically allowed channels from kinematically accessible ones. For the dilute hierarchy \(m_a>\omega_{\rm pl}\), the nonzero positive-positive difference condition cannot be reached at finite \(k\), whereas the Mathieu and MAS sum-frequency collisions remain accessible. Reversing the hierarchy permits a signed \(N=-1\) positive-positive collision. These are bare crossings; finite coupling produces the splitting and self-energy shifts.

\subsection{Positive-positive crossing: temporal avoided crossing}

A positive-positive temporal crossing is the compact local channel: a positive-frequency axion shell and a positive-frequency photon shell become nearly degenerate at the same $k$. From the ladder~\eqref{eq:CartesianTemporalRec}, this condition is
\begin{equation}
    \varepsilon_F+n_a m_a\simeq +\omega_a(k),\qquad
    \varepsilon_F+n_\gamma m_a\simeq +\omega_\gamma(k).
\end{equation}
Eliminating \(\varepsilon_F\) gives the detuning
\begin{equation}
    \Delta_{a\gamma}^{(+,+;N)}(k)
    =
    \omega_a(k)-\omega_\gamma(k)-N m_a.
    \label{eq:detPP}
\end{equation}
This notation will be used for all temporal channels: for a collision between species \(i,j\in\{a,\gamma\}\) with frequency signs \(\sigma_i,\sigma_j=\pm\),
\begin{equation}
    \Delta_{ij}^{(\sigma_i,\sigma_j;N)}(k)
    \equiv
    \sigma_i\omega_i(k)-\sigma_j\omega_j(k)-N m_a,
    \qquad
    N\equiv n_i-n_j .
    \label{eq:temporalUnifiedDetuning}
\end{equation}
The corresponding action-normalized endpoint coupling is denoted by
\(\mu_{ij}^{(\sigma_i,\sigma_j;N)}\). In each channel, the general quantities \(\Delta_t\) and \(\mu_{\rm eff}\) introduced in Section~\ref{sec:floquetBloch} are identified with these channel-specific detunings and couplings.
For the dilute-plasma hierarchy discussed above, this detuning cannot vanish at finite \(k\) for nonzero \(N\). The \(N=0\) case is the ordinary same-\(k\) axion-photon mixing channel, becoming exactly degenerate only in the large-\(k\) limit or at a conventional level crossing in an inhomogeneous medium. In a dense-plasma hierarchy, a negative-integer difference condition is reached, as illustrated in Fig.~\ref{fig:FoldedShells}. The compact two-mode reduction below is the local normal form for such same-sign near-degeneracies.

The endpoint construction is derived explicitly in Appendix~\ref{subsec:PPChainApp}. For the nearest-rung orientation \(N=+1\), the retained endpoints are \(a_0\) and \(A_{y,-1}\), connected through the off-resonant photon \(A_{x,0}\). The opposite orientation \(N=-1\), used in the dense-plasma example of Fig.~\ref{fig:FoldedShells}, follows by reversing the photon rung direction. Eliminating the intermediate rung gives
\begin{equation}
    \Xi_{a\gamma}^{(+,+;1)}
    =e^{\mathrm i\phi_1}\frac{\mathcal G_B\tau_R}{D_\gamma(0)},\qquad
    \Sigma_a^{(+,+;1)}=\frac{\mathcal G_B^2}{D_\gamma(0)},\qquad
    \Sigma_\gamma^{(+,+;1)}=\frac{\tau_R^2}{D_\gamma(0)} .
\end{equation}
Here \(\Xi\) and \(\Sigma\) specialize the off-diagonal endpoint matrix element and diagonal self-energy defined in Eq.~\eqref{eq:endpointHierarchy}; \(\mathcal G_B\) is the Cartesian same-rung bridge, \(\tau_R\) is the rephased real hopping, and \(\phi_1\) absorbs the endpoint phase convention.

For \(|N|>1\), the eliminated photon path in the selected parity sector is collected into a projected harmonic weight \(W_N\). For \(N>0\), the large-detuning expansion has the form
\begin{equation}
    W_N\sim e^{\mathrm i\phi_N}
    \frac{\tau_R^N}{D_\gamma(0)D_\gamma(-1)\cdots D_\gamma(-(N-1))},
    \qquad
    \Xi_{a\gamma}^{(+,+;N)}\equiv \mathcal G_B W_N ,
    \label{eq:ProjectedPhotonWeightPP}
\end{equation}
with the reversed chain used for \(N<0\). The shell-space coupling has dimensions of frequency squared. Linearizing the two endpoint denominators and action-normalizing their coordinates gives
\begin{equation}
    \mu_{a\gamma}^{(+,+;N)}
    \sim
    \frac{\Xi_{a\gamma}^{(+,+;N)}}{2\sqrt{\omega_a\omega_\gamma}} .
\end{equation}
The self-energy terms shift the crossing and are absorbed into its effective detuning. Because both endpoints have positive frequency, the projected two-mode problem has a definite action norm.

When \(|\Delta_{a\gamma}^{(+,+;N)}|\) is comparable to or smaller than the coupling strength, the two positive-frequency modes mix non-perturbatively. Projecting onto their slowly varying envelopes gives the compact normal form of Section~\ref{subsec:localTwoMode},
\begin{equation}
    \mathrm i\partial_t
    \begin{pmatrix}u\\v\end{pmatrix}
    =
    \begin{pmatrix}
    \Delta_{a\gamma}^{(+,+;N)}/2 & \mu_{a\gamma}^{(+,+;N)}\\
    \mu_{a\gamma}^{(+,+;N)*} & -\Delta_{a\gamma}^{(+,+;N)}/2
    \end{pmatrix}
    \begin{pmatrix}u\\v\end{pmatrix},
    \label{eq:SU2temporal}
\end{equation}
where \(u\) and \(v\) are the slowly varying envelopes of the two resonant modes and \(\mu_{a\gamma}^{(+,+;N)}\) is the action-normalized effective coupling obtained by eliminating the off-resonant rungs. This system preserves the definite norm
\begin{equation}
    \partial_t\bigl(|u|^2+|v|^2\bigr)=0,
\end{equation}
and its local quasi-frequency shifts are
\begin{equation}
    \delta\varepsilon_\pm=\pm\sqrt{
    \bigl|\mu_{a\gamma}^{(+,+;N)}\bigr|^2
    +\bigl(\Delta_{a\gamma}^{(+,+;N)}/2\bigr)^2},
\end{equation}
which are purely real. Equivalently, the multipliers are \(\lambda_\pm=\lambda_*\exp(-\mathrm i\delta\varepsilon_\pm T)\), so they remain on the unit circle. The crossing produces a temporal avoided crossing with bounded amplitude exchange---a coherent beating between the axion and photon modes rather than exponential growth.

The cross-species $(+,+)$ temporal channel and the forward spatial channel of Section~\ref{sec:transfer} share the compact local normal form. Section~\ref{sec:pheno} gives the additional WKB mapping to the driven conversion studied in Ref.~\cite{Yao:2026yez}. The temporal channel concerns eigenfrequency stability at globally conserved $k$; the driven-conversion problem instead pulls the temporal pump phase back along a propagation ray and evolves local fixed-carrier envelopes in \(z\).

Closely related heterodyne and upconversion searches also use positive-positive photon-mode conversion, but those systems require an enlarged electromagnetic sector and input-output boundary conditions. We return to this comparison in Section~\ref{sec:pheno}.

\subsection{Positive-negative crossing: Mathieu parametric instability}

A positive-negative temporal crossing occurs when a positive-frequency mode and a folded negative-frequency mode of the same species become nearly degenerate. For the photon sector, the condition is
\begin{equation}
    \varepsilon_F+n_1 m_a\simeq +\omega_\gamma(k),\qquad
    \varepsilon_F+n_2 m_a\simeq -\omega_\gamma(k),
\end{equation}
which yields
\begin{equation}
    2\omega_\gamma(k)\simeq N m_a,\qquad N\equiv n_1-n_2>0.
    \label{eq:MathieuCond}
\end{equation}
The local same-species detuning is
\begin{equation}
    \Delta_{\gamma\gamma}^{(+,-;N)}(k)
    =
    2\omega_\gamma(k)-N m_a .
\end{equation}
This is the familiar parametric resonance condition of the Mathieu equation.
For a vacuum photon, Eq.~\eqref{eq:MathieuCond} reduces to \(k\simeq N m_a/2\). This is the origin of the ``forbidden momentum bands'' discussed in Refs.~\cite{Espriu:2011vj,Espriu:2014lma} for photons in a spatially homogeneous cold axion background. In the present language those bands are temporal Floquet gaps: \(k\) is a conserved parameter, and the special values \(k\simeq N m_a/2\) mark folded positive-negative photon collisions in the quasi-frequency problem.

These temporal Floquet gaps differ from the fixed-\(\omega\) spatial Bragg stop bands discussed in Section~\ref{sec:transfer}: the latter are periodic in \(z\), and \(Q\), rather than \(m_a\), supplies their reciprocal momentum.
Before the endpoint projection, the resonant photon coordinate is written as
\begin{equation}
    A_\gamma(t)
    =
    u(t)e^{-\mathrm i\omega_\gamma t}
    +v^*(t)e^{+\mathrm i\omega_\gamma t},
    \qquad
    |\dot u|,|\dot v|\ll\omega_\gamma |u|,\omega_\gamma |v|.
    \label{eq:MathieuEnvelopeAnsatz}
\end{equation}
The two envelopes multiply opposite-frequency oscillator components. In the local notation of Eq.~\eqref{eq:signatureNormalForm}, the conjugate negative-frequency envelope \(v^*\) represents the second endpoint amplitude. The displayed equation therefore uses the Bogoliubov convention of Eq.~\eqref{eq:bogoliubovConvention}. The pump harmonic with \(N m_a\simeq2\omega_\gamma\) phase matches the positive-frequency component to its conjugate negative-frequency partner, so the Mathieu resonance retains both. Eliminating the off-resonant rungs and projecting onto the resonant envelope pair in Eq.~\eqref{eq:MathieuEnvelopeAnsatz} gives the slow-envelope system
\begin{equation}
    \partial_t
    \begin{pmatrix}u\\v^*\end{pmatrix}
    =
    \begin{pmatrix}
    -\mathrm i\Delta_{\gamma\gamma}^{(+,-;N)}/2 & \mu_{\gamma\gamma}^{(+,-;N)}\\
    \mu_{\gamma\gamma}^{(+,-;N)*} & +\mathrm i\Delta_{\gamma\gamma}^{(+,-;N)}/2
    \end{pmatrix}
    \begin{pmatrix}u\\v^*\end{pmatrix},
    \label{eq:BogoliubovTemporal}
\end{equation}
where \(\mu_{\gamma\gamma}^{(+,-;N)}\) is the action-normalized same-species endpoint coupling. This system preserves the indefinite norm
\begin{equation}
    \partial_t\bigl(|u|^2-|v|^2\bigr)=0,
\end{equation}
and its envelope growth exponents are
\begin{equation}
    s_\pm=\pm\sqrt{\bigl|\mu_{\gamma\gamma}^{(+,-;N)}\bigr|^2
    -\bigl(\Delta_{\gamma\gamma}^{(+,-;N)}/2\bigr)^2}.
    \label{eq:MathieuGrowth}
\end{equation}
When \(|\Delta_{\gamma\gamma}^{(+,-;N)}|<2|\mu_{\gamma\gamma}^{(+,-;N)}|\), the exponents are real. Equivalently, \(\delta\varepsilon_\pm=\pm\mathrm i |s_+|\), and the Floquet multipliers leave the unit circle as a reciprocal pair. The growing and decaying solutions are Bogoliubov normal modes of one positive/negative-frequency photon pair, not independently growing and decaying photon modes. A generic small fluctuation projects onto the growing normal mode, whose two frequency components in Eq.~\eqref{eq:MathieuEnvelopeAnsatz} are generated together. The prescribed axion background is not depleted in the present linearized problem; including backreaction identifies this amplification with stimulated decay of the coherent axion field into photon pairs~\cite{Tkachev:2014dpa,Hertzberg:2018zte,Caputo:2018vmy,Arza:2019nta}.

\subsection{Positive-negative crossing: MAS sum-frequency instability}

The cross-species positive-negative channel couples a positive-frequency axion mode to a folded negative-frequency photon sideband. The resonance condition is the sum-frequency relation
\begin{equation}
    \omega_a(k)+\omega_\gamma(k)\simeq N m_a,
    \label{eq:MAScond}
\end{equation}
This resonance was identified in the MAS analysis through the instability structure of the coupled axion-photon system and appears as a tongue in the corresponding Ince--Strutt chart~\cite{Masaki:2019ggg}. The local detuning is
\begin{equation}
    \Delta_{a\gamma}^{(+,-;N)}(k)
    =
    \omega_a(k)+\omega_\gamma(k)-N m_a.
    \label{eq:MASdetuning}
\end{equation}
For a plasma with \(\omega_{\rm pl}>0\), the \(N=1\) channel is kinematically inaccessible, since \(\omega_a\ge m_a\) and \(\omega_\gamma\ge\omega_{\rm pl}\) imply \(\omega_a+\omega_\gamma>m_a\). In vacuum the equality is reached only at the non-propagating boundary \(k=0\), where \(\omega_\gamma=0\) and the slow-envelope normalization is singular. Propagating MAS tongues therefore begin at \(N\ge2\), unless medium effects modify the dispersion. This is why the numerical example in Section~\ref{sec:numerics} uses \(N=2\).

The bare condition~\eqref{eq:MAScond} locates the uncoupled folded-branch crossing. At finite magnetic and sideband coupling, the same virtual-rung elimination that generates the endpoint coupling also produces diagonal self-energy shifts of the two endpoints. Equivalently, the detuning entering the local two-mode problem is an effective detuning
\begin{equation}
    \Delta_{a\gamma,{\rm eff}}^{(+,-;N)}(k)
    =
    \Delta_{a\gamma}^{(+,-;N)}(k)
    +\delta\Delta_{\rm self}^{(N)}(k),
    \label{eq:MASeffDetuning}
\end{equation}
where $\delta\Delta_{\rm self}^{(N)}$ is higher order in the off-resonant magnetic and sideband couplings. In the perturbative regime this correction is small and the bare condition~\eqref{eq:MAScond} gives the leading location of the resonance.

The selected rung-parity chain of off-resonant photon sidebands mediates the effective coupling between the axion and photon endpoints. As in the positive-positive channel, \(W_N\) packages the eliminated photon path. Because the recursion~\eqref{eq:CartesianTemporalRec} is a second-order shell equation, eliminating the intermediate rungs first produces the frequency-squared endpoint matrix element
\begin{equation}
    \Xi_{a\gamma}^{(+,-;N)}\equiv \mathcal G_B W_N .
    \label{eq:XiWN}
\end{equation}
The first-order slow-envelope equations instead use \(\mu_{a\gamma}^{(+,-;N)}\). This action-normalized coupling has dimensions of frequency.

In the large-detuning limit, Schur-complement elimination of the intermediate rungs gives
\begin{equation}
    W_N
    \sim e^{\mathrm i\phi_N}
    \frac{\tau_R^N}
    {D_\gamma(0)D_\gamma(-1)\cdots D_\gamma(-(N-1))},
    \label{eq:XiScaling}
\end{equation}
where the denominators follow the virtual photon chain and \(\tau_R^N\) is the magnitude of the photon-hop product; \(\phi_N\) contains its phase.

In the co-rotating organization of Appendix~\ref{app:JacobiAnger}, Bessel harmonics of the rotated magnetic vertex resum the same-branch part of this dressing. The photon modulation index is \(\alpha_\gamma\equiv\alpha_a k/(2\omega_\gamma)\). In either formulation, the positive-negative-frequency pump block and diagonal self-energy shifts must also be retained. The shortest ladder path reproduces the leading small-\(\alpha_\gamma\) Bessel coefficient only in the additional high-carrier/eikonal limit \(|N|m_a\ll\omega_\gamma\). In that limit, the intermediate denominators reduce to their leading \(2\omega_\gamma\) factors. Outside this limit, the channel-dependent denominators in Eq.~\eqref{eq:XiScaling} must be retained. Canonical projection converts \(\Xi_{a\gamma}^{(+,-;N)}\) to
\begin{equation}
    \mu_{a\gamma}^{(+,-;N)}
    =
    e^{\mathrm i\varphi_N}
    \frac{\Xi_{a\gamma}^{(+,-;N)}}{2\sqrt{\omega_a\omega_\gamma}},
    \label{eq:mutoxi}
\end{equation}
within the reduced two-mode rotating-wave problem. The phase \(\varphi_N\) depends on the positive/negative-frequency basis convention and drops out of the growth rate.
The reduced slow dynamics has the same Bogoliubov structure as Eq.~\eqref{eq:BogoliubovTemporal}, with the replacements
\(\mu_{\gamma\gamma}^{(+,-;N)}\rightarrow\mu_{a\gamma}^{(+,-;N)}\) and
\(\Delta_{\gamma\gamma}^{(+,-;N)}\rightarrow\Delta_{a\gamma,{\rm eff}}^{(+,-;N)}\). The temporal growth rate is
\begin{equation}
    \gamma_{a\gamma}^{(N)}
    =\sqrt{\bigl|\mu_{a\gamma}^{(+,-;N)}\bigr|^2
           -\bigl(\Delta_{a\gamma,{\rm eff}}^{(+,-;N)}/2\bigr)^2},
    \label{eq:MASgrowth}
\end{equation}
and the instability tongue is open where
\(|\Delta_{a\gamma,{\rm eff}}^{(+,-;N)}|<2|\mu_{a\gamma}^{(+,-;N)}|\).
In the perturbative estimates below, \(\delta\Delta_{\rm self}^{(N)}\) is neglected so that
\(\Delta_{a\gamma,{\rm eff}}^{(+,-;N)}\simeq\Delta_{a\gamma}^{(+,-;N)}\).

At a given resonance \(k_*\) satisfying \(\Delta_{a\gamma}^{(+,-;N)}(k_*)=0\), the leading momentum-space half-width, defined by \(|k-k_*|<\delta k_{1/2}\), is
\begin{equation}
    \delta k_{1/2}\sim
    \frac{2\bigl|\mu_{a\gamma}^{(+,-;N)}\bigr|}
         {\bigl|k_*/\omega_a+k_*/\omega_\gamma\bigr|}.
    \label{eq:tongueWidth}
\end{equation}
The corresponding full width is \(\Delta k_{\rm full}\sim2\delta k_{1/2}\).
The details behind Eqs.~\eqref{eq:XiWN}--\eqref{eq:tongueWidth}, including the sideband-ladder elimination and the action-normalization step from \(\Xi_{a\gamma}^{(+,-;N)}\) to \(\mu_{a\gamma}^{(+,-;N)}\), are derived in Appendix~\ref{app:MAS}.

The phenomenological size of this coupling, and its comparison with the bounded conversion channels, is discussed in Section~\ref{sec:couplingScales}.

\subsection{Summary of temporal channels}

The homogeneous temporal pump therefore supports three principal fixed-$k$ channels. The positive-positive axion-photon collision gives a compact avoided crossing with bounded beating, while the Mathieu and MAS positive-negative collisions give photon-pair and cross-species parametric instabilities. All three are diagnosed by the Floquet multipliers of the temporal monodromy matrix. Section~\ref{sec:transfer} develops the corresponding spatial flux classification, and the unified channel comparison appears in Table~\ref{tab:pheno}.

%% file: sections/spatial_fixed_frequency.tex
\section{Spatial \texorpdfstring{fixed-$\omega$}{fixed-frequency} transfer problem}
\label{sec:transfer}

\subsection{Two sources of spatial periodicity}

After factoring out the harmonic time dependence $e^{-\mathrm i\omega t}$, a fixed-$\omega$ transfer problem requires coefficients that are independent of $t$ and constant or periodic in $z$. The mere presence of a spatial Fourier component in the axion field does not meet this requirement. A freely propagating massive axion component,
\begin{equation}
    \bar a(t,z)\sim a_0\cos(\Omega_{\rm p}t-Q_{\rm p}z),
\end{equation}
produces frequency sidebands separated by $\Omega_{\rm p}$. A single timelike pump can be represented in its rest frame as a purely temporal modulation, but this representation does not define the laboratory fixed-$\omega$ scattering problem. The external fields, plasma frame, and incoming/outgoing boundary conditions also transform under the boost, as summarized in Appendix~\ref{app:boost}. Two controlled constructions provide spatial transfer descriptions.

The first is an exact stationary medium with a genuine spatial modulation,
\begin{equation}
    \bar a(z)=a_0\cos Qz,
\end{equation}
for which the coefficients are periodic in $z$ and the transfer matrix at fixed $\omega$ is well defined. In the notation of Eq.~\eqref{eq:monodromy}, the one-cell spatial transfer matrix is \(F_z(\omega)=\Phi_z(L;\omega)\), with \(L=2\pi/Q\). When this construction is used as a proxy for the equal-time spatial structure of a nonrelativistic axion component, the relevant true spatial Fourier wavenumber is
\begin{equation}
    Q_{\rm real}\equiv m_a v_a.
    \label{eq:Qreal}
\end{equation}
This is the wavenumber assigned to that static proxy. The underlying freely moving axion component remains time dependent and does not define an exact stationary fixed-\(\omega\) problem. The phenomenological size and limitations of the proxy are discussed in Section~\ref{sec:phenoQ}.

The second is a ray-projected approximation to a temporally oscillating pump. A homogeneous pump $\bar a(t)=a_0\cos(m_a t)$ becomes an effective spatial modulation when its phase is pulled back along a forward WKB ray. If the trajectory satisfies $t\simeq\int^z\mathrm d z'/v_g(z')$, the phase along the ray is
\begin{equation}
    \Phi_{\rm pump}(z)=m_a t_{\rm ray}(z),
    \qquad
    Q_{\rm eff}(z)\equiv\frac{\mathrm d\Phi_{\rm pump}}{\mathrm d z}\simeq\frac{m_a}{v_g(z)},
    \label{eq:Qeff}
\end{equation}
where $v_g$ is the local group velocity. For relativistic propagation $v_g\simeq1$, $Q_{\rm eff}\simeq m_a$, which is comparable to the axion mass and can efficiently supply the phase matching needed for driven forward conversion. This ray-projected WKB description underlies the driven resonance studied in Ref.~\cite{Yao:2026yez}; its domain of validity excludes plasma cutoffs, turning points, and regions where $v_g\to0$.

After their couplings are projected and flux-normalized separately, these two constructions can produce the same local signature normal form. They nevertheless differ in the origin and scale of the phase mismatch and in their boundary conditions. Section~\ref{sec:pheno} collects the corresponding phenomenological consequences.

\subsection{Spatial reduction and flux signature}

In a fixed-$\omega$ description the time dependence is separated by the ansatz
\begin{equation}
    \delta\Psi(t,z)=\Psi_\omega(z)\,e^{-\mathrm i\omega t},
\end{equation}
and the dynamics reduces to a set of ordinary differential equations in $z$. For a static spatial modulation we construct the real block in Appendix~\ref{app:block}, where the transfer generator is shown to lie in \(\mathfrak{sp}(6,\mathbb R)\). This establishes the spatial flux signature used below. For ray-projected temporal pumps, the same two-mode equations are local WKB normal forms along a chosen branch, not a global monochromatic scattering problem.

In the decoupled limit a free branch has the plane-wave form
\begin{equation}
    \phi_j(z)\sim e^{+\mathrm i k_j z}\quad\text{(forward)},\qquad
    \phi_j(z)\sim e^{-\mathrm i k_j z}\quad\text{(backward)},
\end{equation}
where the signs label the direction of propagation along $z$. The conserved spatial current for a single scalar channel is
\begin{equation}
    \mathcal F_j=\operatorname{Im}\bigl(\phi_j^*\,\partial_z\phi_j\bigr),
\end{equation}
giving $\mathcal F_j=+k_j$ for forward waves and $\mathcal F_j=-k_j$ for backward waves. In the coupled system the corresponding symplectic current generalizes this Wronskian to the full spatial phase space. Thus a spatial $(+,-)$ crossing is a forward/backward flux collision; the compact or non-compact local algebra is the flux-sign specialization of the normal form in Section~\ref{subsec:localTwoMode}.

\subsection{Spatial ladder and mode coupling}

When the spatial coefficients are periodic with period $L=2\pi/Q$, the Floquet-Bloch expansion of Section~\ref{sec:floquetBloch} applies with the spatial coordinate $z$ as the evolution variable. The spatial Bloch label $\kappa$ plays the spectral role analogous to that of the temporal quasi-frequency $\varepsilon_F$, and the sideband rungs are labeled by the wavenumbers
\begin{equation}
    k_n=\kappa+nQ,\qquad n\in\mathbb Z.
    \label{eq:spatialLadder}
\end{equation}
The shell denominators become functions of the conserved frequency $\omega$:
\begin{equation}
    D_\gamma(n;\omega)\equiv\omega^2-(\kappa+nQ)^2-\omega_{\rm pl}^2,\qquad
    D_a(n;\omega)\equiv\omega^2-(\kappa+nQ)^2-m_a^2.
\end{equation}
The explicit \(\omega\) argument distinguishes these fixed-frequency spatial denominators from the temporal functions \(D_{\gamma,a}(n)\) in Section~\ref{sec:floquet}.
For the static modulation \(\bar a(z)=a_0\cos Qz\), define
\begin{equation}
    \rho(z)\equiv \omega g\bar a'(z)=-\rho_Q\sin Qz,\qquad
    \rho_Q\equiv \omega g a_0 Q,\qquad
    \mathcal B_z\equiv\omega gB .
\end{equation}
As in the temporal problem, we keep the Cartesian polarizations and choose relative Fourier phases that display the rung connectivity:
\begin{align}
    A_x(z)&=\mathrm i\sum_{n\in\mathbb Z}
             A_{x,n}e^{\mathrm i(\kappa+nQ)z},\\
    A_y(z)&=-\sum_{n\in\mathbb Z}
             A_{y,n}e^{\mathrm i(\kappa+nQ)z},\\
    a(z)&=\sum_{n\in\mathbb Z}
             a_ne^{\mathrm i(\kappa+nQ)z}.
    \label{eq:CartesianSpatialExpansion}
\end{align}
With
\begin{equation}
    \tau_z\equiv\frac{\rho_Q}{2\mathrm i},
\end{equation}
the stationary spatial ladder is
\begin{subequations}
\label{eq:CartesianSpatialRec}
\begin{align}
    D_\gamma(n;\omega)A_{x,n}
    -\tau_z\bigl(A_{y,n-1}-A_{y,n+1}\bigr)
    &=\mathcal B_z a_n,\\
    D_\gamma(n;\omega)A_{y,n}
    -\tau_z\bigl(A_{x,n-1}-A_{x,n+1}\bigr)
    &=0,\\
    D_a(n;\omega)a_n&=\mathcal B_z A_{x,n}.
\end{align}
\end{subequations}
The spatial pump therefore alternates the Cartesian photon polarization between neighboring momentum rungs, while the external magnetic field mixes \(a_n\) with \(A_{x,n}\) on the same rung. The ladder again separates into two parity sectors,
\begin{equation}
    \mathcal C_0^{(z)}
    =\{a_{2\ell},A_{x,2\ell},A_{y,2\ell+1}\},
    \qquad
    \mathcal C_1^{(z)}
    =\{a_{2\ell+1},A_{x,2\ell+1},A_{y,2\ell}\},
    \qquad \ell\in\mathbb Z .
    \label{eq:spatialParityChains}
\end{equation}
For a cross-species resonance anchored at \(a_0\), the selected path is
\begin{equation}
    a_0\;\longleftrightarrow\;A_{x,0}
    \;\longleftrightarrow\;A_{y,-1}
    \;\longleftrightarrow\;A_{x,-2}
    \;\longleftrightarrow\;\cdots .
\end{equation}
Rung rephasings can make the scalar hopping matrix symmetric; the phase-matching conditions and reduced coupling magnitudes are convention independent.
For a cross-species path with signed harmonic difference \(N\), let \(W_N^{(z)}\) denote the projected product of spatial photon hops and off-resonant shell denominators. The shell-space endpoint coupling and its flux-normalized counterpart are
\begin{equation}
    \Xi_{a\gamma,z}^{(N)}\equiv \mathcal B_z W_N^{(z)},
    \qquad
    g_z^{(N)}
    \simeq
    \frac{\Xi_{a\gamma,z}^{(N)}}{2\sqrt{k_a k_\gamma}},
    \label{eq:spatialCouplingNormalization}
\end{equation}
up to the phase convention of the endpoint modes. The first quantity has dimensions of wavenumber squared and appears in the second-order shell equation; the second has dimensions of inverse length and appears in the first-order transfer equation. The same normalization applies to the magnitude of an opposite-flux endpoint coupling, with the backward mode represented by its flux-normalized amplitude.

The near-resonance conditions derived from this ladder are phase-matching conditions at fixed $\omega$. For a ray-projected temporal pump, replacing \(Q\) by the local \(Q_{\rm eff}\) reproduces the local phase mismatch. The coupling in Eq.~\eqref{eq:spatialCouplingNormalization} follows separately from projecting the temporal gradient interaction onto the chosen WKB rays. The resulting local two-mode system is then matched to the appropriate incoming and outgoing boundary conditions.

\subsection{Forward-forward conversion: the \texorpdfstring{$(+,+)$}{(+,+)} spatial channel}

A spatial $(+,+)$ crossing occurs when a forward axion branch and a forward photon branch become nearly degenerate. From the spatial ladder~\eqref{eq:spatialLadder}, the phase-matching condition is
\begin{equation}
    k_a(\omega)-k_\gamma(\omega)\simeq NQ,
    \label{eq:forwardMatch}
\end{equation}
with the local detuning
\begin{equation}
    \Delta_z^{(+,+)}(\omega)\equiv k_a(\omega)-k_\gamma(\omega)-NQ.
    \label{eq:detZpp}
\end{equation}
Here $k_a(\omega)=\sqrt{\omega^2-m_a^2}$ and $k_\gamma(\omega)=\sqrt{\omega^2-\omega_{\rm pl}^2}$ are the on-shell wavenumbers at the fixed frequency $\omega$.

Eliminating the off-resonant rungs and projecting onto the two forward-propagating modes yields the compact two-level system
\begin{equation}
    \mathrm i\partial_z
    \begin{pmatrix}u\\v\end{pmatrix}
    =
    \begin{pmatrix}
    \Delta_z/2 & g_z\\
    g_z^* & -\Delta_z/2
    \end{pmatrix}
    \begin{pmatrix}u\\v\end{pmatrix},
    \label{eq:spatialSU2}
\end{equation}
where \(u\) and \(v\) are the flux-normalized slowly varying envelopes and \(g_z\) is the spatially normalized endpoint coupling. This system preserves the definite norm \( |u|^2+|v|^2 \) and describes bounded spatial conversion without exponential growth or decay. At exact resonance, the first complete conversion occurs at \(\ell=\pi/(2|g_z|)\), while the probability oscillation period is \(\pi/|g_z|\).

For a finite, approximately uniform conversion region this system gives the usual bounded Rabi probability; the general estimate and its Landau--Zener inhomogeneous limit are collected in Section~\ref{sec:couplingScales}. Thus the spatial \((+,+)\) channel describes coherent forward scattering, not a temporal instability.

When the spatial periodicity is supplied by the ray-projected temporal pump with $Q=Q_{\rm eff}=m_a/v_g$, condition~\eqref{eq:forwardMatch} becomes the driven resonance of Ref.~\cite{Yao:2026yez}. The local WKB connection to the temporal detuning and the contrast with \(Q_{\rm real}\) are summarized in Section~\ref{sec:pheno}.

\subsection{Bragg and backscattering: the \texorpdfstring{$(+,-)$}{(+,-)} spatial channels}

A spatial $(+,-)$ crossing couples a forward branch to a backward branch. Two classes are relevant:
\begin{align}
    2k_\gamma(\omega) &\simeq NQ,
    &&\text{photon Bragg channel},
    \label{eq:photonBragg}\\
    k_a(\omega)+k_\gamma(\omega) &\simeq NQ,
    &&\text{cross-species Bragg channel}.
    \label{eq:crossBragg}
\end{align}
In the following local normal form, \(\Delta_z\) denotes the corresponding Bragg mismatch, either \(2k_\gamma-NQ\) or \(k_a+k_\gamma-NQ\). Projecting onto the resonant forward and backward modes gives the opposite-flux version of the Bogoliubov convention in Eq.~\eqref{eq:bogoliubovConvention},
\begin{equation}
    \partial_z
    \begin{pmatrix}u\\v\end{pmatrix}
    =
    \begin{pmatrix}
    -\mathrm i\Delta_z/2 & g_z\\
    g_z^* & +\mathrm i\Delta_z/2
    \end{pmatrix}
    \begin{pmatrix}u\\v\end{pmatrix},
    \label{eq:spatialSU11}
\end{equation}
which preserves the indefinite norm $|u|^2-|v|^2$. The spatial Bloch exponent is
\begin{equation}
    \gamma_z=\sqrt{|g_z|^2-(\Delta_z/2)^2}.
    \label{eq:zGrowth}
\end{equation}
When $|\Delta_z|<2|g_z|$, $\gamma_z$ is real and the spatial transfer spectrum has a stop band. The growing and decaying solutions form the local evanescent Bloch basis; the growing solution is the transfer-matrix partner of the decaying one, not a temporal instability. In a finite slab, the incoming wave and the outgoing or regularity condition on the far side select their physical combination. Transmission is then exponentially suppressed with length while reflection is enhanced, as in a photonic-crystal or distributed-Bragg stop band~\cite{Kogelnik:1972dfb,Yablonovitch:1987zz,John:1987zz}. Observable reflection and transmission amplitudes follow from these boundary conditions.

\begin{figure}[t]
    \centering
    \includegraphics[width=0.86\textwidth]{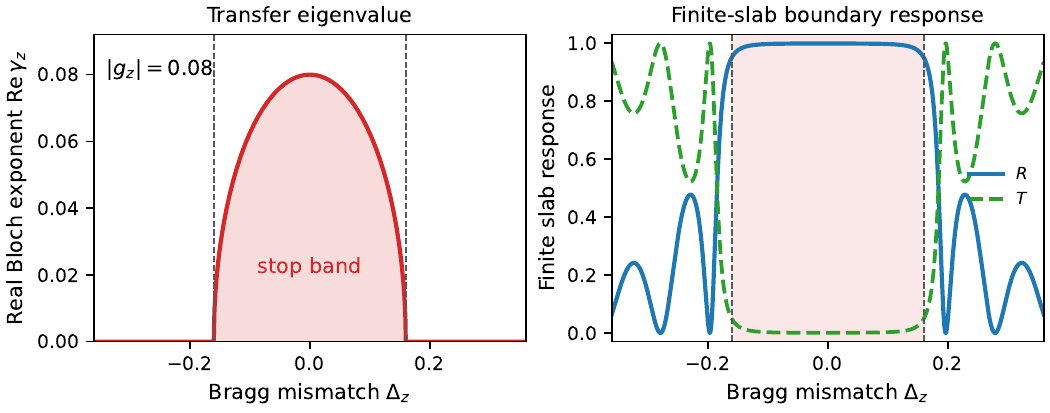}
    \caption{Two-mode check of the stationary spatial stop-band normal form. Left: real part of the Bloch exponent for a uniform forward/backward flux collision with \(|g_z|=0.08\), plotted as zero outside the interval where \(\gamma_z\) is real. Right: reflection \(R\) and transmission \(T\) of a finite slab with \(g_zL=4.4\), computed from the same two-mode transfer matrix with no incoming backward wave on the far side. The shaded interval marks \(|\Delta_z|<2|g_z|\), where the transfer eigenvalues are evanescent. The lossless normalization gives \(R+T=1\).}
    \label{fig:SpatialToyStopBand}
\end{figure}

Figure~\ref{fig:SpatialToyStopBand} applies this normal form to a finite slab with \(u(0)=1\), \(v(0)=r\), and no incoming backward wave on the far side, \(v(L)=0\). It shows how the real-\(\gamma_z\) interval identified from the local Bloch spectrum becomes a high-reflection, low-transmission interval after scattering boundary conditions are imposed.

The photon Bragg condition in Eq.~\eqref{eq:photonBragg} resembles the \(k\simeq N m_a/2\) locations of the forbidden photon bands found in Refs.~\cite{Espriu:2011vj,Espriu:2014lma}. The two gaps arise in different reductions: the temporal problem conserves \(k\), whereas the spatial transfer problem conserves \(\omega\). In those works the cold axion background is spatially homogeneous and time periodic, so \(k\) is conserved and the gap is a temporal Floquet gap in quasi-frequency space, equivalently the vacuum limit of Eq.~\eqref{eq:MathieuCond}.

By contrast, the stop band here is a fixed-\(\omega\) transfer effect induced by a spatial period \(Q\): the Bloch or transfer exponent in \(z\) becomes evanescent near a forward/backward flux collision. Thus \(m_a\) appears in Eq.~\eqref{eq:photonBragg} only if the physical construction supplies a spatial grating with \(Q\simeq m_a\). A genuine nonrelativistic axion component instead has true spatial wavenumber \(Q_{\rm real}=m_a v_a\).

The feasibility of these Bragg conditions depends on whether \(Q\) is a true axion spatial wavenumber, an engineered reciprocal vector, or a ray-projected period. Section~\ref{sec:phenoQ} gives the corresponding scale comparison and shows that backward Bragg matching requires a dedicated backward-branch WKB construction with the appropriate boundary conditions.

Relations to axion-polariton gaps and engineered phase-matching media are discussed alongside the broader literature comparison in Section~\ref{sec:pheno}.

\subsection{Summary of spatial channels}

The fixed-$\omega$ transfer problem has a compact forward-forward channel, which gives bounded spatial conversion, and two opposite-flux channels, which give photon and cross-species Bragg stop bands. Here \(Q\) is the spatial wavenumber supplied by the chosen stationary construction. Table~\ref{tab:pheno} compares these channels with their temporal and ray-projected counterparts; Section~\ref{sec:phenoQ} then distinguishes \(Q_{\rm real}\) from \(Q_{\rm eff}\).

%% file: sections/phenomenological_classification.tex
\section{Phenomenological classification}
\label{sec:pheno}

\subsection{Unified channel table}

Table~\ref{tab:pheno} classifies the near-resonant Maxwell-axion channels developed in Sections~\ref{sec:evolution}--\ref{sec:transfer}. Each row is a distinct physical problem, specified by its evolution variable, conserved label, signature, and, for spatial channels, the origin of the period $Q$.

\begin{table}[H]
\centering
\caption{Classification of the near-resonant channels. Each row specifies the evolution problem, signature, detuning, and physical outcome. Here \(N\) is the signed harmonic difference, with the orientation chosen so that the displayed sum-frequency and Bragg conditions have \(N>0\).}
\label{tab:pheno}
\renewcommand{\arraystretch}{1.18}
\scriptsize
\begin{tabularx}{\textwidth}{lcc>{\raggedright\arraybackslash}X>{\raggedright\arraybackslash}X}
\toprule
Channel & $s$; fixed & Sign & Detuning & Outcome \\
\midrule
Temporal \(a\)-\(\gamma\) difference
& \(t;k\) & \((+,+)\)
& \(\omega_a-\omega_\gamma-Nm_a\)
& Avoided crossing and bounded beating \\
Temporal Mathieu
& \(t;k\) & \((+,-)\)
& \(2\omega_\gamma-Nm_a\)
& Photon-pair parametric instability \\
Temporal MAS
& \(t;k\) & \((+,-)\)
& \(\omega_a+\omega_\gamma-Nm_a\)
& Cross-species parametric instability \\
Static/engineered forward transfer
& \(z;\omega\) & \((+,+)\)
& \(k_a-k_\gamma-NQ\)
& Bounded spatial conversion \\
Static/engineered photon Bragg
& \(z;\omega\) & \((+,-)\)
& \(2k_\gamma-NQ\)
& Stop band and distributed reflection \\
Static/engineered cross Bragg
& \(z;\omega\) & \((+,-)\)
& \(k_a+k_\gamma-NQ\)
& Cross-species stop band \\
Ray-projected forward transfer
& \(z;\omega\) (local WKB) & \((+,+)\)
& \(k_a-k_\gamma-NQ_{\rm eff}\)
& Local driven conversion \\
Ray-projected backward transfer
& \(z;\omega\) (local WKB) & \((+,-)\)
& Branch dependent
& Requires a separate backward-ray construction \\
\bottomrule
\end{tabularx}
\end{table}

In the stationary rows, \(Q\) is the wavenumber of an engineered or otherwise genuinely static modulation. Setting \(Q=Q_{\rm real}\) uses the equal-time spatial phase of a moving axion component only as a proxy and does not define an exact stationary fixed-\(\omega\) problem. The stationary and ray-projected rows share a local two-mode form but have different phase mismatches, couplings, and boundary conditions, as quantified in Section~\ref{sec:phenoQ}. Likewise, opposite signs denote temporal parametric growth for Krein signatures but spatial stop-band behavior for flux signatures.

\subsection{The two \texorpdfstring{$Q$}{Q} scales}
\label{sec:phenoQ}

A central distinction in axion electrodynamics is between two quantities that both carry dimensions of wavenumber but have entirely different physical origins:
\begin{align}
    Q_{\rm real}&\equiv m_a v_a,
    &&\text{true spatial Fourier wavenumber of the axion field},
    \label{eq:QrealPheno}\\
    Q_{\rm eff}&\equiv\frac{m_a}{v_g},
    &&\text{ray-projected effective wavenumber of a temporal pump}.
    \label{eq:QeffPheno}
\end{align}
For virial velocities $v_a\sim10^{-3}$, $Q_{\rm real}$ is three orders of magnitude smaller than $m_a$, while $Q_{\rm eff}$ is of order $m_a$ for relativistic propagation. This scale separation leads to different phase-matching regimes.

When the equal-time axion profile is idealized as a static grating, \(Q_{\rm real}\) sets the reciprocal-wavenumber scale of that proxy. In vacuum, the photon condition $2k_\gamma\simeq NQ_{\rm real}$ gives \(k_\gamma\sim N m_a v_a/2\) and hence \(\omega\sim N m_a v_a/2\) for low order. The cross-species condition \(k_a+k_\gamma\simeq NQ_{\rm real}\) is more restrictive: both modes must propagate at the common frequency, so \(\omega\ge\max(m_a,\omega_{\rm pl})\), while their two nonnegative wavenumbers must sum to the much smaller scale \(N m_a v_a\). At low order, this condition can be approached near simultaneous propagation thresholds. Away from those thresholds, it requires compensating medium dispersion. At typical astrophysical photon energies, both static-proxy Bragg channels are inaccessible at low sideband order unless the medium supplies such compensation or \(|N|\) is very large.

$Q_{\rm eff}$ controls forward driven conversion. The condition $k_a-k_\gamma\simeq NQ_{\rm eff}$ can be satisfied when $\omega\gg m_a v_a$, because $Q_{\rm eff}\sim m_a$ provides the necessary phase compensation. This is the regime studied in Ref.~\cite{Yao:2026yez}. Since $Q_{\rm eff}$ is a ray-construction quantity, backward Bragg conditions such as $2k_\gamma\simeq NQ_{\rm eff}$ or $k_a+k_\gamma\simeq NQ_{\rm eff}$ require a dedicated backward-branch WKB construction with the appropriate incoming boundary conditions.

The single-pump formulas also assume a monochromatic coherent component. Virialized axion dark matter is instead a superposition of components with velocity spread \(\delta v\sim v_a\sim10^{-3}\). The resulting pump-frequency spread is
\begin{equation}
    \delta\Omega_{\rm p}\sim m_a v_a\,\delta v,
\end{equation}
and the true spatial Fourier spread is
\begin{equation}
    \delta Q_{\rm real}\sim m_a\,\delta v .
\end{equation}
Thus a resonance whose intrinsic width in detuning is narrower than \(|N|\delta\Omega_{\rm p}\) in a temporal fixed-\(k\) problem, or narrower than \(|N|\delta Q_{\rm real}\) in an equal-time static-proxy ensemble, is broadened or phase averaged by the pump ensemble. The high-order MAS tongues are especially sensitive because their widths scale with products of weak sideband couplings, while compact forward-conversion channels are less fragile when the conversion region samples only one coherent component over its coherence time.

\subsection{The WKB connection between temporal and spatial detunings}

In a strictly homogeneous temporal problem, the wave number $k$ is a global quantum number and the detuning
\begin{equation}
    \Delta_t(k)=\omega_a(k)-\omega_\gamma(k)-Nm_a
\end{equation}
describes same-$k$ temporal avoided crossing. The Floquet multipliers directly diagnose stability.

In an inhomogeneous propagation environment, the plasma frequency $\omega_{\rm pl}(z)$ and the magnetic field $B(z)$ vary in space, breaking spatial translation symmetry. In this setting $k$ is no longer a good quantum number, and a fixed-$\omega$ description with local wavenumbers
\begin{equation}
    k_a(z,\omega)=\sqrt{\omega^2-m_a^2},\qquad
    k_\gamma(z,\omega)=\sqrt{\omega^2-\omega_{\rm pl}^2(z)},
\end{equation}
and the propagation detuning
\begin{equation}
    \Delta_z(z,\omega)=k_a(z,\omega)-k_\gamma(z,\omega)-NQ_{\rm eff}(z)
    \label{eq:inhomDetuning}
\end{equation}
becomes the natural language. Here \(v_{\rm ray}\equiv v_g\) is the group velocity of the common reference ray and \(Q_{\rm eff}(z)=m_a/v_{\rm ray}(z)\) is the corresponding local ray-projected effective wavenumber.

For a narrow forward wave packet whose two branches can be described by a common reference ray near a crossing, let \(\phi\) denote the interaction phase accumulated along that ray. Define the ray-projected temporal mismatch by
\begin{equation}
    \Delta_t^{\rm ray}\equiv\frac{\mathrm d\phi}{\mathrm dt},
    \qquad
    \frac{\mathrm d\phi}{\mathrm dz}\equiv\Delta_z,
\end{equation}
with \(t_{\rm ray}'(z)\simeq1/v_{\rm ray}\) and \(Q_{\rm eff}=m_a/v_{\rm ray}\). The chain rule then gives the local WKB mapping
\begin{equation}
    \Delta_t^{\rm ray}\simeq v_{\rm ray}\,\Delta_z,
    \label{eq:WKBmap}
\end{equation}
so the zeros of the two ray mismatches coincide locally. The quantity \(\Delta_t^{\rm ray}\) is an accumulated phase rate along the selected trajectory, not the global fixed-\(k\) Floquet detuning \(\Delta_t(k)\) defined above. Equation~\eqref{eq:WKBmap} therefore provides a local correspondence between the temporal and spatial descriptions rather than an identity between their globally defined detunings. Under this common-ray approximation, a temporal $(+,+)$ avoided crossing maps locally to a spatial forward conversion.

The mapping~\eqref{eq:WKBmap} has well-defined limits of validity. It fails near plasma cutoffs and turning points ($v_{\rm ray}\to0$), where the WKB approximation itself breaks down and a turning-point analysis, typically based on local Airy matching, is required. A Landau--Zener reduction applies instead to an isolated avoided crossing between well-defined propagating asymptotic modes. The common-ray mapping also requires negligible differential group delay across the resonant region. If \(v_{g,a}\) and \(v_{g,\gamma}\) differ substantially, the axion and photon branches must be propagated on separate rays, and the local two-mode system is then a coupled-mode approximation rather than a direct image of the temporal detuning.

\subsection{Coupling scales and observable sizes}
\label{sec:couplingScales}

The local classification fixes the algebraic type of a crossing, but its observable size is set by the same endpoint coupling that appears in the reduced two-mode problem. For cross-species axion-photon sideband channels, the perturbative path contains one projected magnetic axion-photon bridge and \(|N|\) pump-induced photon sideband hops. In the temporal normalization of Section~\ref{sec:floquet}, and suppressing the frequency-sign labels,
\begin{equation}
    \Xi_{a\gamma}^{(N)}\sim \mathcal G_B W_N,
    \qquad
    \mu_{a\gamma}^{(N)}
    \sim
    \frac{\Xi_{a\gamma}^{(N)}}{2\sqrt{\omega_a\omega_\gamma}},
    \label{eq:phenoMuFromXi}
\end{equation}
where \(\mathcal G_B\propto gB\) is the Cartesian same-rung magnetic entry and \(W_N\) is the order-\(|N|\) projected photon harmonic generated by the axion pump. The ladder evaluates \(W_N\) as a product of sideband hops and off-resonant denominators. The same elimination separately generates the diagonal self-energy shifts. The co-rotating construction of Appendix~\ref{app:JacobiAnger} instead resums the same-branch polarization dressing into \(J_N(\alpha_\gamma)\), while leaving the positive-negative-frequency pump block explicit. Selecting one isolated harmonic and its two endpoints gives the schematic magnetic-vertex scaling
\begin{equation}
    |\mu_{a\gamma,\mathrm{co}}^{(N)}|
    \simeq
    \frac{gB}{2}\sqrt{\frac{\omega_a}{\omega_\gamma}}\,
    |J_N(\alpha_\gamma)|
    \simeq
    \frac{gB}{2}\sqrt{\frac{\omega_a}{\omega_\gamma}}\,
    \frac{|\alpha_\gamma|^{|N|}}{2^{|N|}|N|!},
    \label{eq:gBalphaN}
\end{equation}
where \(\alpha_\gamma=\alpha_a k/(2\omega_\gamma)\), and the second estimate assumes \(|\alpha_\gamma|\ll1\). Equation~\eqref{eq:MuCoRotating} restores the parity phase and branch label suppressed here. For same-branch dressing, the Bessel coefficient resums both the shortest ladder path and paths with additional backtracking hops. Its leading coefficient agrees with the product-denominator expression in the high-carrier/eikonal overlap \(|N|m_a\ll\omega_\gamma\); outside that limit, the explicit shell denominators must be retained. The product-denominator and Bessel expressions encode the same polarization dressing in two different representations. Temporal opposite-sign channels also require the residual branch-changing pump block and its self-energy corrections in either representation.

The sideband order counts powers of the pump amplitude rather than repeated magnetic conversions. The generic cross-species scaling is therefore \(gB\,\alpha_\gamma^{|N|}\), not \((gB)^{|N|}\), although the coefficient remains channel dependent outside the eikonal overlap. This counting applies to temporal same-sign conversion, the MAS channel, and ray-projected forward conversion when one magnetic bridge is dressed by \(|N|\) pump harmonics. Same-species Mathieu and photon Bragg channels contain no magnetic bridge and are controlled by the photon pump block.

Once the appropriate local coupling is known, the observable estimate follows from the algebraic type. For a compact conversion channel with approximately constant detuning over length \(\ell\),
\begin{equation}
    P_{a\to\gamma}
    \sim
    \frac{|g_z^{(N)}|^2}{|g_z^{(N)}|^2+\Delta_z^2/4}
    \sin^2\!\left[
    \sqrt{|g_z^{(N)}|^2+\Delta_z^2/4}\,\ell
    \right],
    \label{eq:phenoRabiEstimate}
\end{equation}
so \(P_{a\to\gamma}\simeq |g_z^{(N)}|^2\ell^2\) in the weak, phase-matched limit. A detuning swept through an isolated layer instead gives the Landau--Zener passage derived below. For temporal opposite-sign channels, the maximal growth rate at exact resonance is
\begin{equation}
    \gamma_{\rm max}^{(N)}\simeq |\mu_{a\gamma}^{(N)}|,
    \label{eq:phenoGrowthEstimate}
\end{equation}
and the momentum-space half-width follows by dividing the detuning boundary \(2|\mu|\) by the slope of the relevant detuning, as in Eq.~\eqref{eq:tongueWidth}. For spatial opposite-flux channels, the same estimate gives the stop-band exponent \(\gamma_z\simeq |g_z^{(N)}|\); in a uniform slab on resonance, reflection becomes order unity when \(|g_z^{(N)}|L\gtrsim1\).

For a benchmark local density \(\rho_a\simeq0.4\,{\rm GeV\,cm^{-3}}\), mass \(m_a=10^{-22}\,{\rm eV}\), and coupling \(g=10^{-10}\,{\rm GeV}^{-1}\), the coherent amplitude gives
\begin{equation}
    \alpha_a\equiv g a_0\simeq \frac{g\sqrt{2\rho_a}}{m_a}\sim1 .
\end{equation}
A laboratory-scale magnetic field \(B=10\,{\rm T}\) gives \(gB\sim2\times10^{-16}\,{\rm eV}\), corresponding to a frequency scale of order \(10^{-1}\,{\rm s}^{-1}\). For a resonant \(N=2\) cross-species channel with an unsuppressed harmonic factor, this value sets only a parametric upper scale for the local rate; the actual rate is determined by the channel-specific projected weight \(W_N\) and the endpoint normalization.

The validity of the small-\(\alpha_\gamma\) expansion and the isolated-harmonic reduction is set by \(\alpha_\gamma\) and the harmonic separation rather than by the benchmark \(\alpha_a\sim1\) alone. When several harmonics contribute, the quantitative treatment uses the full Bessel series or the corresponding ladder truncation. The enlarged couplings used in the numerical figures are visualization parameters for multiplier topology, finite-coupling self-energy shifts, and weak-coupling scaling; a quantitative prediction requires the relevant \(W_N\), coherence time, geometry, and boundary conditions.

\subsection{Inhomogeneous plasma: Floquet sidebands as generalized level crossing}

The standard axion-photon conversion problem in an inhomogeneous magnetized plasma is usually formulated as a static level crossing: resonant conversion occurs where the local wavenumbers satisfy $k_a(\omega)=k_\gamma(\omega)$~\cite{Raffelt:1987im,Yoshimura:1987ma,Lai:2006af,Pshirkov:2007st,Hook:2018iia}. For the dispersion relations in Eq.~\eqref{eq:inhomDetuning}, this condition is equivalently
\begin{equation}
    \omega_{\rm pl}(z_*)=m_a,
    \label{eq:standardLevelCrossing}
\end{equation}
independent of the external frequency as long as both modes are propagating. The associated conversion probability is often estimated with the Landau--Zener formula~\cite{Zener:1932ws,Carenza:2023nck}. This is the $N=0$ member of the Floquet-Bloch sideband family.

The oscillating axion background extends this picture by providing discrete momentum quanta $Nm_a/v_g(z)$ that can compensate a nonzero wavenumber mismatch. Resonant conversion can then occur even when $k_a\neq k_\gamma$, at positions $z_*$ satisfying
\begin{equation}
    k_a(z_*,\omega)-k_\gamma(z_*,\omega)\simeq N\,\frac{m_a}{v_g(z_*)},
    \qquad N\in\mathbb Z.
    \label{eq:FloquetLevelCrossing}
\end{equation}
These Floquet-Bloch sidebands open the driven mixing channels identified in Ref.~\cite{Yao:2026yez} away from the conventional level-crossing point. Averaging over the pump phase removes these coherent sideband terms because they arise from the explicitly time-dependent drive rather than from an adiabatic static correction.

Ref.~\cite{Yao:2026yez} projects onto a single near-resonant Floquet harmonic. In this rotating-wave approximation, Eq.~\eqref{eq:spatialSU2} describes bounded Rabi exchange between the two forward modes while a wave packet remains within one resonance band. In an inhomogeneous plasma, the changing profile also determines how the packet enters and exits the band. As $\Delta_z(z)$ passes through zero along a ray, the net conversion across an isolated layer becomes a spatial Landau--Zener passage through the Floquet-shifted avoided crossing.

In the present framework, each \(N\neq0\) driven-conversion channel considered here is a spatial $(+,+)$ forward-forward crossing with $Q=Q_{\rm eff}$. To fix the normalization, write the local crossing in the flux-normalized spatial form
\begin{equation}
    \mathrm i\partial_z
    \binom{u}{v}
    =
    \begin{pmatrix}
    \Delta_z(z)/2 & g_z\\
    g_z^* & -\Delta_z(z)/2
    \end{pmatrix}
    \binom{u}{v},
    \label{eq:SpatialLZSystem}
\end{equation}
where \(g_z\) is the coupling that appears in the \(z\)-evolution equation. The single-crossing Landau--Zener probability~\cite{Zener:1932ws,Carenza:2023nck} is then
\begin{equation}
    P_{a\to\gamma}\simeq
    1-\exp\!\left(
    -\frac{2\pi |g_z|^2}
          {|\mathrm d\Delta_z/\mathrm d z|_{z_*}}
    \right).
    \label{eq:LZ}
\end{equation}
The factor of \(2\pi\) follows from the convention in Eq.~\eqref{eq:SpatialLZSystem}. If a time-normalized ray coupling \(\mu_t^{\rm ray}\) is used together with \(\Delta_t^{\rm ray}\simeq v_{\rm ray}\Delta_z\), it must first be converted to the spatial normalization, \(g_z\simeq \mu_t^{\rm ray}/v_{\rm ray}\) for slowly varying \(v_{\rm ray}\).

Equation~\eqref{eq:LZ} assumes a detuning that is approximately linear near \(z_*\), slowly varying \(g_z\) and \(v_{\rm ray}\), and asymptotic regions sufficiently far from resonance. Magnetic-field variation and finite boundary placement can invalidate this single-crossing estimate in realistic plasmas~\cite{Carenza:2023nck}.

If several Floquet orders are relevant, or a non-monotonic profile gives several roots of Eq.~\eqref{eq:FloquetLevelCrossing}, the result is a Landau--Zener--St\"uckelberg problem. Successive crossing matrices must then be combined with the phase evolution between them; only in the weak or phase-averaged limit does this reduce to an approximate probability cascade. Strongly magnetized or anisotropic plasmas, including Euler--Heisenberg/QED corrections to the photon dispersion in neutron-star fields, require ray tracing and dielectric-response ingredients beyond this local scalar model~\cite{Witte:2021arp,Millar:2021gzs,Long:2024qvd}. Equations~\eqref{eq:FloquetLevelCrossing} and~\eqref{eq:LZ} describe the single-layer limit; more general profiles require a full profile-dependent transfer calculation.

\subsection{Relation to existing work}

The classification above locates several established results within their respective evolution problems. The Mathieu instability and the MAS sum-frequency channel are opposite-sign temporal collisions~\cite{Arza:2018dcy,Masaki:2019ggg}. The driven resonance of Ref.~\cite{Yao:2026yez} is a ray-projected forward channel with \(Q=Q_{\rm eff}\), while standard static level-crossing conversion is the \(N=0\) member of the same fixed-\(\omega\) crossing family~\cite{Raffelt:1987im,Yoshimura:1987ma,Lai:2006af,Pshirkov:2007st,Hook:2018iia} and underlies neutron-star and transient-signal applications~\cite{Foster:2020pgt,Edwards:2020afl}.

Heterodyne and upconversion axion searches exhibit the same reduced algebra but do not directly realize the fixed-$k$ Maxwell-axion problem solved here. In those settings the axion background can mediate photon-photon difference-frequency conversion between two electromagnetic modes, or between a strongly occupied carrier and a signal sideband, with \(\omega_2-\omega_1\simeq m_a\) or \(\omega_{\rm sig}\simeq\Omega_{\rm LO}\pm m_a\), where \(\Omega_{\rm LO}\) is the local-oscillator frequency~\cite{Berlin:2020vrk,Thomson:2021zvq,Li:2025pyi}. After projection onto two resonant photon envelopes the local system is positive-positive and produces bounded mode conversion, but the microscopic reduction requires additional cavity or carrier degrees of freedom and input-output boundary conditions; practical implementations must also control ordinary electromagnetic mode-conversion backgrounds~\cite{Ueki:2024sfk}.

Axion polaritons provide a stationary hybridization analogue in which a propagating material axion mode mixes with electromagnetic waves~\cite{Qi:2008ew,Li:2009tla,Wang:2015ava,Zhu:2022fvl,Marsh:2018dlj}. Their bulk dispersion has the schematic form
\begin{equation}
    D_\gamma(\omega,k)D_a(\omega,k)-|\mathcal G(\omega,k)|^2=0,
\end{equation}
where \(\mathcal G\) is the magnetic-field-induced mixing element. At fixed real frequency, ordinary polaritonic attenuation occurs when the hybridized bulk dispersion has no real wavenumber. This differs from the local \((+,+)\) equation for two real co-propagating branches, which gives bounded conversion, and from the forward/backward sideband collision that produces a Bragg stop band.

Engineered structures supply the missing momentum by external rather than axion-background Fourier components. Dielectric and multilayer haloscopes control transfer phases and impedance matching~\cite{Caldwell:2016dcw,Baryakhtar:2018doz}. Crystal Bragg-Primakoff conversion instead uses a reciprocal-lattice vector~\cite{Dent:2023gzl,Thompson:2023jbo}. Modulated fields, quasi-phase matching, and axion magnetic resonance compensate the dispersion mismatch with structured external fields~\cite{VanBibber:1987rq,Arias:2016zqu,Seong:2023ran}. These systems share the fixed-\(\omega\) transfer viewpoint but differ in the physical source of the sideband momentum.

%% file: sections/numerical_illustrations.tex
\section{Numerical illustrations}
\label{sec:numerics}

We illustrate the temporal fixed-$k$ instabilities by solving the real block of Appendix~\ref{app:block} directly, without a rotating-wave or sideband truncation in the time-domain evolution. The sideband ladder predicts the relevant collision and reduced coupling; the numerical monodromy tests the complete periodic system. Ref.~\cite{Yao:2026yez} treated ray-projected forward conversion through WKB propagation and numerical envelope evolution in inhomogeneous media. Here we instead focus on the homogeneous temporal pump, for which the Mathieu and MAS channels are direct Floquet initial-value problems.

The numerical integration is performed in the canonical phase-space variables \(X=(q,p)^{\mathsf T}\), with \(p=\dot q+G_Rq/2\). For each value of \(k\) we integrate the fundamental matrix
\begin{equation}
    \dot\Phi(t;k)=M_{\rm can}(t;k)\Phi(t;k),\qquad \Phi(0;k)=I_6,
\end{equation}
over one pump period $T=2\pi/m_a$. The monodromy matrix is
\begin{equation}
    F_t(k)=\Phi(T;k),
\end{equation}
or equivalently
\begin{equation}
    F_t(k)=\mathcal T\exp\!\left[\int_0^T M_{\rm can}(t;k)\,\mathrm dt\right],
    \label{eq:monodromyNumerics}
\end{equation}
where \(\mathcal T\) denotes time ordering and
\begin{equation}
    M_{\rm can}(t;k)=
    \begin{pmatrix}
    -G_R/2 & I\\
    -V_R(t)+G_R^2/4 & -G_R/2
    \end{pmatrix}.
\end{equation}
In components, if \(e_j\) is the \(j\)th basis vector of \(\mathbb R^6\), the \(j\)th column of \(F_t(k)\) is the solution at \(t=T\) evolved from the canonical initial condition \(X(0)=e_j\). Thus the numerical construction is six simultaneous initial-value problems for the canonical first-order system. The six eigenvalues of \(F_t(k)\) are the Floquet multipliers $\lambda_j(k)$. The growth rate plotted below is
\begin{equation}
    \gamma(k)=\frac{1}{T}\max_j\log|\lambda_j(k)|,
\end{equation}
with negative numerical roundoff clipped to zero. The multiplier branches are tracked across the $k$ scan by nearest-neighbor matching in the complex plane. The colored trajectories are slightly displaced along their local normal direction for visibility; this display-only displacement does not enter the growth calculation.
For the MAS quartet, the highlighted branches are assigned colors after sorting by their local angular position \(\arg(\lambda_j-\lambda_*)\) around the bare collision point; this keeps the color convention fixed across the coupling-scale panels.

The left panels in the figures show one-dimensional cuts through the corresponding instability tongues. A standard Ince--Strutt diagram displays stable and unstable regions in a two-parameter plane, for example detuning versus modulation amplitude in the Mathieu problem. Here we keep the pump amplitude and magnetic coupling fixed and scan the conserved momentum \(k\). The shaded interval denotes the numerically resolved growth region, defined for display by \(\gamma\ge 0.005\,\gamma_{\max}\), while the middle and right panels show the corresponding monodromy-multiplier motion along the same cut. The coupling-scaling test below uses the full width at half maximum (FWHM) of the growth curve, so the scaling test does not depend on this visual threshold.

The plots diagnose the instability through the multiplier topology: unit-circle multipliers collide and leave the circle as reciprocal pairs. The analytic endpoint projection identifies these collisions as opposite-Krein-sign channels. The stable unit-circle branches also permit a direct numerical sign check through $h(v,v)=\mathrm i v^\dagger Jv$, although this diagnostic is not plotted.

\subsection{Mathieu channel}

\begin{figure}[t]
    \centering
    \begin{subfigure}{0.45\textwidth}
        \centering
        \includegraphics[width=\linewidth]{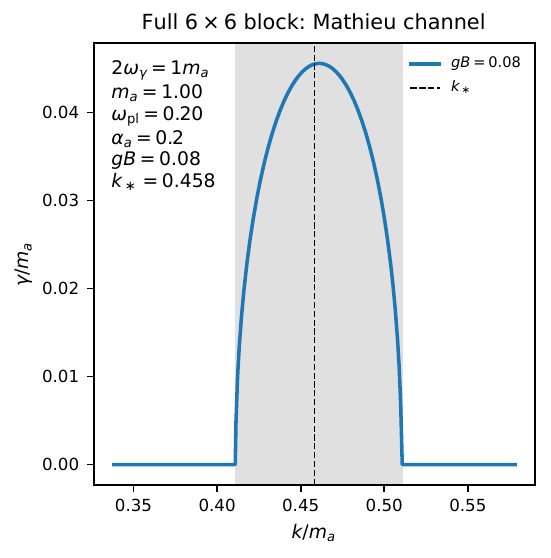}
        \caption{Growth}
    \end{subfigure}
    \hfill
    \begin{subfigure}{0.45\textwidth}
        \centering
        \includegraphics[width=\linewidth]{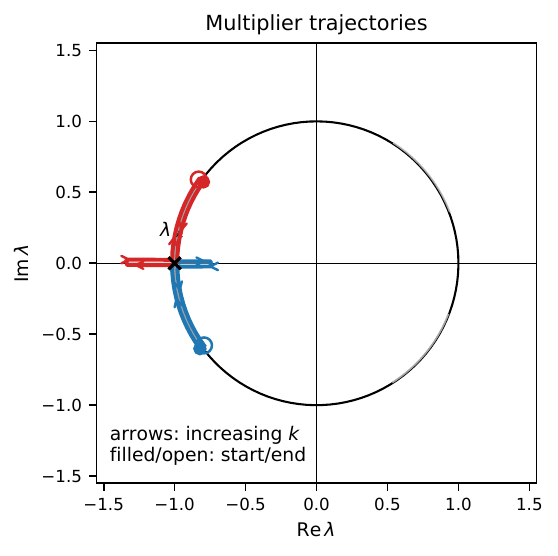}
        \caption{Global spectrum}
    \end{subfigure}
    \hfill\\
    \begin{subfigure}{0.45\textwidth}
        \centering
        \includegraphics[width=\linewidth]{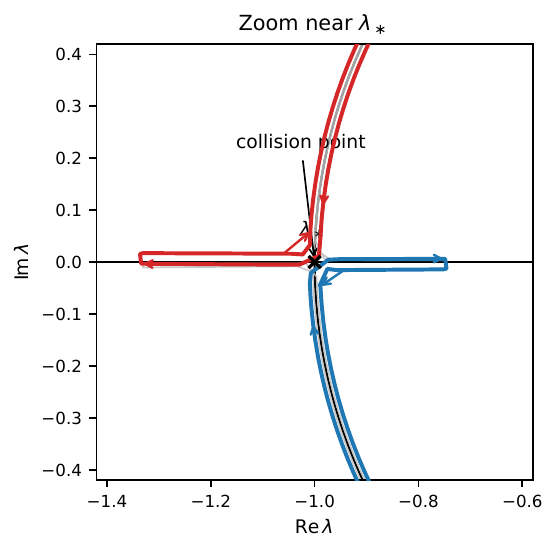}
        \caption{Collision zoom}
    \end{subfigure}
    \caption{Numerical Mathieu instability in the full temporal $6\times6$ real block. Left: one-dimensional fixed-coupling scan through the instability tongue, with growth rate obtained from the canonical monodromy multipliers. Middle: global multiplier trajectories in the complex plane. Right: zoom near the bare collision point $\lambda_*=-1$.}
    \label{fig:MathieuNumerics}
\end{figure}

Figure~\ref{fig:MathieuNumerics} shows the single-species Mathieu channel in the full $6\times6$ block, realizing the stimulated-decay instability of a homogeneous background. The parameters are $m_a=1$, $\omega_{\rm pl}=0.2$, $\alpha_a=0.2$, $gB=0.08$, and $N=1$. The bare resonance is determined by $2\omega_\gamma(k_*)=m_a$, giving $k_*\simeq0.458$. The growth curve peaks close to this value, and the multiplier trajectories show the expected positive-negative collision near $\lambda_*=-1$. In this channel the two highlighted branches approach the collision point from opposite sides of the unit circle and then split radially into a reciprocal pair, the direct multiplier-plane signature of the Mathieu instability described by Eq.~\eqref{eq:MathieuGrowth}.

\subsection{MAS channel}

\begin{figure}[t]
    \centering
    \begin{subfigure}{0.45\textwidth}
        \centering
        \includegraphics[width=\linewidth]{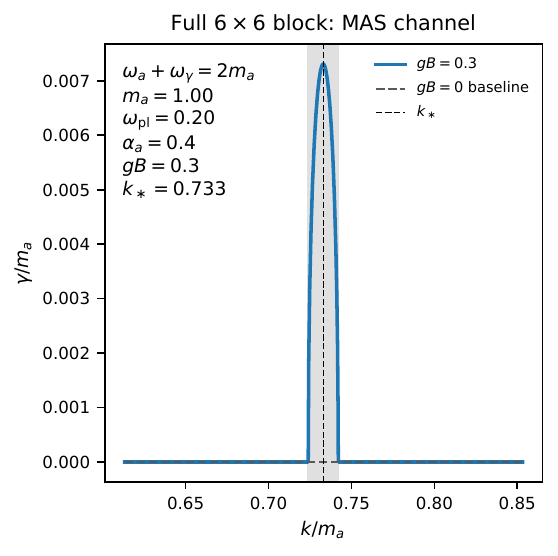}
        \caption{Growth}
    \end{subfigure}
    \hfill
    \begin{subfigure}{0.45\textwidth}
        \centering
        \includegraphics[width=\linewidth]{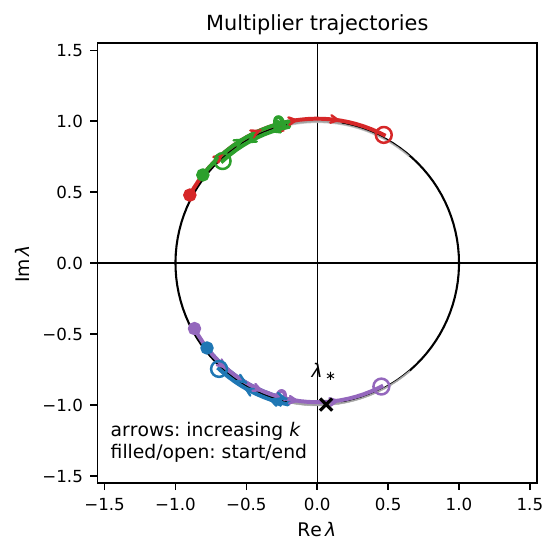}
        \caption{Global spectrum}
    \end{subfigure}
    \hfill\\
    \begin{subfigure}{0.45\textwidth}
        \centering
        \includegraphics[width=\linewidth]{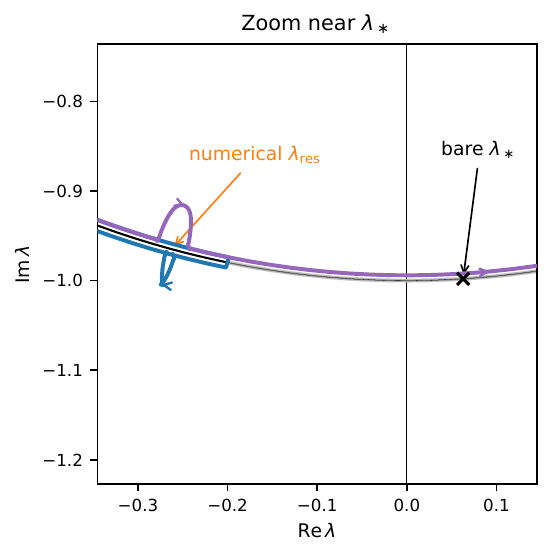}
        \caption{Phase zoom}
    \end{subfigure}
    \caption{Finite-coupling MAS instability in the full temporal $6\times6$ real block. Left: one-dimensional scan through the MAS tongue; the shading marks the resolved-growth interval defined in the text. Middle: global multiplier trajectories, including the reciprocal quartet associated with the cross-species positive-negative collision. Right: zoom comparing the leading-order bare marker \(\lambda_*\) with the numerical resonance phase \(\lambda_{\rm res}\), obtained from the unstable multiplier at maximal growth.}
    \label{fig:MASNumerics}
\end{figure}

Figure~\ref{fig:MASNumerics} shows the cross-species MAS sum-frequency channel with $m_a=1$, $\omega_{\rm pl}=0.2$, $\alpha_a=0.4$, $gB=0.3$, and $N=2$. These couplings are deliberately enlarged so that the reciprocal multiplier quartet is visible. The bare crossing is fixed by $\omega_a(k_*)+\omega_\gamma(k_*)=2m_a$, giving $k_*\simeq0.733$ and the bare multiplier marker
\begin{equation}
    \lambda_*=\exp[-\mathrm i\omega_a(k_*)T].
\end{equation}
To display the finite-coupling phase, we evaluate a highlighted unstable multiplier at the momentum of maximal growth and project it radially onto the unit circle; the resulting marker is denoted by \(\lambda_{\rm res}\).

The angular displacement of \(\lambda_{\rm res}\) from \(\lambda_*\) is sensitive to the endpoint self-energy shifts discussed in Section~\ref{sec:floquet} and Appendix~\ref{app:MAS}. Because the marker is evaluated at maximal growth, its displacement can also contain higher-order effects from the momentum dependence of the reduced coupling and self-energies; the weak-coupling scaling below isolates the leading behavior. Since $\arg\lambda=-\varepsilon_F T$ modulo $2\pi$, a modest quasi-frequency shift can produce a prominent angular displacement without a comparable shift of the tongue center in $k$.

The separation is especially visible for the \(N=2\) MAS channel because the off-diagonal coupling scales as \(\Xi_{a\gamma}^{(+,-;2)}\sim\mathcal G_B\tau_R^2\), while the leading diagonal shifts scale as \(\Sigma_a\sim\mathcal G_B^2/D_\gamma(0)\) and \(\Sigma_\gamma\sim\tau_R^2/D_\gamma(-1)\). Here \(\mathcal G_B\) is the same-rung Cartesian magnetic bridge and \(\tau_R\) is the rephased real photon hopping defined near Eq.~\eqref{eq:GBtauDefs}. Under a common weak-coupling rescaling, the tongue width therefore decreases faster than the resonance-phase displacement. Appendix~\ref{app:MAS} gives the corresponding Schur-complement formulas. In the positive-positive and Mathieu channels, analogous diagonal terms are normally absorbed into the effective detuning.

\subsection{Coupling dependence}

\begin{figure}[t]
    \centering
    \begin{subfigure}{0.45\textwidth}
        \centering
        \includegraphics[width=\linewidth]{figures/fig_mas_canonical_panel_multipliers_zoom.pdf}
        \caption{$\zeta=1$}
    \end{subfigure}
    \hfill
    \begin{subfigure}{0.45\textwidth}
        \centering
        \includegraphics[width=\linewidth]{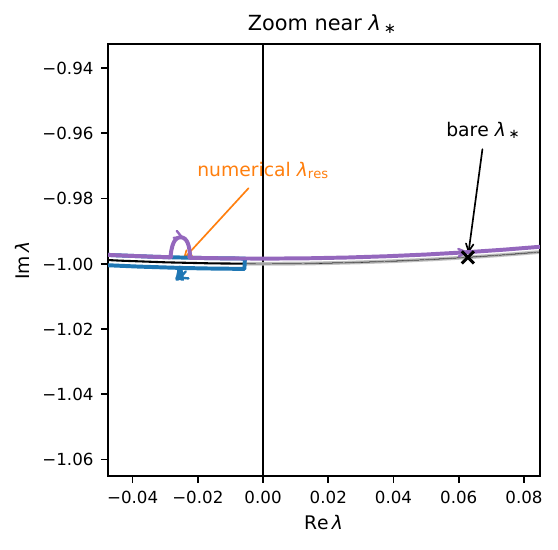}
        \caption{$\zeta=0.5$}
    \end{subfigure}
    \hfill\\
    \begin{subfigure}{0.45\textwidth}
        \centering
        \includegraphics[width=\linewidth]{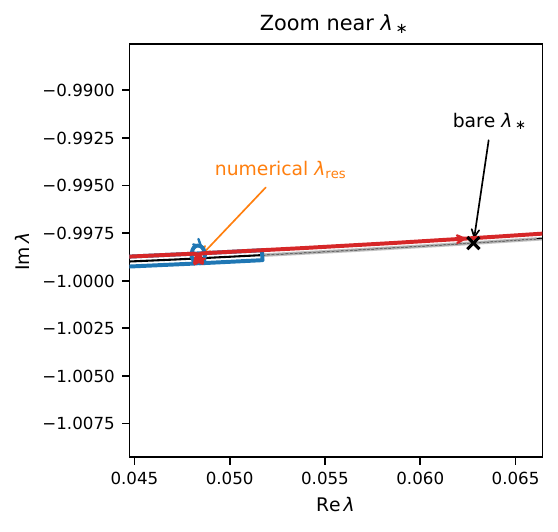}
        \caption{$\zeta=0.2$}
    \end{subfigure}
    \hfill
    \begin{subfigure}{0.45\textwidth}
        \centering
        \includegraphics[width=\linewidth]{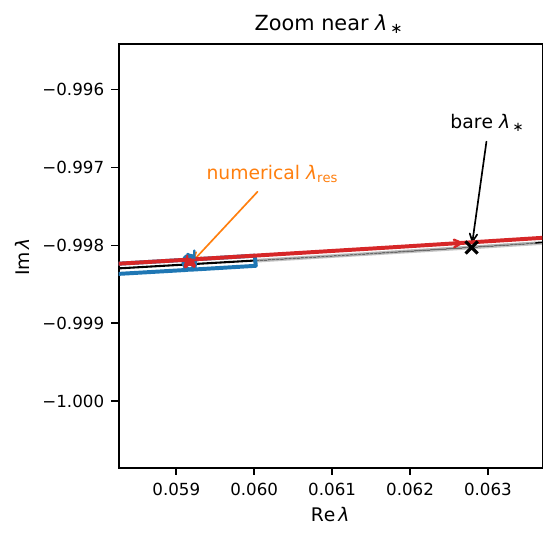}
        \caption{$\zeta=0.1$}
    \end{subfigure}
    \caption{Weak-coupling convergence of the MAS resonance phase. As the common coupling scale $\zeta$ is reduced, the numerical phase marker \(\lambda_{\rm res}\) moves toward the bare folded-branch marker $\lambda_*$. Each panel uses an adaptive zoom to keep the increasingly small displacement visible.}
    \label{fig:MASScaling}
\end{figure}

To test the finite-coupling displacement and tongue-width scalings, we repeat the MAS calculation while scaling the pump and magnetic couplings together,
\begin{equation}
    \alpha_a\rightarrow \zeta\,\alpha_a,\qquad gB\rightarrow \zeta\,gB.
\end{equation}
The $N=2$ MAS coupling is generated by one magnetic endpoint vertex and two sideband hops, so the physical tongue width decreases as \(\zeta^3\) at leading order. The diagonal self-energy shifts instead begin at order \(\zeta^2\). For the multiplier-plane plots, the numerical \(k\) scan must cover the broader phase displacement while resolving the narrower tongue. We therefore narrow the scan window with a \(\zeta^2\) envelope and retain a dense grid. Figure~\ref{fig:MASScaling} shows the zoomed multiplier-plane panels for $\zeta=1,0.5,0.2,0.1$; the numerical resonance phase approaches the bare marker as the coupling decreases.

\begin{figure}[t]
    \centering
    \includegraphics[width=0.92\textwidth]{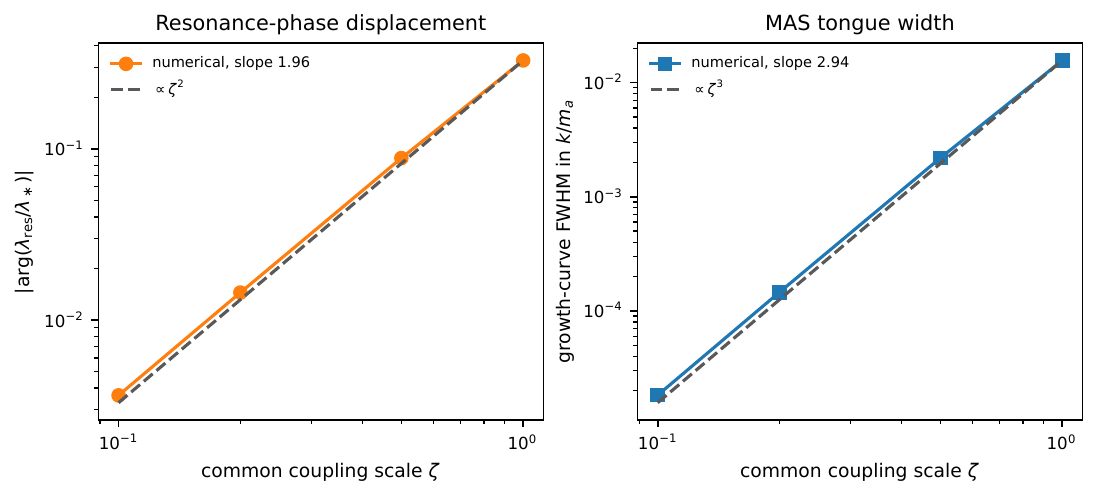}
    \caption{Weak-coupling scaling of the \(N=2\) MAS channel. Left: angular displacement of the numerical resonance phase from the bare folded marker. Right: FWHM of the momentum-space growth curve. Least-squares fits to the four displayed points give slopes \(1.96\) and \(2.94\), consistent with the leading self-energy and endpoint-coupling scalings; dashed lines show the reference powers \(\zeta^2\) and \(\zeta^3\).}
    \label{fig:MASScalingLoglog}
\end{figure}

For a quantitative comparison, define the phase displacement
\begin{equation}
    \Delta\phi_{\rm res}
    \equiv
    \left|\arg\!\left(\frac{\lambda_{\rm res}}{\lambda_*}\right)\right|
\end{equation}
and characterize the tongue width by the FWHM of the numerical growth curve. Figure~\ref{fig:MASScalingLoglog} shows both quantities on logarithmic axes. Linear least-squares fits in log-log space over the four displayed coupling scales give \(\Delta\phi_{\rm res}\propto\zeta^{1.96}\) and \(\Delta k_{\rm FWHM}\propto\zeta^{2.94}\), consistent with the expected \(\zeta^2\) diagonal shifts and \(\zeta^3\) endpoint coupling. These deterministic fits are scaling diagnostics rather than statistical measurements; their robustness is assessed below by repeating the fits on the three weak-coupling points.

Excluding the strongest-coupling point \(\zeta=1\) gives slopes \(1.98\) and \(2.97\), respectively. The approach to the expected powers therefore persists within the three-point weak-coupling subset.


%% file: sections/conclusions.tex
\section{Conclusions}
\label{sec:conclusions}

We have formulated axion electrodynamics in a coherent periodic background as a sideband problem. The pump generates a Floquet-Bloch ladder, and isolated folded degeneracies reduce to two-mode crossings classified by the conserved symplectic signature. This separates the universal local algebra from its physical interpretation, which is fixed by the evolution variable and initial or boundary data.

In temporal fixed-\(k\) evolution, the positive-positive axion-photon channel gives a stable avoided crossing, whereas the Mathieu and MAS positive-negative channels give instability tongues. Sideband-chain elimination determines their endpoint couplings and self-energy shifts, and the full monodromy calculations reproduce both the multiplier topology and the predicted weak-coupling scaling. In stationary fixed-\(\omega\) transfer, forward-forward crossings give bounded conversion while forward-backward crossings give Bragg stop bands and distributed reflection. Ray-projected forward conversion shares the compact local form but uses the effective scale \(Q_{\rm eff}\), not the true axion momentum \(Q_{\rm real}\).

The ladder, external-field diagrams, and co-rotating Bessel basis provide complementary organizations of the same periodic system: the ladder resolves virtual rungs; the diagrams display momentum bookkeeping; and the Bessel basis resums same-branch polarization dressing. The present treatment assumes prescribed backgrounds and a lossless local plasma response; ray projection also excludes cutoffs and turning points. Absorptive media require a non-Hermitian extension, while nonlocal response requires an extended dispersive formulation. Separately, backward-ray scattering with realistic boundaries requires a full profile-dependent scattering calculation.

A broader direction is to extend the sideband construction beyond axion electrodynamics. Axion couplings to fermions, including neutrinos, or to additional gauge sectors can generate periodic coefficients in the equations governing spin, flavor, and hidden-sector excitations. Linearizing the corresponding equations should again produce frequency-momentum sidebands, although the relevant conserved inner product and crossing signature must be derived for each system. The pump itself can also be generalized to other coherently oscillating dark-matter candidates. Scalar or modulus backgrounds can periodically modulate particle masses and effective couplings, while vector dark matter can introduce polarization- and direction-dependent mixing. Multicomponent backgrounds would further generate multifrequency or quasiperiodic ladders rather than the single-frequency Floquet structure studied here. A natural extension is to determine which of these systems admit compact or non-compact two-mode reductions and how their resonances manifest in temporal evolution, propagation, or scattering observables.

%% file: appendix.tex
\appendix

\input{appendices/real_block_symplectic}
\input{appendices/pump_rest_frame}

\input{appendices/mas_sideband_chain}
\input{appendices/jacobi_anger}

%% file: appendices/real_block_symplectic.tex
\section{Real-block symplectic construction}
\label{app:block}

Section~\ref{sec:evolution} established that the parent $s$-evolution system admits a real symplectic structure when $G^{\mathsf T}=-G$ and $V^{\mathsf T}=V$. This appendix provides the explicit Lagrangian and Hamiltonian constructions for the temporal and spatial real blocks.

\subsection{Temporal real block}

For the homogeneous temporal pump ($Q_{\rm p}=0$) at fixed $k$, the field perturbations are expanded in standing-wave coordinates:
\begin{align}
    \delta A_x(t,z) &= X_c(t)\cos kz+X_s(t)\sin kz,\\
    \delta A_y(t,z) &= Y_c(t)\cos kz+Y_s(t)\sin kz,\\
    \delta a(t,z)   &= \alpha_c(t)\cos kz+\alpha_s(t)\sin kz.
\end{align}
Two equivalent real three-component blocks are
\begin{equation}
    q_1=(X_c,\,Y_s,\,\alpha_c)^{\mathsf T},\qquad
    q_2=(X_s,\,-Y_c,\,\alpha_s)^{\mathsf T}.
\end{equation}
Each block satisfies a second-order system. For $q_1$, substituting into the Cartesian fixed-$k$ equations yields the three coupled equations
\begin{align}
    \ddot X_c+\omega_\gamma^2 X_c+\sigma(t)Y_s-gB\dot\alpha_c &=0,\\
    \ddot Y_s+\omega_\gamma^2 Y_s+\sigma(t)X_c &=0,\\
    \ddot\alpha_c+\omega_a^2\alpha_c+gB\dot X_c &=0,
\end{align}
with $\omega_\gamma^2\equiv k^2+\omega_{\rm pl}^2$, $\omega_a^2\equiv k^2+m_a^2$, and $\sigma(t)\equiv g k\dot{\bar a}(t)$. Collecting these as $\ddot q_1+G_R\dot q_1+V_R(t)q_1=0$ gives the explicit matrices
\begin{equation}
    G_R=
    \begin{pmatrix}
    0&0&-gB\\
    0&0&0\\
    gB&0&0
    \end{pmatrix},\qquad
    V_R(t)=
    \begin{pmatrix}
    \omega_\gamma^2 & \sigma(t) & 0\\
    \sigma(t) & \omega_\gamma^2 & 0\\
    0 & 0 & \omega_a^2
    \end{pmatrix}.
\end{equation}
The matrices satisfy $G_R^{\mathsf T}=-G_R$ and $V_R^{\mathsf T}=V_R$. A different standing-wave phase convention can reverse the sign of $\sigma(t)$ through a corresponding basis rotation. With the present \(q_1\) and $\sigma(t)=gk\dot{\bar a}$ conventions, $V_{12}=V_{21}=+\sigma(t)$. The other block $q_2$ satisfies the same matrix equation.

The real quadratic Lagrangian
\begin{equation}
    L_t=\frac12\dot q^{\mathsf T}\dot q-\frac12q^{\mathsf T}G_R\dot q-\frac12q^{\mathsf T}V_R(t)q
\end{equation}
produces the equations of motion through the Euler--Lagrange equations. The canonical momentum is
\begin{equation}
    p\equiv\frac{\partial L_t}{\partial\dot q}=\dot q+\frac12 G_R q,
\end{equation}
and the phase-space variable $X\equiv(q,p)^{\mathsf T}\in\mathbb R^6$ satisfies
\begin{equation}
    \dot X=M_t(t)X,\qquad
    M_t(t)=JK_t(t),\qquad
    K_t(t)^{\mathsf T}=K_t(t),
\end{equation}
with
\begin{equation}
    J=
    \begin{pmatrix}
    0&I_3\\
    -I_3&0
    \end{pmatrix},\qquad
    K_t(t)=
    \begin{pmatrix}
    V_R(t)-G_R^2/4 & G_R/2\\
    -G_R/2 & I_3
    \end{pmatrix}.
\end{equation}
The generator $M_t(t)$ lies in the symplectic Lie algebra $\mathfrak{sp}(6,\mathbb R)$, and the one-period map
\begin{equation}
    F_t\equiv\Phi_t(T)\in\operatorname{Sp}(6,\mathbb R)
\end{equation}
is exactly symplectic. The conserved Krein form $h(X,Y)=\mathrm i X^\dagger J Y$ evaluated on each Floquet eigenmode yields the temporal Krein signature: positive for positive-frequency branches and negative for negative-frequency branches.

\subsection{Spatial real block}

The analogous construction exists for the fixed-$\omega$ spatial transfer problem. For a static spatially periodic background $\bar a(z)=a_0\cos Qz$ and a magnetic field $\bar{\mathbf B}=B\hat x$, the field perturbations are taken as
\begin{equation}
    A_x(t,z)=A_x(z)e^{-\mathrm i\omega t},\quad
    A_y(t,z)=A_y(z)e^{-\mathrm i\omega t},\quad
    a(t,z)=a(z)e^{-\mathrm i\omega t},
\end{equation}
with $\omega$ the fixed external frequency. Defining the effective wavenumbers
\begin{equation}
    k_\gamma^2(\omega)\equiv\omega^2-\omega_{\rm pl}^2,\qquad
    k_a^2(\omega)\equiv\omega^2-m_a^2,
\end{equation}
and the spatial modulation
\begin{equation}
    \rho(z)\equiv\omega g\bar a'(z),\qquad
    \beta\equiv\omega gB,
\end{equation}
the linearized equations reduce to
\begin{align}
    A_x''+k_\gamma^2 A_x &= \mathrm i\beta\,a+\mathrm i\rho(z)A_y,\\
    A_y''+k_\gamma^2 A_y &= -\mathrm i\rho(z)A_x,\\
    a''+k_a^2 a &= -\mathrm i\beta\,A_x,
\end{align}
where the prime denotes $\mathrm d/\mathrm d z$. These equations still contain both forward and backward spatial branches; no forward-only propagation approximation has been made.

To obtain a real system, split the complex amplitudes into two equivalent
real quadrature blocks. One convenient choice is
\begin{equation}
    q(z)=
    \begin{pmatrix}q_1\\q_2\\q_3\end{pmatrix},\qquad
    A_x=q_1,\;\; A_y=-\mathrm i q_2,\;\; a=-\mathrm i q_3.
\end{equation}
Equivalently, the physical real fields in this block are
\begin{equation}
    \delta A_x(t,z)=q_1(z)\cos\omega t,\quad
    \delta A_y(t,z)=-q_2(z)\sin\omega t,\quad
    \delta a(t,z)=-q_3(z)\sin\omega t.
\end{equation}
Substituting yields the purely real second-order spatial system
\begin{equation}
    q''+V_z(z;\omega)q=0,
\end{equation}
with the symmetric matrix
\begin{equation}
    V_z(z;\omega)=
    \begin{pmatrix}
    k_\gamma^2 & -\rho(z) & -\beta\\
    -\rho(z) & k_\gamma^2 & 0\\
    -\beta & 0 & k_a^2
    \end{pmatrix}.
\end{equation}
One has $V_z^{\mathsf T}=V_z$ and $V_z(z+L)=V_z(z)$ with $L=2\pi/Q$.

The second real quadrature block follows from the complementary
components
\begin{equation}
    \tilde q(z)=
    \begin{pmatrix}\tilde q_1\\\tilde q_2\\\tilde q_3\end{pmatrix},
    \qquad
    A_x=\mathrm i\tilde q_1,\;\; A_y=\tilde q_2,\;\; a=\tilde q_3,
\end{equation}
or, in real fields,
\begin{equation}
    \delta A_x(t,z)=\tilde q_1(z)\sin\omega t,\quad
    \delta A_y(t,z)=\tilde q_2(z)\cos\omega t,\quad
    \delta a(t,z)=\tilde q_3(z)\cos\omega t.
\end{equation}
It satisfies the identical equation
\begin{equation}
    \tilde q''+V_z(z;\omega)\tilde q=0.
\end{equation}
Thus the fixed-\(\omega\) spatial problem has the same doubling as the
temporal standing-wave construction: the two real blocks are related by a
global time-quadrature shift of the monochromatic field and carry the same
one-cell transfer spectrum. In the following we write only one block.

The spatial Lagrangian is
\begin{equation}
    L_z=\frac12 q'^{\mathsf T}q'-\frac12 q^{\mathsf T}V_z(z;\omega)q,
\end{equation}
giving the canonical momentum $p=q'$. Defining $X=(q,p)^{\mathsf T}\in\mathbb R^6$, the $z$-evolution is
\begin{equation}
    \frac{\mathrm d X}{\mathrm d z}=M_z(z)X,\qquad
    M_z(z)=
    \begin{pmatrix}
    0&I_3\\
    -V_z(z)&0
    \end{pmatrix}
    =J K_z(z),\qquad
    K_z(z)=
    \begin{pmatrix}
    V_z(z)&0\\
    0&I_3
    \end{pmatrix}.
\end{equation}
Since $K_z^{\mathsf T}=K_z$, the generator satisfies $M_z^{\mathsf T}J+J M_z=0$ and $M_z(z)\in\mathfrak{sp}(6,\mathbb R)$. The one-period transfer matrix
\begin{equation}
    F_z\equiv\Phi_z(L)\in\operatorname{Sp}(6,\mathbb R)
\end{equation}
is symplectic. The conserved quantity is the spatial flux rather than the temporal Krein norm: the sign distinguishes forward from backward propagation branches.

In the decoupled limit $\rho=\beta=0$, each branch satisfies $q_j''+k_j^2 q_j=0$ with plane-wave solutions $q_j\sim e^{\pm\mathrm i k_j z}$. The corresponding phase-space vectors are $\nu_{j,\pm}=(e_j,\pm\mathrm i k_j e_j)^{\mathsf T}$. The Hermitian symplectic form $h_z(u,u)=\mathrm i u^\dagger J u$ evaluates to opposite signs for forward and backward branches,
\begin{equation}
    \operatorname{sgn}h_z(\nu_{j,+})=-\operatorname{sgn}h_z(\nu_{j,-}),
\end{equation}
and, with the conventions above, satisfies \(h_z=-2\mathcal F\). The fixed overall minus sign does not affect the definite/indefinite classification; throughout the main text the labels \((+,-)\) in the spatial problem refer to the physical flux signs.

\subsection{Variable first-derivative coupling}

If the fixed-$\omega$ system contains a spatially varying first-derivative coupling $G_z(z)q'$, as can arise from the $\partial_t\bar a\times(\nabla\times\delta\mathbf A)$ term for a general spacetime pump, the Lagrangian generalizes to
\begin{equation}
    L_z=\frac12 q'^{\mathsf T}q'-\frac12 q^{\mathsf T}G_z(z)q'-\frac12 q^{\mathsf T}V_z(z)q,
\end{equation}
with $G_z^{\mathsf T}=-G_z$ and $V_z^{\mathsf T}=V_z$. The Euler--Lagrange equation becomes
\begin{equation}
    q''+G_z(z)q'+\Bigl[V_z(z)+\frac12 G_z'(z)\Bigr]q=0.
\end{equation}
If the original equation is written as $q''+G_z(z)q'+W_z(z)q=0$, the condition for a real quadratic Lagrangian origin is
\begin{equation}
    G_z^{\mathsf T}=-G_z,\qquad
    W_z(z)-\frac12 G_z'(z)=V_z(z),\qquad
    V_z^{\mathsf T}=V_z.
\end{equation}
The antisymmetric first-derivative coupling's spatial variation contributes a $\frac12 G_z'$ correction to the effective potential. The canonical momentum generalizes to $p=q'+\frac12 G_z q$, and the transfer matrix remains symplectic under these conditions.

\subsection{Limitations}

The symplectic construction requires $\omega\in\mathbb R$ and real, lossless coefficients. If absorption or gain is present (e.g.\ $\omega_{\rm pl}^2\to\omega_{\rm pl}^2+\mathrm i\Gamma$), $V_z$ is no longer real symmetric and the transfer matrix generally leaves $\operatorname{Sp}(6,\mathbb R)$. Furthermore, the transfer matrix describes the periodic medium's internal Bloch/transfer behavior; observable reflection and transmission amplitudes require additional specification of incoming and outgoing boundary conditions.

%% file: appendices/pump_rest_frame.tex
\section{Pump-rest-frame representative}
\label{app:boost}

For a single coherent axion component, the traveling-wave pump~\eqref{eq:pump} is simplified by a Lorentz boost to the frame in which the pump phase is purely temporal. This appendix records the transformation and its physical limitations.

\subsection{Lorentz boost derivation}

The pump four-momentum is $K^\mu=(\Omega_{\rm p},Q_{\rm p})$ with $K_\mu K^\mu=m_a^2>0$, so $K^\mu$ is timelike. Define
\begin{equation}
    v\equiv\frac{Q_{\rm p}}{\Omega_{\rm p}},\qquad
    \gamma\equiv\frac{1}{\sqrt{1-v^2}}=\frac{\Omega_{\rm p}}{m_a}.
\end{equation}
A Lorentz boost along the $z$-direction with velocity $v$ gives the transformed coordinates
\begin{align}
    t'&=\gamma(t-vz)=\frac{\Omega_{\rm p}t-Q_{\rm p}z}{m_a}=\frac{\xi}{m_a},\\
    z'&=\gamma(z-vt)=\frac{\Omega_{\rm p}z-Q_{\rm p}t}{m_a}=-\frac{\eta}{m_a}.
\end{align}
In the boosted frame the pump becomes purely temporal:
\begin{equation}
    \bar a'(t',z')=a_0\cos(m_a t'+\varphi),
\end{equation}
and the Floquet-Bloch ladder simplifies to
\begin{equation}
    \omega_n'=(\nu+n)m_a,\qquad
    k_n'=\lambda_\eta m_a,
\end{equation}
where $\lambda_\eta$ is the conserved Fourier label conjugate to the continuous symmetry direction $\eta$. Thus for any single timelike pump, a representative exists in which the sideband ladder is organized by a purely temporal frequency index. This statement concerns only the pump phase and the selection rules; it establishes that the rung label $n$ can always be interpreted as a frequency sideband index.

\subsection{Boundary-value limitations}

The boost simplifies the sideband ladder but does \emph{not} identify a laboratory fixed-$\omega$ scattering problem with a pump-rest-frame fixed-$k$ temporal initial-value problem. Three reasons underlie this restriction.

First, the laboratory external magnetic field and plasma rest frame transform nontrivially under the boost. What appears as a simple static magnetic field $\bar{\mathbf B}=B\hat x$ and a stationary plasma in the laboratory frame becomes a different electromagnetic and medium configuration in the boosted frame. The equations of motion in the two frames are not identical in form unless the full system of external fields is boosted as well.

Second, realistic virialized axion dark matter consists of a superposition of many nearly degenerate timelike Fourier components, each with a slightly different four-velocity. No single global rest frame exists that simultaneously removes the spatial phase of all components.

Third, fixed-$\omega$ scattering observables depend on incoming and outgoing boundary conditions---the specification of which modes carry energy toward or away from the scattering region. These boundary conditions are frame-dependent: a monochromatic incident wave in the laboratory frame is not a monochromatic incident wave in the pump rest frame. The boost is a ladder-level simplification, not a physical equivalence of boundary-value problems.

The Lorentz boost is therefore a convenient representative for the Floquet-Bloch decomposition itself---it exhibits the ladder in its simplest form---without reducing the physical phenomenology of different $s$-evolution problems to a single global eigenvalue problem.

%% file: appendices/mas_sideband_chain.tex
\section{Sideband-chain Schur complement and channel specializations}
\label{app:MAS}
\label{app:sidebandChain}

This appendix separates the channel-independent Schur reduction from its two cross-species temporal specializations. Before the resonant endpoints are chosen, any finite virtual path through off-resonant sideband rungs has the same Schur-complement form. The endpoint frequency signs enter when the shell denominators are linearized and the retained coordinates are converted to action-normalized slow envelopes. We first carry out this step for the positive-positive axion-photon crossing, obtaining the compact avoided-crossing normal form, and then for the positive-negative MAS crossing, obtaining the Bogoliubov growth system.

We therefore begin with an abstract chain. Let \(x_L\) and \(x_R\) denote the two shells retained in the resonant block, and let \(y_1,\ldots,y_M\) denote the \(M\) off-resonant shells connecting them. The symbols \(D_L,D_R,D_j\) are the corresponding shell denominators, while \(u_L,u_R,t_j\) are nearest-neighbor vertices along the virtual path. At this stage these symbols do not specify whether the endpoint shells are axion or photon, positive or negative frequency, or same-species or cross-species.

For axion-photon applications the vertices are later identified with one magnetic bridge and a product of photon sideband hoppings. The Cartesian recursion~\eqref{eq:CartesianTemporalRec} alternates the two photon polarizations between adjacent rungs and separates into the two rung-parity sectors in Eq.~\eqref{eq:temporalParityChains}. For a resonance anchored at \(a_0\), the scalar path is
\begin{equation}
    a_0\;\longleftrightarrow\;A_{x,0}
    \;\longleftrightarrow\;A_{y,-1}
    \;\longleftrightarrow\;A_{x,-2}
    \;\longleftrightarrow\;\cdots ,
\end{equation}
with Cartesian same-rung bridge \(\mathcal G_B=gB\varepsilon_F\) and rephased real photon hopping \(\tau_R=|\epsilon|/2\). The complementary parity sector is invariant and does not enter this endpoint reduction.

\subsection{Schur chain from the temporal ladder}

\subsubsection{Minimal one-rung elimination}

We retain three amplitudes:
\begin{equation}
    \mathcal V_{\rm ch}=(x_L,\;y_1,\;x_R)^{\mathsf T}.
\end{equation}
Phase redefinitions of the chain amplitudes bring the truncated shell-space algebraic system to the real symmetric form
\begin{equation}
    \begin{pmatrix}
    D_L & u_L & 0\\
    u_L & D_1 & u_R\\
    0 & u_R & D_R
    \end{pmatrix}
    \begin{pmatrix}x_L\\y_1\\x_R\end{pmatrix}=0.
    \label{eq:MASchainTrunc}
\end{equation}
If the intermediate shell is far off shell, \(|D_1|\gg |u_L|,|u_R|\), we eliminate it algebraically:
\begin{equation}
    y_1=-\frac{u_L x_L+u_R x_R}{D_1}.
\end{equation}
Substituting back yields the effective two-endpoint system
\begin{equation}
    \mathcal M_{\rm eff}
    \binom{x_L}{x_R}=0,\qquad
    \mathcal M_{\rm eff}=
    \begin{pmatrix}
    D_L-\dfrac{u_L^2}{D_1} & -\dfrac{u_Lu_R}{D_1}\\[2ex]
    -\dfrac{u_Ru_L}{D_1} & D_R-\dfrac{u_R^2}{D_1}
    \end{pmatrix}.
    \label{eq:M2eff}
\end{equation}
The off-diagonal entry defines the algebraic chain strength
\begin{equation}
    \Xi^{(1)}=-\frac{u_Lu_R}{D_1}.
    \label{eq:Xi1}
\end{equation}
The diagonal entries receive the self-energy shifts \(u_L^2/D_1\) and \(u_R^2/D_1\). No channel classification has entered this step.

Only after \(x_L\) and \(x_R\) are identified as particular folded shells does this endpoint matrix become a compact avoided crossing or a non-compact Bogoliubov reduction. For example, choosing \(x_L=a_0\), \(y_1=A_{x,0}\), and \(x_R=A_{y,-1}\), with \(u_L=-\mathcal G_B\) and \(u_R=-\tau_R\), gives the positive-positive \(a\)-\(\gamma\) endpoint derivation used in Section~\ref{sec:floquet}. Reversing the frequency sign of the right endpoint gives the MAS endpoint algebra before canonical projection.

\subsubsection{General \texorpdfstring{$M$}{M}-shell Schur complement}

For a longer virtual chain, the endpoint amplitudes \(x_L,x_R\) couple through the off-resonant rungs
\begin{equation}
    y_1,\ y_2,\ \ldots,\ y_M,
\end{equation}
so the full virtual path is
\begin{equation}
    x_L\;\leftrightarrow\;y_1\;\leftrightarrow\;y_2\;\leftrightarrow\;\cdots\;\leftrightarrow\;y_M\;\leftrightarrow\;x_R.
\end{equation}
Here \(M\) is the positive number of eliminated shells, introduced separately from the signed harmonic difference \(N\) used in the main text. This convention reproduces the minimal case above when \(M=1\).

Let \(H_M\) denote the \(M\times M\) internal chain,
\begin{equation}
    (H_M)_{jj}=D_j,\qquad
    (H_M)_{j,j+1}=(H_M)_{j+1,j}=t_j,
    \qquad j=1,\ldots,M-1 .
\end{equation}
Thus \(t_j\) is the hopping between the neighboring eliminated rungs \(y_j\) and \(y_{j+1}\). The endpoint couplings \(u_L\) and \(u_R\) connect \(x_L\) to \(y_1\) and \(y_M\) to \(x_R\), respectively. With the amplitudes ordered as \((x_L,x_R;y_1,\ldots,y_M)\), the full block matrix is
\begin{equation}
    \mathcal M_{\rm ch}^{(M)}
    =
    \begin{pmatrix}
    A & C\\
    C^{\mathsf T} & H_M
    \end{pmatrix},
    \qquad
    A=
    \begin{pmatrix}
    D_L & 0\\
    0 & D_R
    \end{pmatrix},
    \qquad
    C=
    \begin{pmatrix}
    u_L & 0 & \cdots & 0\\
    0 & \cdots & 0 & u_R
    \end{pmatrix}.
\end{equation}
Here \(C\) is a \(2\times M\) endpoint-to-internal coupling block; equivalently, \(C_{11}=u_L\), \(C_{2M}=u_R\), and all other entries vanish.
The determinant \(\mathcal D_M=\det H_M\) obeys
\begin{equation}
    \mathcal D_0=1,\qquad
    \mathcal D_1=D_1,
\end{equation}
with the continuant recurrence
\begin{equation}
    \mathcal D_\ell
    =D_\ell\,\mathcal D_{\ell-1}-t_{\ell-1}^2\mathcal D_{\ell-2},
    \qquad \ell\ge2.
    \label{eq:continuantDN}
\end{equation}
Provided every eliminated rung remains far off shell and no internal continuant nearly vanishes, Schur-complement elimination gives the effective two-endpoint shell matrix
\begin{equation}
    \mathcal M_{\rm eff}^{(M)}(\varepsilon_F)
    \binom{x_L}{x_R}=0,
    \qquad
    \mathcal M_{\rm eff}^{(M)}=
    A-C H_M^{-1}C^{\mathsf T}
    =
    \begin{pmatrix}
    D_L-\Sigma_L^{(M)} & \Xi^{(M)}\\[1ex]
    \tilde\Xi^{(M)} & D_R-\Sigma_R^{(M)}
    \end{pmatrix},
\end{equation}
where
\begin{equation}
    \begin{pmatrix}
    \Sigma_L^{(M)} & -\Xi^{(M)}\\
    -\tilde\Xi^{(M)} & \Sigma_R^{(M)}
    \end{pmatrix}
    =
    C\,H_M^{-1}C^{\mathsf T}.
\end{equation}
The endpoint coupling is therefore
\begin{equation}
    \Xi^{(M)}
    =e^{\mathrm i\phi_M}
    \frac{u_L\,t_1t_2\cdots t_{M-1}\,u_R}{\mathcal D_M},
    \label{eq:XiGen}
\end{equation}
where the product over \(t_j\) is absent for \(M=1\), reproducing Eq.~\eqref{eq:Xi1}, and the phase \(\phi_M\) absorbs sign and sideband phase conventions. In the large-detuning limit this reduces to the product-denominator estimate
\begin{equation}
    \Xi^{(M)}
    \simeq e^{\mathrm i\phi_M}
    \frac{u_L\,t_1t_2\cdots t_{M-1}\,u_R}{D_1D_2\cdots D_M}.
    \label{eq:XiProd}
\end{equation}
For an \(a\)-\(\gamma\) path of signed harmonic difference \(N\), the number of eliminated photon shells is \(M=|N|\). For the orientation \(N>0\), rephase the selected parity chain so that \(u_L=\mathcal G_B\), \(u_R=\tau_R\), and \(t_j=\tau_R\), with \(D_j=D_\gamma(-(j-1))\). The phase of the original directed hoppings is then carried by \(\phi_M\), not by the diagonal self-energies.

The main text denotes the photon factor multiplying \(\mathcal G_B\) by \(W_N\). Thus \(\Xi_{a\gamma}^{(+,-;N)}=\mathcal G_B W_N\) for the MAS channel, while \(\Xi_{a\gamma}^{(+,+;N)}=\mathcal G_B W_N\) for the same-sign axion-photon channel.

For \(M=|N|=1\), the formula gives \(|\Xi|\sim |\mathcal G_B|\tau_R/|D_\gamma(0)|\). This result fixes the indexing and avoids an extra denominator from double-counting the first off-resonant rung.

\subsection{Positive-positive axion-photon specialization}
\label{subsec:PPChainApp}

Consider first the nearest-rung orientation \(N=+1\). The retained positive-frequency endpoints are \(a_0\) and \(A_{y,-1}\), while \(A_{x,0}\) is off resonance:
\begin{equation}
    D_a(0)\simeq0,\qquad
    D_\gamma(-1)\simeq0,\qquad
    |D_\gamma(0)|\gg|\mathcal G_B|,\tau_R.
\end{equation}
After rephasing the selected parity chain, its shell-space equation is
\begin{equation}
    \begin{pmatrix}
    D_a(0) & -\mathcal G_B & 0\\
    -\mathcal G_B & D_\gamma(0) & -\tau_R\\
    0 & -\tau_R & D_\gamma(-1)
    \end{pmatrix}
    \begin{pmatrix}a_0\\A_{x,0}\\A_{y,-1}\end{pmatrix}=0 .
    \label{eq:PPchainTrunc}
\end{equation}
Here \(\mathcal G_B\) is the Cartesian same-rung magnetic bridge and \(\tau_R\) is the real symmetric hopping obtained after rephasing the selected parity sector. The complementary sector does not enter this Schur complement.

The intermediate amplitude is
\begin{equation}
    A_{x,0}
    =
    \frac{\mathcal G_B a_0+\tau_R A_{y,-1}}{D_\gamma(0)} .
\end{equation}
Substitution into the endpoint equations gives
\begin{equation}
    \begin{pmatrix}
    D_a(0)-\dfrac{\mathcal G_B^2}{D_\gamma(0)}
    &
    -\dfrac{\mathcal G_B\tau_R}{D_\gamma(0)}
    \\[2ex]
    -\dfrac{\tau_R\mathcal G_B}{D_\gamma(0)}
    &
    D_\gamma(-1)-\dfrac{\tau_R^2}{D_\gamma(0)}
    \end{pmatrix}
    \binom{a_0}{A_{y,-1}}=0 .
    \label{eq:PPchainEff}
\end{equation}
Thus, in this phase convention,
\begin{equation}
    \Xi_{a\gamma}^{(+,+;1)}
    =-\frac{\mathcal G_B\tau_R}{D_\gamma(0)},\qquad
    \Sigma_a^{(+,+;1)}
    =\frac{\mathcal G_B^2}{D_\gamma(0)},\qquad
    \Sigma_\gamma^{(+,+;1)}
    =\frac{\tau_R^2}{D_\gamma(0)} .
    \label{eq:PPAppShifts}
\end{equation}
For \(|N|>1\), Eq.~\eqref{eq:XiGen} replaces the single denominator by the corresponding continuant; reversing the rung orientation gives the \(N<0\) chain.

The channel-specific distinction appears when the two endpoint shells are linearized. For the positive-frequency endpoints,
\begin{align}
    D_a(0)
    &\simeq
    -2\omega_a(\varepsilon_F-\omega_a),\\
    D_\gamma(-1)
    &\simeq
    -2\omega_\gamma(\varepsilon_F-m_a-\omega_\gamma).
    \label{eq:PPPositiveSlopes}
\end{align}
Both shell slopes are negative. With the diagonal entries written as \(D_a-\Sigma_a\) and \(D_\gamma-\Sigma_\gamma\), the shifted folded frequencies are therefore
\begin{equation}
    \varepsilon_a^{\rm eff}
    \simeq
    \omega_a-\frac{\Sigma_a}{2\omega_a},
    \qquad
    \varepsilon_\gamma^{\rm eff}
    \simeq
    m_a+\omega_\gamma-\frac{\Sigma_\gamma}{2\omega_\gamma},
\end{equation}
where the self-energies are evaluated at the bare crossing to the order retained. The effective positive-positive detuning is
\begin{equation}
    \Delta_{a\gamma,{\rm eff}}^{(+,+;1)}
    =
    \omega_a-\omega_\gamma-m_a
    -\frac{\Sigma_a}{2\omega_a}
    +\frac{\Sigma_\gamma}{2\omega_\gamma}.
    \label{eq:PPEffDetuningApp}
\end{equation}

Both endpoint slopes have the same sign. Dividing by their magnitudes and action-normalizing the oscillator coordinates gives
\begin{equation}
    \mu_{a\gamma}^{(+,+;1)}
    =
    e^{\mathrm i\varphi_1}
    \frac{\Xi_{a\gamma}^{(+,+;1)}}{2\sqrt{\omega_a\omega_\gamma}} .
    \label{eq:PPCanonicalProj}
\end{equation}
The resulting slow-envelope generator is Hermitian,
\begin{equation}
    \mathrm i\frac{\mathrm d}{\mathrm dt}
    \binom{c_a}{c_\gamma}
    =
    \begin{pmatrix}
    \Delta_{a\gamma,{\rm eff}}^{(+,+;1)}/2
    & \mu_{a\gamma}^{(+,+;1)}\\
    \mu_{a\gamma}^{(+,+;1)*}
    & -\Delta_{a\gamma,{\rm eff}}^{(+,+;1)}/2
    \end{pmatrix}
    \binom{c_a}{c_\gamma},
    \label{eq:PPCompactApp}
\end{equation}
and preserves \(|c_a|^2+|c_\gamma|^2\). Its eigenvalue shifts are real,
\begin{equation}
    \delta\varepsilon_\pm
    =
    \pm\sqrt{
    |\mu_{a\gamma}^{(+,+;1)}|^2
    +
    \bigl(\Delta_{a\gamma,{\rm eff}}^{(+,+;1)}/2\bigr)^2},
\end{equation}
so the Schur-induced endpoint coupling opens an avoided crossing rather than an instability.

\subsection{MAS specialization}

The MAS sum-frequency channel uses the same Schur chain but changes the frequency sign of the photon endpoint. The algebraic path is still
\begin{equation}
    a_0\;\leftrightarrow\;A_{x,0}
    \;\leftrightarrow\;A_{y,-1}
    \;\leftrightarrow\;A_{x,-2}
    \;\leftrightarrow\;\cdots ,
\end{equation}
but its rung-\(-N\) endpoint is now the folded negative-frequency photon branch. The Schur complement supplies the same shell-space coupling
\begin{equation}
    \Xi_{a\gamma}^{(+,-;N)}=\mathcal G_B W_N,
\end{equation}
and the same algebraic form of the diagonal self-energy shifts. Unlike Eq.~\eqref{eq:PPPositiveSlopes}, however, the negative-frequency photon shell has the opposite slope. The MAS-specific step is the projection of this endpoint matrix onto a positive-frequency axion amplitude and the conjugate negative-frequency photon amplitude.

The diagonal shifts enter the MAS Floquet multiplier phase through the endpoint entries
\begin{equation}
    D_a-\Sigma_a,\qquad D_\gamma(-N)-\Sigma_\gamma,
\end{equation}
and expanding near the folded positive-negative crossing,
\begin{align}
    D_a&=\omega_a^2-\varepsilon_F^2
        \simeq -2\omega_a(\varepsilon_F-\omega_a),\\
    D_\gamma(-N)&=\omega_\gamma^2-(\varepsilon_F-Nm_a)^2
        \simeq 2\omega_\gamma(\varepsilon_F-Nm_a+\omega_\gamma),
\end{align}
so the two endpoint slopes have opposite signs. Their product is therefore negative, in contrast to the positive product in Eq.~\eqref{eq:PPPositiveSlopes}; this is the shell-space origin of the indefinite signature that appears after canonical projection. Solving the shifted endpoint conditions gives
\begin{equation}
    \varepsilon_a^{\rm eff}
    \simeq
    \omega_a-\frac{\Sigma_a}{2\omega_a},
    \qquad
    \varepsilon_\gamma^{\rm eff}
    \simeq
    Nm_a-\omega_\gamma+\frac{\Sigma_\gamma}{2\omega_\gamma}.
    \label{eq:EffQuasiFreq}
\end{equation}
With the self-energies defined by the diagonal entries \(D_a-\Sigma_a\) and \(D_\gamma-\Sigma_\gamma\), the effective MAS detuning is therefore
\begin{equation}
    \Delta_{a\gamma,{\rm eff}}^{(+,-;N)}
    =
    \varepsilon_a^{\rm eff}-\varepsilon_\gamma^{\rm eff}
    =
    \omega_a+\omega_\gamma-Nm_a
    -\frac{\Sigma_a}{2\omega_a}
    -\frac{\Sigma_\gamma}{2\omega_\gamma},
    \label{eq:EffDetuningSigma}
\end{equation}
The finite-coupling collision phase follows from the center quasi-frequency
\begin{equation}
    \varepsilon_*^{\rm eff}
    \simeq
    \frac{\varepsilon_a^{\rm eff}+\varepsilon_\gamma^{\rm eff}}{2},
    \qquad
    \lambda_*^{\rm eff}=\exp(-\mathrm i\varepsilon_*^{\rm eff}T).
    \label{eq:EffCollisionPhase}
\end{equation}
In the weak-coupling limit $\Sigma_a,\Sigma_\gamma\to0$, this reduces to the bare folded-branch phase used in the leading-order selection rule.

\subsubsection{Explicit \texorpdfstring{$N=2$}{N=2} MAS chain}

The numerical MAS example in Section~\ref{sec:numerics} uses the first kinematically accessible vacuum/plasma channel, \(N=2\). After rephasing the selected parity-chain amplitudes, the relevant four-shell chain has the symmetric tridiagonal form
\begin{equation}
    \mathcal V_{N=2}
    =(a_0,\;A_{x,0},\;A_{y,-1},\;A_{x,-2})^{\mathsf T},
\end{equation}
with algebraic matrix
\begin{equation}
    \mathcal M_{N=2}=
    \begin{pmatrix}
    D_a(0) & -\mathcal G_B & 0 & 0\\
    -\mathcal G_B & D_\gamma(0) & -\tau_R & 0\\
    0 & -\tau_R & D_\gamma(-1) & -\tau_R\\
    0 & 0 & -\tau_R & D_\gamma(-2)
    \end{pmatrix}.
    \label{eq:MASN2Matrix}
\end{equation}
The original directed-hopping phases have been absorbed into the rung amplitudes. The remaining \(\tau_R\) is real and nonnegative; the overall endpoint-coupling phase still depends on the sideband convention, while the diagonal shifts are invariant.

The internal block to be eliminated is
\begin{equation}
    H_2=
    \begin{pmatrix}
    D_\gamma(0) & -\tau_R\\
    -\tau_R & D_\gamma(-1)
    \end{pmatrix},
    \qquad
    \mathcal D_2=\det H_2
    =D_\gamma(0)D_\gamma(-1)-\tau_R^2 .
    \label{eq:D2explicit}
\end{equation}
Its inverse is
\begin{equation}
    H_2^{-1}
    =
    \frac{1}{\mathcal D_2}
    \begin{pmatrix}
    D_\gamma(-1) & \tau_R\\
    \tau_R & D_\gamma(0)
    \end{pmatrix}.
\end{equation}
Taking the two endpoints to be \(a_0\) and \(A_{x,-2}\), the endpoint--internal coupling matrix is
\begin{equation}
    C=
    \begin{pmatrix}
    -\mathcal G_B & 0\\
    0 & -\tau_R
    \end{pmatrix}.
\end{equation}
The Schur complement gives
\begin{equation}
    \mathcal M_{\rm eff}^{(2)}
    =
    \begin{pmatrix}
    D_a(0) & 0\\
    0 & D_\gamma(-2)
    \end{pmatrix}
    -C H_2^{-1}C^{\mathsf T},
\end{equation}
or explicitly
\begin{equation}
    \mathcal M_{\rm eff}^{(2)}
    =
    \begin{pmatrix}
    D_a(0)-\dfrac{\mathcal G_B^2D_\gamma(-1)}{\mathcal D_2}
    &
    -\dfrac{\mathcal G_B\tau_R^2}{\mathcal D_2}
    \\[2ex]
    -\dfrac{\mathcal G_B\tau_R^2}{\mathcal D_2}
    &
    D_\gamma(-2)-\dfrac{\tau_R^2D_\gamma(0)}{\mathcal D_2}
    \end{pmatrix}.
    \label{eq:MASN2Meff}
\end{equation}
Thus, in this phase convention,
\begin{align}
    \Xi_{a\gamma}^{(+,-;2)}
    &=-\,\frac{\mathcal G_B\tau_R^2}{D_\gamma(0)D_\gamma(-1)-\tau_R^2},
    \label{eq:XiN2explicit}\\
    \Sigma_a^{(2)}
    &=\frac{\mathcal G_B^2D_\gamma(-1)}
           {D_\gamma(0)D_\gamma(-1)-\tau_R^2},
    \label{eq:SigmaaN2}\\
    \Sigma_\gamma^{(2)}
    &=\frac{\tau_R^2D_\gamma(0)}
           {D_\gamma(0)D_\gamma(-1)-\tau_R^2}.
    \label{eq:SigmagN2}
\end{align}
In the large-detuning limit, \(\tau_R^2\ll|D_\gamma(0)D_\gamma(-1)|\), these expressions reduce to
\begin{equation}
    \Xi_{a\gamma}^{(+,-;2)}
    \simeq
    -\,\frac{\mathcal G_B\tau_R^2}{D_\gamma(0)D_\gamma(-1)},
    \qquad
    \Sigma_a^{(2)}\simeq\frac{\mathcal G_B^2}{D_\gamma(0)},
    \qquad
    \Sigma_\gamma^{(2)}\simeq\frac{\tau_R^2}{D_\gamma(-1)}.
    \label{eq:N2LargeDetuning}
\end{equation}
After canonical projection,
\begin{equation}
    \mu_{a\gamma}^{(+,-;2)}
    =
    e^{\mathrm i\varphi_2}
    \frac{\Xi_{a\gamma}^{(+,-;2)}}{2\sqrt{\omega_a\omega_\gamma}},
    \label{eq:muN2explicit}
\end{equation}
with an unobservable phase \(\varphi_2\).

Equations~\eqref{eq:XiN2explicit}--\eqref{eq:muN2explicit} also explain the coupling scaling used in Fig.~\ref{fig:MASScaling}. Since \(\mathcal G_B\propto gB\) and \(\tau_R\propto|\epsilon|\propto|\alpha_a|\), a common rescaling
\begin{equation}
    \alpha_a\rightarrow\zeta\alpha_a,
    \qquad
    gB\rightarrow\zeta gB
\end{equation}
gives \(|\mu_{a\gamma}^{(+,-;2)}|\propto\zeta^3\) at leading order. By contrast, the diagonal self-energy shifts begin at order \(\Sigma_a^{(2)}\sim\mathcal G_B^2/D_\gamma(0)\propto\zeta^2\) and \(\Sigma_\gamma^{(2)}\sim\tau_R^2/D_\gamma(-1)\propto\zeta^2\). The width of the instability tongue is therefore controlled by the \(\zeta^3\) endpoint coupling, while the finite-coupling displacement of the folded multiplier phase can decrease only as \(\zeta^2\) in the same weak-coupling limit.

\subsubsection{MAS canonical projection and growth rate}

The Schur complement above is an algebraic statement about the second-order shell equations. After the off-resonant rungs have been eliminated, the retained endpoints obey a two-coordinate equation of the form
\begin{align}
    \bigl(\partial_t^2+\omega_a^2+\cdots\bigr)q_a
    +\Xi_{a\gamma}^{(+,-;N)}e^{-\mathrm iNm_at}q_\gamma&=0,\\
    \bigl(\partial_t^2+\omega_\gamma^2+\cdots\bigr)q_\gamma
    +\tilde\Xi_{a\gamma}^{(+,-;N)}e^{+\mathrm iNm_at}q_a&=0,
\end{align}
where the omitted diagonal terms are the same self-energy shifts that enter
\(\Delta_{a\gamma,{\rm eff}}^{(+,-;N)}\). This equation is still written in coordinate normalization: \(\Xi\) has dimensions of frequency squared. The canonical projection used in the main text is the conversion from these coordinates to action-normalized oscillator amplitudes, followed by the resonant rotating-wave truncation.

For an endpoint oscillator, write
\begin{equation}
    q_j(t)=\frac{1}{\sqrt{2\omega_j}}
           \bigl(c_j(t)e^{-\mathrm i\omega_j t}+c_j^*(t)e^{+\mathrm i\omega_j t}\bigr),
    \qquad |\dot c_j|\ll\omega_j|c_j|.
\end{equation}
The second-order operator then reduces to
\begin{equation}
    \Bigl(\frac{\mathrm d^2}{\mathrm d t^2}+\omega_j^2\Bigr)q_j
    \simeq\frac{-2\mathrm i\omega_j\dot c_j}{\sqrt{2\omega_j}}e^{-\mathrm i\omega_j t}
    +\text{c.c.}
\end{equation}

For the MAS sum-frequency channel, the pump phase connecting the endpoints makes the resonant part of the \(a\) equation proportional to the negative-frequency component of the photon endpoint:
\begin{equation}
    \Xi_{a\gamma}^{(+,-;N)}e^{-\mathrm i N m_a t}q_\gamma
    \supset
    \frac{\Xi_{a\gamma}^{(+,-;N)}c_\gamma^*}{\sqrt{2\omega_\gamma}}
    e^{-\mathrm i(N m_a-\omega_\gamma)t}.
\end{equation}
This term is slow when \(\omega_a+\omega_\gamma\simeq Nm_a\). Matching the coefficient of \(e^{-\mathrm i\omega_at}\) gives, with the factor of \(-\mathrm i\) absorbed into the convention-dependent phase,
\begin{equation}
    \dot c_a
    =
    e^{\mathrm i\varphi_N}
    \frac{\Xi_{a\gamma}^{(+,-;N)}}{2\sqrt{\omega_a\omega_\gamma}}\,
    c_\gamma^*
\end{equation}
in the reduced rotating-wave system. The conjugate endpoint equation gives the corresponding evolution of \(c_\gamma^*\). Thus the action-normalized envelope coupling is
\begin{equation}
    \mu_{a\gamma}^{(+,-;N)}
    =
    e^{\mathrm i\varphi_N}
    \frac{\Xi_{a\gamma}^{(+,-;N)}}{2\sqrt{\omega_a\omega_\gamma}},
    \label{eq:CanonicalProj}
\end{equation}
where \(\varphi_N\) is fixed by the positive/negative-frequency basis convention and has no invariant meaning for the growth exponent. This is the normalization factor within the two-endpoint rotating-wave reduction; it remains subject to the prior finite-chain truncation and the neglect of nonresonant fast terms.

Including the effective detuning by distributing it symmetrically between the two slow phases gives the local Bogoliubov system in the same convention as Eq.~\eqref{eq:BogoliubovTemporal}:
\begin{equation}
    \frac{\mathrm d}{\mathrm dt}
    \begin{pmatrix}c_a\\ c_\gamma^*\end{pmatrix}
    =
    \begin{pmatrix}
    -\mathrm i\Delta_{a\gamma,{\rm eff}}^{(+,-;N)}/2
    & \mu_{a\gamma}^{(+,-;N)}\\
    \mu_{a\gamma}^{(+,-;N)*}
    & +\mathrm i\Delta_{a\gamma,{\rm eff}}^{(+,-;N)}/2
    \end{pmatrix}
    \begin{pmatrix}c_a\\ c_\gamma^*\end{pmatrix}.
    \label{eq:MASBogoliubovApp}
\end{equation}
Its envelope growth exponents are
\begin{equation}
    s_{a\gamma,\pm}^{(N)}
    =
    \pm
    \sqrt{\bigl|\mu_{a\gamma}^{(+,-;N)}\bigr|^2
    -\bigl(\Delta_{a\gamma,{\rm eff}}^{(+,-;N)}/2\bigr)^2}.
\end{equation}
The physical growth rate is
\begin{equation}
    \gamma_{a\gamma}^{(N)}=\sqrt{\bigl|\mu_{a\gamma}^{(+,-;N)}\bigr|^2
                              -\bigl(\Delta_{a\gamma,{\rm eff}}^{(+,-;N)}/2\bigr)^2},
\end{equation}
when the expression under the square root is positive.
The detuning in this formula is the effective detuning of
Eq.~\eqref{eq:EffDetuningSigma}. If the self-energy shifts are neglected,
it reduces to
\begin{equation}
    \Delta_{a\gamma}^{(+,-;N)}
    \equiv
    \omega_a+\omega_\gamma-N m_a .
\end{equation}
The minimal \(N=1\) large-detuning estimate follows by inserting
Eq.~\eqref{eq:XiProd},
\begin{equation}
    \mu_{a\gamma}^{(+,-;1)}
    \simeq
    e^{\mathrm i\varphi_1}
    \frac{(gB\,\varepsilon_F)\,|\epsilon|}
    {4D_\gamma(0)\sqrt{\omega_a\omega_\gamma}},
\end{equation}
where the phase \(\varphi_1\) includes the original directed-hopping and rung-orientation conventions. This last expression is an estimate for the off-resonant chain strength; the normalization relation~\eqref{eq:CanonicalProj} and the Bogoliubov growth formula above are the reduced two-mode result.

\subsubsection{MAS tongue half-width in momentum space}

For phenomenological estimates, the momentum-space half-width of the instability tongue is useful. If \(k_*\) satisfies the resonance condition \(\Delta_{a\gamma,{\rm eff}}^{(+,-;N)}(k_*)=0\), expand the effective detuning linearly:
\begin{equation}
    \Delta_{a\gamma,{\rm eff}}^{(+,-;N)}(k)\simeq
    \Delta_{a\gamma,{\rm eff}}^{(+,-;N)\prime}(k_*)\,(k-k_*).
\end{equation}
If the self-energy shifts are neglected in this derivative, then
\begin{equation}
    \Delta_{a\gamma,{\rm eff}}^{(+,-;N)\prime}(k_*)
    \simeq
    \frac{\mathrm d\omega_a}{\mathrm d k}
    +\frac{\mathrm d\omega_\gamma}{\mathrm d k}
    =\frac{k}{\omega_a}+\frac{k}{\omega_\gamma}.
\end{equation}
The instability condition
\(|\Delta_{a\gamma,{\rm eff}}^{(+,-;N)}|<2|\mu_{a\gamma}^{(+,-;N)}|\)
then gives \(|k-k_*|<\delta k_{1/2}\), where
\begin{equation}
    \delta k_{1/2}
    \simeq
    \frac{2\bigl|\mu_{a\gamma}^{(+,-;N)}(k_*)\bigr|}
         {\bigl|\Delta_{a\gamma,{\rm eff}}^{(+,-;N)\prime}(k_*)\bigr|}
    \simeq
    \frac{2\bigl|\mu_{a\gamma}^{(+,-;N)}(k_*)\bigr|}
         {\bigl|k_*/\omega_a+k_*/\omega_\gamma\bigr|}.
    \label{eq:MASTongueWidth}
\end{equation}
The full width is \(\Delta k_{\rm full}\simeq2\delta k_{1/2}\). The final form is therefore a leading-order half-width estimate: the exact two-mode instability condition uses the effective detuning, while the displayed denominator uses the bare dispersion slope.

%% file: appendices/jacobi_anger.tex
\section{Co-rotating reorganization of the temporal sidebands}
\label{app:JacobiAnger}

The temporal ladder of Section~\ref{sec:floquet} works directly with the Fourier coefficients of the second-order field equations. A complementary organization proceeds in two steps: first, it rewrites the same equations as a doubled first-order system that retains both frequency signs; second, it absorbs the polarization rotation within each sign sector into a co-rotating basis. The resulting magnetic vertices contain trigonometric functions of a sinusoidal phase and can be expanded in Bessel harmonics. This appendix develops that construction, shows how eliminating non-resonant Bessel harmonics generates diagonal self-energy shifts, and identifies the approximations used when the system is reduced to a single resonant harmonic.

\subsection{Doubled first-order temporal system}

Write the Cartesian photon amplitude as
\(\mathbf A=(A_x,A_y)^{\mathsf T}\), define
\begin{equation}
    \sigma_2=
    \begin{pmatrix}
    0&-\mathrm i\\
    \mathrm i&0
    \end{pmatrix},
    \qquad
    \mathbf e_x=
    \begin{pmatrix}1\\0\end{pmatrix},
\end{equation}
and use \(\sigma(t)=-\epsilon\sin(m_at)\) from Section~\ref{sec:floquet}. Equations~\eqref{eq:complexFixedKx}--\eqref{eq:complexFixedKy} become
\begin{equation}
    \ddot{\mathbf A}+\omega_\gamma^2\mathbf A
    =
    \sigma(t)\sigma_2\mathbf A
    +gB\,\dot a\,\mathbf e_x .
    \label{eq:CartesianPhotonVector}
\end{equation}
Introduce positive- and negative-frequency coordinates for the photon and axion oscillators,
\begin{align}
    \mathbf c_{\gamma,\pm}
    &\equiv
    \sqrt{\frac{\omega_\gamma}{2}}\,\mathbf A
    \mathbin{\pm}
    \frac{\mathrm i}{\sqrt{2\omega_\gamma}}\,\dot{\mathbf A},
    \label{eq:PhotonNambuCoordinates}\\
    c_{a,\pm}
    &\equiv
    \sqrt{\frac{\omega_a}{2}}\,a
    \mathbin{\pm}
    \frac{\mathrm i}{\sqrt{2\omega_a}}\,\dot a .
    \label{eq:AxionNambuCoordinates}
\end{align}
The transformation is invertible: the original phase-space variables are recovered as
\begin{equation}
    \mathbf A=
    \frac{\mathbf c_{\gamma,+}+\mathbf c_{\gamma,-}}
         {\sqrt{2\omega_\gamma}},
    \qquad
    \dot{\mathbf A}
    =-\mathrm i\sqrt{\frac{\omega_\gamma}{2}}
    \bigl(\mathbf c_{\gamma,+}-\mathbf c_{\gamma,-}\bigr).
\end{equation}
Thus \((\mathbf c_{\gamma,+},\mathbf c_{\gamma,-})\) repackages
\((\mathbf A,\dot{\mathbf A})\) as frequency-sign coordinates rather than introducing additional photon degrees of freedom.

To display the pump structure, temporarily omit the magnetic terms and define
\begin{equation}
    \delta_\gamma(t)\equiv\frac{\sigma(t)}{2\omega_\gamma}
    =-\alpha_\gamma m_a\sin(m_at),
    \qquad
    \alpha_\gamma\equiv
    \frac{\epsilon}{2\omega_\gamma m_a}
    =\frac{\alpha_a k}{2\omega_\gamma}.
    \label{eq:alphaDef}
\end{equation}
Differentiating Eq.~\eqref{eq:PhotonNambuCoordinates} and using Eq.~\eqref{eq:CartesianPhotonVector} to replace \(\ddot{\mathbf A}\) turns the single second-order vector equation into the coupled first-order system
\begin{equation}
    \mathrm i\partial_t
    \begin{pmatrix}
    \mathbf c_{\gamma,+}\\
    \mathbf c_{\gamma,-}
    \end{pmatrix}
    =
    \begin{pmatrix}
    \omega_\gamma\mathbb{I}_2-\delta_\gamma\sigma_2
    &
    -\delta_\gamma\sigma_2\\
    \delta_\gamma\sigma_2
    &
    -\omega_\gamma\mathbb{I}_2+\delta_\gamma\sigma_2
    \end{pmatrix}
    \begin{pmatrix}
    \mathbf c_{\gamma,+}\\
    \mathbf c_{\gamma,-}
    \end{pmatrix}.
    \label{eq:PhotonNambuSystem}
\end{equation}
The diagonal blocks rotate the Cartesian polarization within a fixed frequency-sign sector. The off-diagonal blocks mix the positive- and negative-frequency sectors and are the first-order representation of the photon parametric channel.

\subsection{Co-rotating Cartesian frame}

Set
\begin{equation}
    \theta(t)\equiv\alpha_\gamma\cos(m_at),
    \qquad
    \begin{pmatrix}
    \mathbf c_{\gamma,+}\\
    \mathbf c_{\gamma,-}
    \end{pmatrix}
    =
    \mathcal U_\gamma(t)
    \begin{pmatrix}
    \boldsymbol\phi_{\gamma,+}\\
    \boldsymbol\phi_{\gamma,-}
    \end{pmatrix},
    \qquad
    \mathcal U_\gamma(t)=
    \begin{pmatrix}
    e^{+\mathrm i\theta\sigma_2}&0\\
    0&e^{-\mathrm i\theta\sigma_2}
    \end{pmatrix}.
    \label{eq:TemporalCoRotatingTransform}
\end{equation}
Here \(\dot\theta=\delta_\gamma\). The derivative term from the transformation therefore removes the polarization rotation in the two diagonal blocks. The photon system becomes
\begin{equation}
    \mathrm i\partial_t
    \begin{pmatrix}
    \boldsymbol\phi_{\gamma,+}\\
    \boldsymbol\phi_{\gamma,-}
    \end{pmatrix}
    =
    \begin{pmatrix}
    \omega_\gamma\mathbb{I}_2
    &
    -\delta_\gamma\sigma_2e^{-2\mathrm i\theta\sigma_2}\\
    \delta_\gamma\sigma_2e^{+2\mathrm i\theta\sigma_2}
    &
    -\omega_\gamma\mathbb{I}_2
    \end{pmatrix}
    \begin{pmatrix}
    \boldsymbol\phi_{\gamma,+}\\
    \boldsymbol\phi_{\gamma,-}
    \end{pmatrix}.
    \label{eq:CoRotatedPhotonNambuSystem}
\end{equation}
Thus the co-rotating transformation resums the repeated polarization rotation within each frequency branch, while the branch-changing pump terms remain explicit. Ref.~\cite{Yao:2026yez} first used a WKB reduction to obtain a forward first-order propagation equation; the backward or negative-frequency sector is absent from that reduced system. Applying the analogous approximation here amounts to dropping the off-diagonal blocks in Eq.~\eqref{eq:CoRotatedPhotonNambuSystem}. Those blocks must instead be retained when analyzing temporal Mathieu channels or their corrections to a cross-species resonance.

Restore the external magnetic field and label the photon frequency branch by
\(\varsigma=\pm1\), with \(\mathbf c_{\gamma,\varsigma}=\mathbf c_{\gamma,+}\) for \(\varsigma=+1\) and \(\mathbf c_{\gamma,\varsigma}=\mathbf c_{\gamma,-}\) for \(\varsigma=-1\). Before the co-rotating transformation, the magnetic contribution is
\begin{equation}
    \left.\mathrm i\dot{\mathbf c}_{\gamma,\varsigma}\right|_B
    =
    \varsigma\,\mathrm i\,\frac{gB}{2}
    \sqrt{\frac{\omega_a}{\omega_\gamma}}\,
    \bigl(c_{a,+}-c_{a,-}\bigr)\mathbf e_x,
    \qquad \varsigma=\pm1 .
    \label{eq:MagneticNambuVertex}
\end{equation}
The inverse rotation acting on this source replaces \(\mathbf e_x\) by
\begin{equation}
    \mathbf d_\varsigma(t)
    \equiv
    e^{-\mathrm i\varsigma\theta(t)\sigma_2}\mathbf e_x
    =
    \begin{pmatrix}
        \cos\theta(t)\\
        \varsigma\sin\theta(t)
    \end{pmatrix}.
    \label{eq:RotatedMagneticBranchDirection}
\end{equation}
Thus the co-rotating magnetic vertex has a time-dependent direction
\(\mathbf d_\varsigma(t)\) in the Cartesian polarization plane.

\subsection{Jacobi--Anger harmonics and parity}

Using
\begin{equation}
    e^{\mathrm i z\cos\varphi}
    =
    \sum_{n\in\mathbb Z}
    \mathrm i^nJ_n(z)e^{\mathrm in\varphi},
\end{equation}
expand the rotated magnetic direction as
\begin{align}
    \mathbf d_\varsigma(t)
    &=
    \sum_{n\in\mathbb Z}
    \boldsymbol{\mathcal C}_n^{(\varsigma)}
    e^{\mathrm in m_at},
    \nonumber\\
    \mathcal C_{n,x}^{(\varsigma)}
    &=
    \frac{1+(-1)^n}{2}\,
    \mathrm i^nJ_n(\alpha_\gamma),
    \qquad
    \mathcal C_{n,y}^{(\varsigma)}
    =
    \varsigma\,\frac{1-(-1)^n}{2}\,
    \mathrm i^{n-1}J_n(\alpha_\gamma),
    \label{eq:CoRotatingVectorCoefficient}
\end{align}
where \(\mathcal C_{n,r}^{(\varsigma)}
=\mathbf e_r^{\mathsf T}\boldsymbol{\mathcal C}_n^{(\varsigma)}\).
The \(x\) component carries even harmonics and the \(y\) component carries odd harmonics. This is the co-rotating representation of the Cartesian rung-parity sectors in Eq.~\eqref{eq:temporalParityChains}: a path anchored at \(a_0\) reaches \(A_x\) at even rung separation and \(A_y\) at odd rung separation.

The residual branch-changing blocks in Eq.~\eqref{eq:CoRotatedPhotonNambuSystem} contain
\(e^{\pm2\mathrm i\theta\sigma_2}\). In the circular polarization basis, where \(\sigma_2\) is diagonal, these factors become
\(e^{\pm2\mathrm ih\alpha_\gamma\cos(m_at)}\) for \(h=\pm1\). Their Fourier coefficients involve \(J_n(2\alpha_\gamma)\), together with the adjacent-order shifts generated by the prefactor \(\delta_\gamma(t)\propto\sin(m_at)\). The circular basis therefore makes the phase modulation of each polarization transparent, whereas the Cartesian co-rotating basis displays directly how even and odd harmonics connect to the two polarization coefficients used in the ladder.

For \(|\alpha_\gamma|\ll1\), the Bessel coefficients may be expanded as
\begin{equation}
    J_n(\alpha_\gamma)
    \simeq
    \frac{1}{n!}
    \left(\frac{\alpha_\gamma}{2}\right)^{\!n},
    \qquad n\ge0,
    \label{eq:BesselSmall}
\end{equation}
with negative orders fixed by \(J_{-n}(z)=(-1)^nJ_n(z)\). This is the approximation that turns the resummed harmonic weight into an explicit power series in the pump amplitude.

\subsection{Relation to the ladder and resonant reduction}

After moving to the interaction picture of the free frequencies, the \(N\)th harmonic of the rotated magnetic vertex couples a positive-frequency axion branch to a photon branch of sign \(\varsigma=\pm1\) with detuning
\begin{equation}
    \Delta_{a\gamma}^{(+,\varsigma;N)}
    =
    \omega_a-\varsigma\omega_\gamma-Nm_a .
    \label{eq:CoRotatingDetuning}
\end{equation}
Thus \(\varsigma=+1\) gives the same-sign difference-frequency channel, while \(\varsigma=-1\) gives the MAS sum-frequency channel. Keeping only the nearly stationary harmonic and its two resonant endpoint amplitudes is the isolated-channel rotating-wave approximation. For an opposite-sign pair, the action-normalized endpoint equations have the Bogoliubov form of Eq.~\eqref{eq:BogoliubovTemporal}; for a same-sign pair, they have the compact form of Eq.~\eqref{eq:SU2temporal}.

The resonant endpoint polarization is \(r_N=x\) for even \(N\) and
\(r_N=y\) for odd \(N\). The corresponding Fourier component satisfies
\(\lvert\mathcal C_{N,r_N}^{(\varsigma)}\rvert
=\lvert J_N(\alpha_\gamma)\rvert\).
Using the shell normalization of Section~\ref{sec:floquet}, the co-rotating isolated-harmonic contribution is
\begin{align}
    \Xi_{a\gamma,\mathrm{co}}^{(+,\varsigma;N)}
    &=
    e^{\mathrm i\chi_{\varsigma N}}\,
    gB\,\omega_a\,\mathcal C_{N,r_N}^{(\varsigma)},
    \label{eq:XiCoRotating}\\
    \mu_{a\gamma,\mathrm{co}}^{(+,\varsigma;N)}
    &=
    e^{\mathrm i\varphi_{\varsigma N}}\,
    \frac{\Xi_{a\gamma,\mathrm{co}}^{(+,\varsigma;N)}}
         {2\sqrt{\omega_a\omega_\gamma}}
    =
    e^{\mathrm i\widetilde\varphi_{\varsigma N}}\,
    \frac{gB}{2}
    \sqrt{\frac{\omega_a}{\omega_\gamma}}\,
    \mathcal C_{N,r_N}^{(\varsigma)} .
    \label{eq:MuCoRotating}
\end{align}
Here the shell bridge has been evaluated at the bare positive-frequency axion endpoint, \(\varepsilon_F\simeq\omega_a\), and the displayed phases absorb the Nambu, rung, and endpoint conventions.

\paragraph{Comparison of the two representations.}
Before truncation, Eq.~\eqref{eq:TemporalCoRotatingTransform} is a periodic invertible change of variables, so the two representations have the same temporal Floquet spectrum. The ladder basis keeps nearest-neighbor pump hops, individual shell denominators, and virtual paths explicit. It is therefore well suited to low-order crossings, controlled finite-rung truncations, diagonal self-energy shifts, and comparison with the full monodromy matrix, but an order-\(N\) endpoint coupling appears as a product of \(N\) pump insertions and intermediate denominators. The co-rotating basis performs the complementary reorganization: it resums the same-branch polarization dressing into \(J_N(\alpha_\gamma)\), giving the compact all-order endpoint estimate~\eqref{eq:MuCoRotating} and making several sideband orders easy to compare. The transformed Floquet operator then couples multiple Bessel harmonics, while the positive-negative-frequency pump block remains separate. Finite-coupling corrections follow by retaining and eliminating the non-resonant Bessel harmonics.

\subsubsection{Bessel-dressed Schur complement and diagonal shifts}

The action-normalized magnetic vertex carried by the \(\ell\)th harmonic into polarization \(r=x,y\) is
\begin{equation}
    \mu_{\ell,r,\mathrm{co}}^{(\varsigma)}
    \equiv
    e^{\mathrm i\widetilde\varphi_{\varsigma\ell r}}\,
    \frac{gB}{2}
    \sqrt{\frac{\omega_a}{\omega_\gamma}}\,
    \mathcal C_{\ell,r}^{(\varsigma)} .
    \label{eq:CoRotatingHarmonicVertex}
\end{equation}
Equation~\eqref{eq:MuCoRotating} is the component
\(\mu_{N,r_N,\mathrm{co}}^{(\varsigma)}\) selected for the target channel.

Let \(P\) project onto the two resonant endpoints and \(Q=1-P\) onto all other frequency signs, polarizations, and harmonics. For the co-rotating Floquet operator \(\mathcal L_{\mathrm{co}}(\varepsilon_F)\), eliminating the \(Q\) sector gives
\begin{equation}
    \mathcal L_{\mathrm{eff}}^{\mathrm{co}}
    =
    P\mathcal L_{\mathrm{co}}P
    -
    P\mathcal L_{\mathrm{co}}Q
    \bigl(Q\mathcal L_{\mathrm{co}}Q\bigr)^{-1}
    Q\mathcal L_{\mathrm{co}}P .
    \label{eq:CoRotatingSchur}
\end{equation}
This is the same block-elimination identity as Eq.~\eqref{eq:localSchur}; only the basis used to organize the intermediate states has changed.

As an explicit example, consider the positive-positive channel and temporarily retain only the positive-frequency magnetic vertices. The unperturbed axion endpoint has quasi-energy \(E_{a,0}=\omega_a\), while the \(\ell\)th positive-frequency photon replica has
\begin{equation}
    E_{\gamma,\ell}=\omega_\gamma+\ell m_a,
    \qquad
    \Delta_\ell^{(+,+)}
    \equiv
    E_{a,0}-E_{\gamma,\ell}
    =
    \omega_a-\omega_\gamma-\ell m_a .
\end{equation}
For target order \(N\), the pair \(\{a_0,\gamma_N\}\) belongs to \(P\). To second order in the magnetic vertices, Eq.~\eqref{eq:CoRotatingSchur} gives
\begin{align}
    \delta\varepsilon_{a,\mathrm{co}}^{(2)}
    &=
    \sum_{\ell\neq N}
    \sum_{r=x,y}
    \frac{\bigl|\mu_{\ell,r,\mathrm{co}}^{(+)}\bigr|^2}
         {\Delta_\ell^{(+,+)}},
    \label{eq:CoRotatingAxionShift}\\
    \delta\varepsilon_{\gamma,N,\mathrm{co}}^{(2)}
    &=
    -\sum_{\ell\neq N}
    \frac{\bigl|\mu_{\ell,r_N,\mathrm{co}}^{(+)}\bigr|^2}
         {\Delta_\ell^{(+,+)}} .
    \label{eq:CoRotatingPhotonShift}
\end{align}
The second line views the retained photon replica as coupling to translated axion replicas, whose denominators are \(-\Delta_\ell^{(+,+)}\).

The axion endpoint couples to alternating \(x\)- and \(y\)-polarized replicas, whereas the retained photon endpoint has the fixed polarization \(r_N\). Its numerator therefore vanishes unless \(\ell\) has the same parity as \(N\). If another denominator becomes small, that harmonic must be moved into \(P\). The polarization-summed weight of each eliminated harmonic is
\begin{equation}
    \sum_{r=x,y}
    \bigl|\mu_{\ell,r,\mathrm{co}}^{(+)}\bigr|^2
    =
    \frac{(gB)^2}{4}
    \frac{\omega_a}{\omega_\gamma}
    \bigl|J_\ell(\alpha_\gamma)\bigr|^2 .
    \label{eq:BesselWeightInSelfEnergy}
\end{equation}
Thus the target Bessel coefficient enters the off-diagonal endpoint coupling linearly, while the eliminated harmonics enter the diagonal shifts quadratically and are weighted by their detunings.

For the positive-positive shell convention of Appendix~\ref{subsec:PPChainApp}, the corresponding shell-space self-energies are
\begin{equation}
    \Sigma_{a,\mathrm{co}}^{(+,+;N)}
    =
    -2\omega_a\,
    \delta\varepsilon_{a,\mathrm{co}}^{(2)},
    \qquad
    \Sigma_{\gamma,\mathrm{co}}^{(+,+;N)}
    =
    -2\omega_\gamma\,
    \delta\varepsilon_{\gamma,N,\mathrm{co}}^{(2)} .
    \label{eq:CoRotatingShellSelfEnergies}
\end{equation}
These signs follow from the negative slopes of both positive-frequency shell denominators. A single-harmonic rotating-wave truncation discards all \(\ell\neq N\) before forming Eq.~\eqref{eq:CoRotatingSchur}; it therefore retains the resonant coupling but sets these off-resonant self-energy shifts to zero.

For the MAS channel, the magnetic-only contribution has the analogous structure
\begin{align}
    \delta\varepsilon_{a,B}^{(+,-;N)}
    &=
    \sum_{\ell\neq N}
    \sum_{r=x,y}
    \frac{\bigl|\mu_{\ell,r,\mathrm{co}}^{(-)}\bigr|^2}
         {\Delta_{a\gamma}^{(+,-;\ell)}},
    \label{eq:CoRotatingMASMagneticShift}\\
    \delta\varepsilon_{\gamma,N,B}^{(+,-;N)}
    &=
    -\sum_{\ell\neq N}
    \frac{\bigl|\mu_{\ell,r_N,\mathrm{co}}^{(-)}\bigr|^2}
         {\Delta_{a\gamma}^{(+,-;\ell)}},
    \nonumber\\
    \Delta_{a\gamma}^{(+,-;\ell)}
    &=
    \omega_a+\omega_\gamma-\ell m_a .
    \nonumber
\end{align}
Converting these quasi-frequency shifts to the shell convention gives
\begin{align}
    \Sigma_{a,\mathrm{co}}^{(+,-;N)}
    &=
    -2\omega_a\delta\varepsilon_{a,B}^{(+,-;N)},\\
    \Sigma_{\gamma,\mathrm{co}}^{(+,-;N)}
    &=
    +2\omega_\gamma\delta\varepsilon_{\gamma,N,B}^{(+,-;N)} .
\end{align}
The second sign differs from Eq.~\eqref{eq:CoRotatingShellSelfEnergies} because the negative-frequency photon shell has the opposite slope. A complete MAS reduction must also retain the branch-changing blocks of Eq.~\eqref{eq:CoRotatedPhotonNambuSystem} inside \(Q\mathcal L_{\mathrm{co}}Q\). Their Fourier coefficients involve \(J_q(2\alpha_\gamma)\) and the adjacent orders generated by \(\delta_\gamma(t)\). Equations~\eqref{eq:CoRotatingMASMagneticShift} give only the Bessel-dressed magnetic contribution; Appendix~\ref{app:MAS} gives the complete sparse-ladder endpoint coupling and diagonal shifts.

The weak-modulation overlap provides a direct perturbative check. For \(N\ge0\),
\(J_N(\alpha_\gamma)\simeq(\alpha_\gamma/2)^N/N!\).
At a same-sign difference-frequency crossing, the intermediate ladder denominators contain \(2\omega_\gamma+\ell m_a\). When \(|N|m_a\ll\omega_\gamma\), replacing these factors by \(2\omega_\gamma\) reduces the shortest path in Eq.~\eqref{eq:XiProd} to the same leading Bessel coefficient; the higher terms in \(J_N\) resum paths with additional backtracking hops. Outside this overlap, a finite ladder truncation retains channel-dependent denominators that are not contained in the simple Bessel asymptotic. The two formulas provide alternative representations of the same untruncated Floquet problem.